\documentclass[12pt]{article}
\usepackage{amsmath}

\allowdisplaybreaks
\renewcommand{\theequation}{\arabic{section}.\arabic{equation}}

\begin{document}

\author{C. Bizdadea\thanks{%
E-mail address: bizdadea@central.ucv.ro}, E. M. Cioroianu\thanks{%
E-mail address: manache@central.ucv.ro}, I.
Negru,\\
S. O. Saliu\thanks{%
E-mail address: osaliu@central.ucv.ro},
S. C. S\u{a}raru\thanks{%
E-mail address: scsararu@central.ucv.ro} \\
Faculty of Physics, University of Craiova,\\
13 A. I. Cuza Str., Craiova 200585, Romania}
\title{On the generalized Freedman-Townsend model}
\date{}
\maketitle

\begin{abstract}
Consistent interactions that can be added to a free, Abelian gauge theory
comprising a finite collection of BF models and a finite set of two-form
gauge fields (with the Lagrangian action written in first-order form as a
sum of Abelian Freedman-Townsend models) are constructed from the
deformation of the solution to the master equation based on specific
cohomological techniques. Under the hypotheses of smoothness in the coupling
constant, locality, Lorentz covariance, and Poincar\'{e} invariance of the
interactions, supplemented with the requirement on the preservation of the
number of derivatives on each field with respect to the free theory, we
obtain that the deformation procedure modifies the Lagrangian action, the
gauge transformations as well as the accompanying algebra. The interacting
Lagrangian action contains a generalized version of non-Abelian
Freedman-Townsend model. The consistency of interactions to all orders in
the coupling constant unfolds certain equations, which are shown to have
solutions.

\end{abstract}

\section{Introduction}

The power of the BRST formalism was strongly increased by its cohomological
development, which allowed, among others, a useful investigation of many
interesting aspects related to the perturbative renormalization problem~\cite%
{4a,4b,4c,4d,5}, anomaly-tracking mechanism~\cite{5,6a,6b,6c,6d,6e},
simultaneous study of local and rigid invariances of a given theory~\cite{7}
as well as to the reformulation of the construction of consistent
interactions in gauge theories~\cite{7a1,7a2,7a3,7a4,7a5} in terms of the
deformation theory~\cite{8b1,8b2,8b3} or, actually, in terms of the
deformation of the solution to the master equation~\cite{def,contempmath}.

The scope of this paper is to investigate the consistent interactions that
can be added to a free, Abelian gauge theory consisting of a finite
collection of BF models and a finite set of two-form gauge fields (described
by a sum of Abelian Freedman-Townsend actions). Each BF model from the
collection comprises a scalar field, a two-form and two sorts of one-forms.
We work under the hypotheses that the interactions are smooth in the
coupling constant, local, Lorentz covariant, and Poincar\'{e} invariant,
supplemented with the requirement on the preservation of the number of
derivatives on each field with respect to the free theory. Under these
hypotheses, we obtain the most general form of the theory that describes the
cross-couplings between a collection of BF models and a set of two-form
gauge fields. The resulting interacting model is accurately formulated in
terms of a gauge theory with gauge transformations that close according to
an open algebra (the commutators among the deformed gauge transformations
only close on the stationary surface of deformed field equations).

Topological BF models ~\cite{birmingham91} are important in view of the fact
that certain interacting, non-Abelian versions are related to a Poisson
structure algebra~\cite{stroblspec} present in various versions of Poisson
sigma models~\cite%
{psmikeda94,psmstrobl95,psmstroblCQG961,psmstroblCQG962,psmstrobl97,psmcattaneo2000,psmcattaneo2001}%
, which are known to be useful at the study of two-dimensional gravity~\cite%
{grav2teit83,grav2jackiw85,grav2katanaev86,grav2brown88,grav2katanaev90,grav2schmidt,grav2solod,grav2ikedaizawa90,grav2strobl94,grav2grumvassil02}
(for a detailed approach, see~\cite{grav2strobl00}). It is well known that
pure three-dimensional gravity is just a BF theory. Moreover, in higher
dimensions general relativity and supergravity in Ashtekar formalism may
also be formulated as topological BF theories with some extra constraints
\cite{ezawa,freidel,smolin,ling}. Due to these results, it is important to
know the self-interactions in BF theories as well as the couplings between
BF models and other theories. This problem has been considered in literature
in relation with self-interactions in various classes of BF models~\cite%
{defBFizawa2000,defBFmpla,defBFikeda00,defBFikeda01,defBFijmpa,defBFjhep,defBFijmpajuvi06}
and couplings to matter fields~\cite{defBFepjc} and vector fields~\cite%
{defBFikeda03,defBFijmpajuvi04} by using the powerful BRST cohomological
reformulation of the problem of constructing consistent interactions. Other
aspects concerning interacting, topological BF models can be found in~\cite%
{otherBFikeda02,otherBFikedaizawa04,otherBFikeda06}. On the other hand,
models with $p$-form gauge fields play an important role in string and
superstring theory as well as in supergravity. Based on these
considerations, the study of interactions between BF models and two-forms
appears as a topic that might enlighten certain aspects in both gravity and
supergravity theories.

Our strategy goes as follows. Initially, we determine in Section
\ref{free} the antifield-BRST symmetry of the free model, which
splits as the sum between the Koszul-Tate differential and the
exterior derivative along the gauge orbits, $s=\delta +\gamma $.
Then, in Section \ref{defrev} we briefly present the reformulation
of the problem of constructing consistent interactions in gauge
field theories in terms of the deformation of the solution to the
master equation. Next, in Section \ref{int} we determine the
consistent deformations of the solution to the master equation for
the model under consideration. The first-order deformation belongs
to the local cohomology $H^{0}(s\vert d)$, where $d$ is the exterior
spacetime derivative. The computation of the cohomological space
$H^{0}(s\vert d)$ proceeds by expanding the co-cycles according to
the antighost number and further using the cohomological groups
$H(\gamma )$ and $H(\delta \vert d)$. We find that the first-order
deformation is parameterized by $11$ types of smooth functions
of the undifferentiated scalar fields, which become restricted to fulfill $%
19 $ kinds of equations in order to produce a deformation that is
consistent to all orders in the coupling constant. With the help of
these equations we show that the remaining deformations, of orders
$2$ and higher, can be taken to vanish. The
identification of the interacting model is developed in Section \ref{lagint}%
. All the interaction vertices are derivative-free. Among the
cross-couplings between the collection of BF models and the set of two-form
gauge fields we find a generalized version of non-Abelian Freedman-Townsend
vertex. (By `generalized' we mean that its form is identical with the
standard non-Abelian Freedman-Townsend vertex up to the point that the
structure constants of a Lie algebra are replaced here with some functions
depending on the undifferentiated scalar fields from the BF sector.)
Meanwhile, both the gauge transformations corresponding to the coupled model
and their algebra are deformed with respect to the initial Abelian theory in
such a way that the new gauge algebra becomes open and the reducibility
relations only close on-shell (on the stationary surface of deformed field
equations). It is interesting to mention that by contrast to the standard
non-Abelian Freedman-Townsend model, where the auxiliary vector fields are
gauge-invariant, here these fields gain nonvanishing gauge transformations,
proportional with some BF gauge parameters. In the end of Section \ref%
{lagint} we comment on several classes of solutions to the equations
satisfied by the various functions of the scalar fields that parameterize
the deformed solution to the master equation. Section \ref{concl} closes the
paper with the main conclusions. The present paper also contains 4
appendices, in which various notations used in the main body of the paper as
well as some formulas concerning the gauge structure of the interacting
model are listed.

\section{Free model: Lagrangian formulation and BRST symmetry\label{free}}

The starting point is given by a free theory in four spacetime dimensions
that describes a finite collection of BF models and a finite set of two-form
gauge fields, with the Lagrangian action
\begin{eqnarray}
S_{0}[A_{\mu }^{a},H_{\mu }^{a},\varphi _{a},B_{a}^{\mu \nu },V_{\mu \nu
}^{A},V_{\mu }^{A}] &=&\int d^{4}x\left( H_{\mu }^{a}\partial ^{\mu }\varphi
_{a}+\tfrac{1}{2}B_{a}^{\mu \nu }\partial _{\lbrack \mu }^{\left. {}\right.
}A_{\nu ]}^{a}\right.  \notag \\
&&\left. +\tfrac{1}{2}V_{A}^{\mu \nu }F_{\mu \nu }^{A}+\tfrac{1}{2}V_{\mu
}^{A}V_{A}^{\mu }\right) .  \label{bfa1}
\end{eqnarray}%
Each of the BF models from the collection (to be indexed by lower case
letters $a$, $b$, etc.) comprises a scalar field $\varphi _{a}$, two kinds
of one-forms $A_{\mu }^{a}$ and $H_{\mu }^{a}$, and a two-form $B_{a}^{\mu
\nu }$. The action for the set of Abelian two-forms decomposes as a sum of
individual two-form actions, indexed via capital Latin letters ($A$, $B$,
etc.). Each two-form action is written in first-order form as an Abelian
Freedman-Townsend action, in terms of a two-form $V_{A}^{\mu \nu }$ and of
an auxiliary vector $V_{\mu }^{A}$, with the Abelian field strength $F_{\mu
\nu }^{A}=\partial _{\lbrack \mu }^{\left. {}\right. }V_{\nu ]}^{A}$. The
collection indices from the two-form sector are lowered with the
(non-degenerate) metric $k_{AB}$ induced by the Lagrangian density $\tfrac{1%
}{2}\left( V_{A}^{\mu \nu }F_{\mu \nu }^{A}+V_{\mu }^{A}V_{A}^{\mu }\right) $
from (\ref{bfa1}) (i.e. $F_{A}^{\mu \nu }=k_{AB}F^{B\mu \nu }$) and are
raised with its inverse, of elements $k^{AB}$. Of course, we consider the
general situation, where the two types of collection indexes run
independently one from each other. Everywhere in this paper the notation $%
\left[ \mu \ldots \nu \right] $ signifies complete antisymmetry with respect
to the (Lorentz) indices between brackets, with the conventions that the
minimum number of terms is always used and the result is never divided by
the number of terms. Action (\ref{bfa1}) is found invariant under the gauge
transformations
\begin{gather}
\delta _{\epsilon }A_{\mu }^{a}=\partial _{\mu }\epsilon ^{a},\quad \delta
_{\epsilon }H_{\mu }^{a}=-2\partial ^{\nu }\epsilon _{\nu \mu }^{a},\quad
\delta _{\epsilon }\varphi _{a}=0,  \label{bfa2} \\
\delta _{\epsilon }B_{a}^{\mu \nu }=-3\partial _{\rho }\epsilon _{a}^{\rho
\mu \nu },\quad \delta _{\epsilon }V_{\mu \nu }^{A}=\varepsilon _{\mu \nu
\rho \lambda }\partial ^{\rho }\epsilon ^{A\lambda },\quad \delta _{\epsilon
}V_{\mu }^{A}=0,  \label{bfa2i}
\end{gather}%
where all the gauge parameters are bosonic, with $\epsilon _{\mu \nu }^{a}$
and $\epsilon _{a}^{\mu \nu \rho }$ completely antisymmetric. It is easy to
see that the above gauge transformations are Abelian and off-shell
(everywhere in the space of field histories, not only on the stationary
surface of field equations for (\ref{bfa1})), second-order reducible.
Indeed, related to the first-order reducibility, we observe that if we make
the transformations $\epsilon _{\mu \nu }^{a}(\theta )=-3\partial ^{\lambda
}\theta _{\lambda \mu \nu }^{a}$, $\epsilon _{a}^{\mu \nu \rho }(\theta
)=-4\partial _{\lambda }\theta _{a}^{\lambda \mu \nu \rho }$, $\epsilon
^{A\lambda }(\theta )=\partial ^{\lambda }\theta ^{A}$, with $\theta $s
arbitrary, bosonic functions, completely antisymmetric (where applicable) in
their Lorentz indices, then the corresponding gauge transformations
identically vanish, $\delta _{\epsilon (\theta )}H_{\mu }^{a}=0$, $\delta
_{\epsilon (\theta )}B_{a}^{\mu \nu }=0$, $\delta _{\epsilon (\theta
)}V_{\mu \nu }^{A}=0$. The last two transformation laws of the gauge
parameters can be further annihilated by trivial transformations only: $%
\epsilon _{a}^{\mu \nu \rho }(\theta )=0$ if and only if $\theta
_{a}^{\lambda \mu \nu \rho }=0$ and $\epsilon ^{A\lambda }(\theta )=0$ if
and only if $\theta ^{A}=0$, so there is no higher-order reducibility
associated with them. By contrast, the first one can be made to vanish
strongly via the transformation $\theta _{\lambda \mu \nu }^{a}(\omega
)=-4\partial ^{\alpha }\omega _{\alpha \lambda \mu \nu }^{a}$, with $\omega
_{\alpha \lambda \mu \nu }^{a}$ an arbitrary, completely antisymmetric,
bosonic function (which indeed produces $\epsilon _{\mu \nu }^{a}(\theta
\left( \omega \right) )=0$), but there is no nontrivial transformation of $%
\omega _{\alpha \lambda \mu \nu }^{a}$ such that $\theta _{\lambda
\mu \nu }^{a}$ becomes zero. Thus, the reducibility of
(\ref{bfa2})--(\ref{bfa2i}) stops at order $2$ and holds off-shell.

In order to construct the BRST symmetry of this free\ theory, we introduce
the field/ghost and antifield spectra
\begin{gather}
\Phi ^{\alpha _{0}}=\left( A_{\mu }^{a},H_{\mu }^{a},\varphi _{a},B_{a}^{\mu
\nu },V_{\mu \nu }^{A},V_{\mu }^{A}\right) ,  \label{bfa6} \\
\Phi _{\alpha _{0}}^{\ast }=\left( A_{a}^{\ast \mu },H_{a}^{\ast \mu
},\varphi ^{\ast a},B_{\mu \nu }^{\ast a},V_{A}^{\ast \mu \nu },V_{A}^{\ast
\mu }\right) ,  \label{bfa6a} \\
\eta ^{\alpha _{1}}=\left( \eta ^{a},C_{\mu \nu }^{a},\eta _{a}^{\mu \nu
\rho },C_{\mu }^{A}\right) ,  \label{bfa7} \\
\eta _{\alpha _{1}}^{\ast }=\left( \eta _{a}^{\ast },C_{a}^{\ast \mu \nu
},\eta _{\mu \nu \rho }^{\ast a},C_{A}^{\ast \mu }\right) ,  \label{bfa7a} \\
\eta ^{\alpha _{2}}=\left( C_{\mu \nu \rho }^{a},\eta _{a}^{\mu \nu \rho
\lambda },C^{A}\right) ,\quad \eta _{\alpha _{2}}^{\ast }=\left( C_{a}^{\ast
\mu \nu \rho },\eta _{\mu \nu \rho \lambda }^{\ast a},C_{A}^{\ast }\right) ,
\label{bfa8} \\
\eta ^{\alpha _{3}}=\left( C_{\mu \nu \rho \lambda }^{a}\right) ,\quad \eta
_{\alpha _{3}}^{\ast }=\left( C_{a}^{\ast \mu \nu \rho \lambda }\right) .
\label{bfa9}
\end{gather}%
The fermionic ghosts $\eta ^{\alpha _{1}}$ respectively correspond to the
bosonic gauge parameters $\epsilon ^{\alpha _{1}}=\left( \epsilon
^{a},\epsilon _{\mu \nu }^{a},\epsilon _{a}^{\mu \nu \rho },\epsilon _{\mu
}^{A}\right) $, the bosonic ghosts for ghosts $\eta ^{\alpha _{2}}$ are due
to the first-order reducibility relations (the $\theta $-parameters from the
previous transformations), while the fermionic ghosts for ghosts for ghosts $%
\eta ^{\alpha _{3}}$ are required by the second-order reducibility relations
(the $\omega $-function from the above). The star variables represent the
antifields of the corresponding fields/ghosts. (Their Grassmann parities are
respectively opposite to those of the associated fields/ghosts, in agreement
with the general rules of the antifield-BRST method.)

Since both the gauge generators and the reducibility functions are
field-independent, it follows that the BRST differential reduces to
\begin{equation}
s=\delta +\gamma ,  \label{desc}
\end{equation}%
where $\delta $ is the Koszul-Tate differential and $\gamma $ denotes the
exterior longitudinal derivative. The Koszul-Tate differential is graded in
terms of the antighost number ($\mathrm{agh}$, $\mathrm{agh}\left( \delta
\right) =-1$) and enforces a resolution of the algebra of smooth functions
defined on the stationary surface of field equations for action (\ref{bfa1}%
), $C^{\infty }\left( \Sigma \right) $, $\Sigma :\delta S_{0}/\delta \Phi
^{\alpha _{0}}=0$. The exterior longitudinal derivative is graded in terms
of the pure ghost number ($\mathrm{pgh}$, $\mathrm{pgh}\left( \gamma \right)
=1$) and is correlated with the original gauge symmetry via its cohomology
at pure ghost number $0$ computed in $C^{\infty }\left( \Sigma \right) $,
which is isomorphic to the algebra of physical observables for the free
theory. These two degrees do not interfere ($\mathrm{agh}\left( \gamma
\right) =0$, $\mathrm{pgh}\left( \delta \right) =0$). The pure ghost number
and antighost number of BRST generators (\ref{bfa6})--(\ref{bfa9}) are
valued as follows:
\begin{gather}
\mathrm{pgh}\left( \Phi ^{\alpha _{0}}\right) =0,\quad \mathrm{pgh}\left(
\eta ^{\alpha _{1}}\right) =1,\quad \mathrm{pgh}\left( \eta ^{\alpha
_{2}}\right) =2,\quad \mathrm{pgh}\left( \eta ^{\alpha _{3}}\right) =3,
\label{bfa10} \\
\mathrm{pgh}\left( \Phi _{\alpha _{0}}^{\ast }\right) =\mathrm{pgh}\left(
\eta _{\alpha _{1}}^{\ast }\right) =\mathrm{pgh}\left( \eta _{\alpha
_{2}}^{\ast }\right) =\mathrm{pgh}\left( \eta _{\alpha _{3}}^{\ast }\right)
=0,  \label{bfa11} \\
\mathrm{agh}\left( \Phi ^{\alpha _{0}}\right) =\mathrm{agh}\left( \eta
^{\alpha _{1}}\right) =\mathrm{agh}\left( \eta ^{\alpha _{2}}\right) =%
\mathrm{agh}\left( \eta ^{\alpha _{3}}\right) =0,  \label{bfa12} \\
\mathrm{agh}\left( \Phi _{\alpha _{0}}^{\ast }\right) =1,\quad \mathrm{agh}%
\left( \eta _{\alpha _{1}}^{\ast }\right) =2,\quad \mathrm{agh}\left( \eta
_{\alpha _{2}}^{\ast }\right) =3,\quad \mathrm{agh}\left( \eta _{\alpha
_{3}}^{\ast }\right) =4,  \label{bfa13}
\end{gather}%
where the actions of $\delta $ and $\gamma $ on them read as
\begin{gather}
\delta \Phi ^{\alpha _{0}}=\delta \eta ^{\alpha _{1}}=\delta \eta ^{\alpha
_{2}}=\delta \eta ^{\alpha _{3}}=0,  \label{bfa15} \\
\delta A_{a}^{\ast \mu }=-\partial _{\nu }B_{a}^{\mu \nu },\quad \delta
H_{a}^{\ast \mu }=-\partial ^{\mu }\varphi _{a},\quad \delta \varphi ^{\ast
a}=\partial ^{\mu }H_{\mu }^{a},  \label{bfa16} \\
\delta B_{\mu \nu }^{\ast a}=-\tfrac{1}{2}\partial _{[\mu }^{\left.
{}\right. }A_{\nu ]}^{a},\quad \delta V_{A}^{\ast \mu \nu }=-\tfrac{1}{2}%
F_{A}^{\mu \nu },\quad \delta V_{A}^{\ast \mu }=-\left( V_{A}^{\mu
}+\partial _{\nu }V_{A}^{\mu \nu }\right) ,  \label{bfa17} \\
\delta \eta _{a}^{\ast }=-\partial _{\mu }A_{a}^{\ast \mu },\quad \delta
C_{a}^{\ast \mu \nu }=\partial _{\left. {}\right. }^{[\mu }H_{a}^{\ast \nu
]},\quad \delta \eta _{\mu \nu \rho }^{\ast a}=\partial _{[\mu }^{\left.
{}\right. }B_{\nu \rho ]}^{\ast a},  \label{bfa18} \\
\delta C_{A}^{\ast \mu }=\varepsilon ^{\mu \nu \rho \lambda }\partial _{\nu
}V_{A\rho \lambda }^{\ast },\quad \delta C_{a}^{\ast \mu \nu \rho
}=-\partial _{\left. {}\right. }^{\left[ \mu \right. }C_{a}^{\ast \nu \rho
]},  \label{bfa19} \\
\delta \eta _{\mu \nu \rho \lambda }^{\ast a}=-\partial _{[\mu }^{\left.
{}\right. }\eta _{\nu \rho \lambda ]}^{\ast a},\quad \delta C_{A}^{\ast
}=\partial _{\mu }C_{A}^{\ast \mu },\quad \delta C_{a}^{\ast \mu \nu \rho
\lambda }=\partial _{\left. {}\right. }^{[\mu }C_{a}^{\ast \nu \rho \lambda
]},  \label{bfa19a} \\
\gamma \Phi _{\alpha _{0}}^{\ast }=\gamma \eta _{\alpha _{1}}^{\ast }=\gamma
\eta _{\alpha _{2}}^{\ast }=\gamma \eta _{\alpha _{3}}^{\ast }=0,
\label{bfa20} \\
\gamma A_{\mu }^{a}=\partial _{\mu }\eta ^{a},\quad \gamma H_{\mu
}^{a}=2\partial ^{\nu }C_{\mu \nu }^{a},\quad \gamma B_{a}^{\mu \nu
}=-3\partial _{\rho }\eta _{a}^{\mu \nu \rho },  \label{bfa21} \\
\gamma \varphi _{a}=0=\gamma V_{\mu }^{A},\quad \gamma V_{\mu \nu
}^{A}=\varepsilon _{\mu \nu \rho \lambda }\partial ^{\rho }C^{A\lambda
},\quad \gamma \eta ^{a}=0,  \label{bfa22} \\
\gamma C_{\mu \nu }^{a}=-3\partial ^{\rho }C_{\mu \nu \rho }^{a},\quad
\gamma \eta _{a}^{\mu \nu \rho }=4\partial _{\lambda }\eta _{a}^{\mu \nu
\rho \lambda },\quad \gamma C_{\mu }^{A}=\partial _{\mu }C^{A},
\label{bfa23} \\
\gamma C_{\mu \nu \rho }^{a}=4\partial ^{\lambda }C_{\mu \nu \rho \lambda
}^{a},\quad \gamma \eta _{a}^{\mu \nu \rho \lambda }=\gamma C^{A}=0,\quad
\gamma C_{\mu \nu \rho \lambda }^{a}=0.  \label{bfa24}
\end{gather}

The overall degree of the BRST complex is named ghost number ($\mathrm{gh}$)
and is defined like the difference between the pure ghost number and the
antighost number, such that $\mathrm{gh}\left( \delta \right) =\mathrm{gh}%
\left( \gamma \right) =\mathrm{gh}\left( s\right) =1$. The BRST symmetry
admits a canonical action $s\cdot =\left( \cdot ,\bar{S}\right) $ in an
antibracket structure $\left( ,\right) $, where its canonical generator is a
bosonic functional of ghost number $0$ ($\varepsilon \left( \bar{S}\right)
=0 $, $\mathrm{gh}\left( \bar{S}\right) =0$) that satisfies the classical
master equation $\left( \bar{S},\bar{S}\right) =0$. In the case of the free
theory under discussion, the solution to the master equation takes the form
\begin{eqnarray}
\bar{S}= S_{0}&+&\int d^{4}x\left( A_{a}^{\ast \mu }\partial _{\mu }\eta
^{a}+2H_{a}^{\ast \mu }\partial ^{\nu }C_{\mu \nu }^{a}-3B_{\mu \nu }^{\ast
a}\partial _{\rho }\eta _{a}^{\mu \nu \rho }\right.  \notag \\
&&+\varepsilon _{\mu \nu \rho \lambda }V^{\ast A\mu \nu }\partial ^{\rho
}C_{A}^{\lambda }-3C_{a}^{\ast \mu \nu }\partial ^{\rho }C_{\mu \nu \rho
}^{a}+4\eta _{\mu \nu \rho }^{\ast a}\partial _{\lambda }\eta _{a}^{\mu \nu
\rho \lambda }  \notag \\
&&\left. +C_{\mu }^{\ast A}\partial ^{\mu }C_{A}+4C_{a}^{\ast \mu \nu \rho
}\partial ^{\lambda }C_{\mu \nu \rho \lambda }^{a}\right)  \label{solfree}
\end{eqnarray}%
and contains pieces of antighost number ranging from $0$ to $3$.

\section{Deformation of the solution to the master equation: a brief review
\label{defrev}}

\setcounter{equation}{0}

We begin with a ``free" gauge theory, described by a Lagrangian action $%
S_{0}^{\mathrm{L}}\left[ \Phi ^{\alpha _{0}}\right] $, invariant under some
gauge transformations $\delta _{\epsilon }\Phi ^{\alpha _{0}}=Z_{\;\;\alpha
_{1}}^{\alpha _{0}}\epsilon ^{\alpha _{1}}$, i.e. $\frac{\delta S_{0}^{%
\mathrm{L}}}{\delta \Phi ^{\alpha _{0}}}Z_{\;\;\alpha _{1}}^{\alpha _{0}}=0$%
, and consider the problem of constructing consistent interactions among the
fields $\Phi ^{\alpha _{0}}$ such that the couplings preserve both the field
spectrum and the original number of gauge symmetries. This matter is
addressed by means of reformulating the problem of constructing consistent
interactions as a deformation problem of the solution to the master equation
corresponding to the ``free" theory~\cite{def,contempmath}. Such a
reformulation is possible due to the fact that the solution to the master
equation contains all the information on the gauge structure of the theory.
If an interacting gauge theory can be consistently constructed, then the
solution $\bar{S}$ to the master equation $\left( \bar{S},\bar{S}\right) =0$
associated with the ``free" theory can be deformed into a solution $S$
\begin{equation}
\bar{S}\rightarrow S=\bar{S}+\lambda S_{1}+\lambda ^{2}S_{2}+\cdots =\bar{S}%
+\lambda \int d^{D}x\,a+\lambda ^{2}\int d^{D}x\,b+\cdots  \label{bff3.1}
\end{equation}%
of the master equation for the deformed theory
\begin{equation}
\left( S,S\right) =0,  \label{bff3.2}
\end{equation}%
such that both the ghost and antifield spectra of the initial theory are
preserved. Equation (\ref{bff3.2}) splits, according to the various orders
in the coupling constant (deformation parameter) $\lambda $, into a tower of
equations:
\begin{eqnarray}
\left( \bar{S},\bar{S}\right) &=&0,  \label{bff3.3} \\
2\left( S_{1},\bar{S}\right) &=&0,  \label{bff3.4} \\
2\left( S_{2},\bar{S}\right) +\left( S_{1},S_{1}\right) &=&0,  \label{bff3.5}
\\
\left( S_{3},\bar{S}\right) +\left( S_{1},S_{2}\right) &=&0,  \label{bff3.6}
\\
&&\vdots  \notag
\end{eqnarray}

Equation (\ref{bff3.3}) is fulfilled by hypothesis. The next equation
requires that the first-order deformation of the solution to the master
equation, $S_{1}$, is a co-cycle of the ``free" BRST differential, $sS_{1}=0$%
. However, only cohomologically nontrivial solutions to (\ref{bff3.4})
should be taken into account, as the BRST-exact ones can be eliminated by
some (in general nonlinear) field redefinitions. This means that $S_{1}$
pertains to the ghost number $0$ cohomological space of $s$, $H^{0}\left(
s\right) $, which is generically nonempty because it is isomorphic to the
space of physical observables of the ``free" theory. It has been shown (by
the triviality of the antibracket map in the cohomology of the BRST
differential) that there are no obstructions in finding solutions to the
remaining equations, namely (\ref{bff3.5}), (\ref{bff3.6}), etc. However,
the resulting interactions may be nonlocal, and obstructions might even
appear if one insists on their locality. The analysis of these obstructions
can be carried out by means of standard cohomological techniques.

\section{Consistent interactions between a collection of topological BF
models and a set of Abelian two-forms\label{int}}

\setcounter{equation}{0}

This section is devoted to the investigation of consistent interactions that
can be introduced between a collection of topological BF models and a set of
Abelian two-forms in four spacetime dimensions. This matter is addressed in
the context of the antifield-BRST deformation procedure briefly addressed in
the above and relies on computing the solutions to equations (\ref%
{bff3.4})--(\ref{bff3.6}), etc., with the help of the free BRST cohomology.

\subsection{Standard material: basic cohomologies}

For obvious reasons, we consider only smooth, local, Lorentz covariant, and
Poincar\'{e} invariant deformations (i.e., we do not allow explicit
dependence on the spacetime coordinates). Moreover, we require the
preservation of the number of derivatives on each field with respect to the
free theory (derivative-order assumption). The smoothness of the
deformations refers to the fact that the deformed solution to the master
equation, (\ref{bff3.1}), is smooth in the coupling constant $\lambda $ and
reduces to the original solution, (\ref{solfree}), in the free limit ($%
\lambda =0$). The preservation of the number of derivatives on each field
with respect to the free theory means here that the following two
requirements must be simultaneously satisfied: (i) the derivative order of
the equations of motion on each field is the same for the free and for the
interacting theory, respectively; (ii) the maximum number of derivatives
allowed within the interaction vertices is equal to $2$, i.e. the maximum
number of derivatives from the free Lagrangian. If we make the notation $%
S_{1}=\int d^{4}x\,a$, with $a$ a local function, then equation (\ref{bff3.4}%
), which we have seen that controls the first-order deformation, takes the
local form
\begin{equation}
sa=\partial _{\mu }m^{\mu },\quad \text{\textrm{gh}}\left( a\right) =0,\quad
\varepsilon \left( a\right) =0,  \label{3.1}
\end{equation}%
for some local $m^{\mu }$. It shows that the nonintegrated density
of the first-order deformation pertains to the local cohomology of
$s$ in ghost number $0$, $a\in H^{0}\left( s\vert d\right) $, where
$d$ denotes the exterior
spacetime differential. The solution to (\ref{3.1}) is unique up to $s$%
-exact pieces plus divergences
\begin{equation}
a\rightarrow a+sb+\partial _{\mu }n^{\mu },\, \text{\textrm{gh}}\left(
b\right) =-1,\, \varepsilon \left( b\right) =1,\, \text{\textrm{gh}}\left(
n^{\mu }\right) =0,\, \varepsilon \left( n^{\mu }\right) =0.  \label{3.1a}
\end{equation}%
At the same time, if the general solution to (\ref{3.1}) is found to be
completely trivial, $a=sb+\partial _{\mu }n^{\mu }$, then it can be made to
vanish $a=0$.

In order to analyze equation (\ref{3.1}) we develop $a$ according to the
antighost number
\begin{equation}
a=\sum\limits_{i=0}^{I}a_{i},\quad \text{\textrm{agh}}\left( a_{i}\right)
=i,\quad \text{\textrm{gh}}\left( a_{i}\right) =0,\quad \varepsilon \left(
a_{i}\right) =0,  \label{3.2}
\end{equation}%
and assume, without loss of generality, that the above decomposition stops
at some finite value of $I$. This can be shown, for instance, like in~\cite%
{gen2} (Section 3), under the sole assumption that the interacting
Lagrangian at the first order in the coupling constant, $a_{0}$, has a
finite, but otherwise arbitrary derivative order. Inserting decomposition (%
\ref{3.2}) into equation (\ref{3.1}) and projecting it on the various values
of the antighost number, we obtain the tower of equations
\begin{eqnarray}
\gamma a_{I} &=&\partial _{\mu }\overset{\left( I\right) }{m}^{\mu },
\label{3.3} \\
\delta a_{I}+\gamma a_{I-1} &=&\partial _{\mu }\overset{\left( I-1\right) }{m%
}^{\mu },  \label{3.4} \\
\delta a_{i}+\gamma a_{i-1} &=&\partial _{\mu }\overset{\left( i-1\right) }{m%
}^{\mu },\quad 1\leq i\leq I-1,  \label{3.5}
\end{eqnarray}%
where $\left( \overset{\left( i\right) }{m}^{\mu }\right) _{i=\overline{0,I}%
} $ are some local currents with $\text{agh}\left( \overset{\left( i\right) }%
{m}^{\mu }\right) =i$. Equation (\ref{3.3}) can be replaced in strictly
positive values of the antighost number by
\begin{equation}
\gamma a_{I}=0,\quad I>0.  \label{3.6}
\end{equation}%
Due to the second-order nilpotency of $\gamma $ ($\gamma ^{2}=0$), the
solution to (\ref{3.6}) is clearly unique up to $\gamma $-exact
contributions
\begin{equation}
a_{I}\rightarrow a_{I}+\gamma b_{I},\quad \text{\textrm{agh}}\left(
b_{I}\right) =I,\quad \text{\textrm{pgh}}\left( b_{I}\right) =I-1,\quad
\varepsilon \left( b_{I}\right) =1.  \label{r68}
\end{equation}%
Meanwhile, if it turns out that $a_{I}$ exclusively reduces to $\gamma $%
-exact terms, $a_{I}=\gamma b_{I}$, then it can be made to vanish, $a_{I}=0$%
. In other words, the nontriviality of the first-order deformation $a$ is
translated at its highest antighost number component into the requirement
that $a_{I}\in H^{I}\left( \gamma \right) $, where $H^{I}\left( \gamma
\right) $ denotes the cohomology of the exterior longitudinal derivative $%
\gamma $ in pure ghost number equal to $I$. So, in order to solve equation (%
\ref{3.1}) (equivalent with (\ref{3.6}) and (\ref{3.4})--(\ref{3.5})), we
need to compute the cohomology of $\gamma $, $H\left( \gamma \right) $, and,
as it will be made clear below, also the local homology of $\delta $, $%
H\left( \delta \vert d\right) $.

On behalf of definitions (\ref{bfa20})--(\ref{bfa24}) it is simple to see
that $H\left( \gamma \right) $ is spanned by
\begin{equation}
F_{\bar{A}}=\left( \varphi _{a},\partial _{\lbrack \mu }^{\left. {}\right.
}A_{\nu ]}^{a},\partial ^{\mu }H_{\mu }^{a},\partial _{\mu }B_{a}^{\mu \nu
},V_{\mu }^{A},\tilde{F}_{\mu \nu \rho }^{A}\right) ,  \label{3.7}
\end{equation}%
the antifields
\begin{equation}
\chi _{\Delta }^{\ast }=\left( \Phi _{\alpha _{0}}^{\ast },\eta _{\alpha
_{1}}^{\ast },\eta _{\alpha _{2}}^{\ast },\eta _{\alpha _{3}}^{\ast }\right)
,  \label{notat}
\end{equation}%
all of their spacetime derivatives as well as by the undifferentiated ghosts
\begin{equation}
\eta ^{\bar{\Upsilon}}=\left( \eta ^{a},C^{A},\eta _{a}^{\mu \nu \rho
\lambda },C_{\mu \nu \rho \lambda }^{a}\right) .  \label{notat1}
\end{equation}%
In formula (\ref{3.7}) we used the notation
\begin{equation}
\tilde{F}_{\mu \nu \rho }^{A}=\partial _{\lbrack \mu }^{\left. {}\right. }%
\tilde{V}_{\nu \rho ]}^{A},\quad \tilde{V}_{\mu \nu }^{A}\equiv \tfrac{1}{2}%
\varepsilon _{\mu \nu \rho \lambda }V^{A\rho \lambda }.  \label{notat2}
\end{equation}%
(The derivatives of the ghosts $\eta ^{\bar{\Upsilon}}$ are removed from $%
H\left( \gamma \right) $ since they are $\gamma $-exact, in agreement with
the first relation from (\ref{bfa21}), the last formula in (\ref{bfa23}),
the second equation in (\ref{bfa23}), and the first definition from (\ref%
{bfa24}).) If we denote by $e^{M}\left( \eta
^{\bar{\Upsilon}}\right) $ the elements with pure ghost number $M$
of a basis in the space of the polynomials in the ghosts
(\ref{notat1}), then it follows that the general solution to
equation (\ref{3.6}) takes the form
\begin{equation}
a_{I}=\alpha _{I}\left( \left[ F_{\bar{A}}\right] ,\left[ \chi _{\Delta
}^{\ast }\right] \right) e^{I}\left( \eta ^{\bar{\Upsilon}}\right) ,
\label{3.8}
\end{equation}%
where $\text{agh}\left( \alpha _{I}\right) =I$ and $\text{pgh}\left(
e^{I}\right) =I$. The notation $f([q])$ means that $f$ depends on $q$ and
its spacetime derivatives up to a finite order. The objects $\alpha _{I}$
(obviously nontrivial in $H^{0}\left( \gamma \right) $) will be called
\textquotedblleft invariant polynomials". The result that we can replace
equation (\ref{3.3}) with the less obvious one (\ref{3.6}) is a nice
consequence of the fact that the cohomology of the exterior spacetime
differential is trivial in the space of invariant polynomials in strictly
positive antighost numbers.

Inserting (\ref{3.8}) in (\ref{3.4}) we obtain that a necessary (but
not sufficient) condition for the existence of (nontrivial)
solutions $a_{I-1}$ is that the invariant polynomials $\alpha _{I}$
are (nontrivial) objects from the local cohomology of Koszul-Tate
differential $H\left( \delta \vert d\right) $ in antighost number
$I>0$ and in pure ghost number $0$,
\begin{equation}
\delta \alpha _{I}=\partial _{\mu }\overset{\left( I-1\right) }{j}^{\mu
},\quad \text{\textrm{agh}}\left( \overset{\left( I-1\right) }{j}^{\mu
}\right) =I-1,\quad \text{\textrm{pgh}}\left( \overset{\left( I-1\right) }{j}%
^{\mu }\right) =0.  \label{3.10a}
\end{equation}%
We recall that the local cohomology $H\left( \delta \vert d\right) $
is completely trivial in both strictly positive antighost
\textit{and} pure ghost numbers (for instance, see~\cite{gen1a},
Theorem 5.4, and~\cite{gen1b} ), so from now on it is understood
that by $H\left( \delta \vert d\right) $ we mean the local
cohomology of $\delta $ at pure ghost number $0$. Using the fact
that the free BF model under study is a linear gauge theory of
Cauchy order equal to $4$ and the general result
from~\cite{gen1a,gen1b}, according to which the local cohomology of
the Koszul-Tate differential is trivial in antighost numbers
strictly greater than its Cauchy order, we can state that
\begin{equation}
H_{J}\left( \delta \vert d\right) =0\quad \text{\textrm{for\
all}}\quad J>4, \label{3.11}
\end{equation}%
where $H_{J}\left( \delta \vert d\right) $ represents the local
cohomology of the Koszul-Tate differential in antighost number $J$.
Moreover, if the invariant polynomial $\alpha _{J}$, with
\textrm{agh}$\left( \alpha _{J}\right) =J\geq 4$, is trivial in
$H_{J}\left( \delta \vert d\right) $, then it can be taken to
be trivial also in $H_{J}^{\text{\textrm{inv}}}\left( \delta \vert d\right) $%
\begin{equation}
\left( \alpha _{J}=\delta b_{J+1}+\partial _{\mu }\overset{(J)}{c}^{\mu },%
\text{\textrm{agh}}\left( \alpha _{J}\right) =J\geq 4\right) \Rightarrow
\alpha _{J}=\delta \beta _{J+1}+\partial _{\mu }\overset{(J)}{\gamma }^{\mu
},  \label{3.12ax}
\end{equation}%
with both $\beta _{J+1}$ and $\overset{(J)}{\gamma }^{\mu }$
invariant polynomials. Here, $H_{J}^{\text{\textrm{inv}}}\left(
\delta \vert d\right) $ denotes the invariant characteristic
cohomology in antighost number $J$ (the local cohomology of the
Koszul-Tate differential in the space of invariant polynomials). (An
element of $H_{I}^{\text{\textrm{inv}}}\left( \delta \vert d\right)
$ is defined via an equation like (\ref{3.10a}), but with the
corresponding current an invariant polynomial.). This result together with (%
\ref{3.11}) ensures that the entire invariant characteristic cohomology in
antighost numbers strictly greater than $4$ is trivial
\begin{equation}
H_{J}^{\text{\textrm{inv}}}\left( \delta \vert d\right) =0\quad
\text{\textrm{for all}}\quad J>4.  \label{3.12x}
\end{equation}

The nontrivial representatives of $H_{J}(\delta \vert d)$ and of $H_{J}^{\mathrm{%
inv}}(\delta \vert d)$ for $J\geq 2$ depend neither on $\left(
\partial _{\lbrack \mu }^{\left. {}\right. }A_{\nu ]}^{a},\partial
^{\mu }H_{\mu }^{a},\partial _{\mu }B_{a}^{\mu \nu },\tilde{F}_{\mu
\nu \rho }^{A}\right) $ nor on the spacetime derivatives of
$F_{\bar{A}}$ defined in (\ref{3.7}), but only on the
undifferentiated scalar fields and auxiliary vector fields from the
two-form sector, $\left( \varphi _{a},V_{\mu }^{A}\right) $. With
the help
of relations (\ref{bfa15})--(\ref{bfa19a}), it can be shown that $H_{4}^{%
\text{\textrm{inv}}}\left( \delta \vert d\right) $ is generated by
the elements
\begin{eqnarray}
\left( P_{\Lambda }\left( W\right) \right) ^{\mu \nu \rho \lambda } &=&\frac{%
\partial W_{\Lambda }}{\partial \varphi _{a}}C_{a}^{\ast \mu \nu \rho
\lambda }+\frac{\partial ^{2}W_{\Lambda }}{\partial \varphi _{a}\partial
\varphi _{b}}\left( H_{a}^{\ast \lbrack \mu }C_{b}^{\ast \nu \rho \lambda
]}+C_{a}^{\ast \lbrack \mu \nu }C_{b}^{\ast \rho \lambda ]}\right)  \notag \\
&&+\frac{\partial ^{3}W_{\Lambda }}{\partial \varphi _{a}\partial \varphi
_{b}\partial \varphi _{c}}H_{a}^{\ast \lbrack \mu }H_{b}^{\ast \nu
}C_{c}^{\ast \rho \lambda ]}  \notag \\
&&+\frac{\partial ^{4}W_{\Lambda }}{\partial \varphi _{a}\partial \varphi
_{b}\partial \varphi _{c}\partial \varphi _{d}}H_{a}^{\ast \mu }H_{b}^{\ast
\nu }H_{c}^{\ast \rho }H_{d}^{\ast \lambda },  \label{3.13}
\end{eqnarray}%
where $W_{\Lambda }=W_{\Lambda }\left( \varphi _{a}\right) $ are arbitrary,
smooth functions depending only on the undifferentiated scalar fields $%
\varphi _{a}$ and $\Lambda $ is some multi-index (composed of internal
and/or Lorentz indices). Indeed, direct computation yields
\begin{equation}
\delta \left( P_{\Lambda }\left( W\right) \right) ^{\mu \nu \rho \lambda
}=\partial _{\left. {}\right. }^{[\mu }\left( P_{\Lambda }\left( W\right)
\right) ^{\nu \rho \lambda ]},\quad \mathrm{agh}\left( \left( P_{\Lambda
}\left( W\right) \right) ^{\nu \rho \lambda }\right) =3,  \label{3.13a}
\end{equation}%
where we made the notation
\begin{eqnarray}
\left( P_{\Lambda }\left( W\right) \right) ^{\mu \nu \rho } &=&\frac{%
\partial W_{\Lambda }}{\partial \varphi _{a}}C_{a}^{\ast \mu \nu \rho }+%
\frac{\partial ^{2}W_{\Lambda }}{\partial \varphi _{a}\partial \varphi _{b}}%
H_{a}^{\ast \lbrack \mu }C_{b}^{\ast \nu \rho ]}  \notag \\
&&+\frac{\partial ^{3}W_{\Lambda }}{\partial \varphi _{a}\partial \varphi
_{b}\partial \varphi _{c}}H_{a}^{\ast \mu }H_{b}^{\ast \nu }H_{c}^{\ast \rho
}.  \label{3.14}
\end{eqnarray}%
It is clear that $\left( P_{\Lambda }\left( W\right) \right) ^{\mu \nu \rho
} $ is an invariant polynomial. By applying the operator $\delta $ on it, we
have that
\begin{equation}
\delta \left( P_{\Lambda }\left( W\right) \right) ^{\mu \nu \rho }=-\partial
_{\left. {}\right. }^{[\mu }\left( P_{\Lambda }\left( W\right) \right) ^{\nu
\rho ]},\quad \mathrm{agh}\left( \left( P_{\Lambda }\left( W\right) \right)
^{\nu \rho }\right) =2,  \label{3.14a}
\end{equation}%
where we employed the convention
\begin{equation}
\left( P_{\Lambda }\left( W\right) \right) ^{\mu \nu }=\frac{\partial
W_{\Lambda }}{\partial \varphi _{a}}C_{a}^{\ast \mu \nu }+\frac{\partial
^{2}W_{\Lambda }}{\partial \varphi _{a}\partial \varphi _{b}}H_{a}^{\ast \mu
}H_{b}^{\ast \nu }.  \label{3.15}
\end{equation}%
Since $\left( P_{\Lambda }\left( W\right) \right) ^{\mu \nu }$ is also an
invariant polynomial, from (\ref{3.14a}) it follows that $\left( P_{\Lambda
}\left( W\right) \right) ^{\mu \nu \rho }$ belongs to $H_{3}^{\text{\textrm{%
inv}}}\left( \delta \vert d\right) $. Moreover, further calculations produce%
\begin{equation}
\delta \left( P_{\Lambda }\left( W\right) \right) ^{\mu \nu }=\partial
_{\left. {}\right. }^{[\mu }\left( P_{\Lambda }\left( W\right) \right) ^{\nu
]},\quad \mathrm{agh}\left( \left( P_{\Lambda }\left( W\right) \right) ^{\nu
}\right) =1,  \label{3.15a}
\end{equation}%
with
\begin{equation}
\left( P_{\Lambda }\left( W\right) \right) ^{\mu }=\frac{\partial W_{\Lambda
}}{\partial \varphi _{a}}H_{a}^{\ast \mu }.  \label{3.16}
\end{equation}%
Due to the fact that $\left( P_{\Lambda }\left( W\right) \right)
^{\mu }$ is an invariant polynomial, we deduce that $\left(
P_{\Lambda }\left( W\right) \right) ^{\mu \nu }$ pertains to
$H_{2}^{\text{\textrm{inv}}}\left( \delta \vert d\right) $. Using
again the actions of $\delta $ on the BRST generators, it can be
proved that $H_{3}^{\text{\textrm{inv}}}\left( \delta \vert d\right)
$ is spanned, beside the elements $\left( P_{\Lambda }\left(
W\right) \right) ^{\mu \nu \rho }$ given in (\ref{3.14}), also by
the objects
\begin{eqnarray}
Q_{\Lambda }\left( f\right) &=&f_{\Lambda }^{A}C_{A}^{\ast }-\left(
P_{\Lambda }^{A}\left( f\right) \right) ^{\mu }C_{A\mu }^{\ast }-\tfrac{1}{2}%
\varepsilon _{\mu \nu \rho \lambda }\left( \tfrac{1}{3}\left( P_{\Lambda
}^{A}\left( f\right) \right) ^{\mu \nu \rho }V_{A}^{\lambda }\right.  \notag
\\
&&\left. +\left( P_{\Lambda }^{A}\left( f\right) \right) ^{\mu \nu
}V_{A}^{\ast \rho \lambda }\right)  \label{p}
\end{eqnarray}%
and by the undifferentiated antifields $\eta _{\mu \nu \rho \lambda }^{\ast
a}$ (according to the first definition from (\ref{bfa19a})). In formula (\ref%
{p}) $f_{\Lambda }^{A}=f_{\Lambda }^{A}\left( \varphi _{a}\right) $ are some
arbitrary, smooth functions of the undifferentiated scalar fields $\varphi
_{a}$ carrying at least an internal index $A$ from the two-form sector and
possibly a supplementary multi-index $\Lambda $. The factors $\left(
P_{\Lambda }^{A}\left( f\right) \right) ^{\mu }$, $\left( P_{\Lambda
}^{A}\left( f\right) \right) ^{\mu \nu }$, and $\left( P_{\Lambda
}^{A}\left( f\right) \right) ^{\mu \nu \rho }$ read as in (\ref{3.16}), (\ref%
{3.15}), and (\ref{3.14}), respectively, with $W_{\Lambda }\left( \varphi
_{a}\right) \rightarrow f_{\Lambda }^{A}\left( \varphi _{a}\right) $.
Concerning $Q_{\Lambda }\left( f\right) $, we have that
\begin{equation}
\delta Q_{\Lambda }\left( f\right) =\partial _{\mu }\left( Q_{\Lambda
}\left( f\right) \right) ^{\mu },\quad \mathrm{agh}\left( \left( Q_{\Lambda
}\left( f\right) \right) ^{\mu }\right) =2,  \label{pa}
\end{equation}%
where we employed the notation
\begin{equation}
\left( Q_{\Lambda }\left( f\right) \right) ^{\mu }=f_{\Lambda
}^{A}C_{A}^{\ast \mu }+\varepsilon ^{\mu \nu \rho \lambda }\left( \left(
P_{\Lambda }^{A}\left( f\right) \right) _{\nu }V_{A\rho \lambda }^{\ast }+%
\tfrac{1}{2}\left( P_{\Lambda }^{A}\left( f\right) \right) _{\nu \rho
}V_{A\lambda }\right) .  \label{pm}
\end{equation}%
With the help of definitions (\ref{bfa15})--(\ref{bfa19a}) it can be checked
that
\begin{equation}
\delta \left( Q_{\Lambda }\left( f\right) \right) ^{\mu }=\partial _{\nu
}\left( Q_{\Lambda }\left( f\right) \right) ^{\mu \nu },\quad \mathrm{agh}%
\left( \left( Q_{\Lambda }\left( f\right) \right) ^{\mu \nu }\right) =1,
\label{3.17}
\end{equation}%
where we made the notation
\begin{equation}
\left( Q_{\Lambda }\left( f\right) \right) ^{\mu \nu }=\varepsilon ^{\mu \nu
\rho \lambda }\left( f_{\Lambda }^{A}V_{A\rho \lambda }^{\ast }+\left(
P_{\Lambda }^{A}\left( f\right) \right) _{\rho }V_{A\lambda }\right) .
\label{3.17a}
\end{equation}%
Direct computation shows that the objects%
\begin{eqnarray}
R_{\Lambda }\left( g\right) &=&g_{\Lambda }^{AB}\left( C_{A}^{\ast \mu
}V_{B\mu }+\tfrac{1}{2}\varepsilon _{\mu \nu \rho \lambda }V_{A}^{\ast \mu
\nu }V_{B}^{\ast \rho \lambda }\right)  \notag \\
&&-\varepsilon _{\mu \nu \rho \lambda }\left( \left( P_{\Lambda
}^{AB}\left( g\right) \right) ^{\mu }V_{A}^{\ast \nu \rho
}+\tfrac{1}{4}\left( P_{\Lambda }^{AB}\left( g\right) \right) ^{\mu
\nu }V_{A}^{\rho }\right) V_{B}^{\lambda }  \label{q}
\end{eqnarray}%
satisfy
\begin{equation}
\delta R_{\Lambda }\left( g\right) =\partial ^{\mu }\left( R_{\Lambda
}\left( g\right) \right) _{\mu },\quad \mathrm{agh}\left( \left( R_{\Lambda
}\left( g\right) \right) _{\mu }\right) =1,  \label{qa}
\end{equation}%
with
\begin{equation}
\left( R_{\Lambda }\left( g\right) \right) _{\mu }=-\varepsilon _{\mu \nu
\rho \lambda }\left( g_{\Lambda }^{AB}V_{A}^{\ast \nu \rho }+\tfrac{1}{2}%
\left( P_{\Lambda }^{AB}\left( g\right) \right) ^{\nu }V_{A}^{\rho }\right)
V_{B}^{\lambda }.  \label{qm}
\end{equation}%
In formulas (\ref{q}) and (\ref{qm}) $g_{\Lambda }^{AB}=g_{\Lambda
}^{AB}\left( \varphi _{a}\right) $ stand for some smooth functions of the
undifferentiated scalar fields that in addition are antisymmetric with
respect to $A$ and $B$%
\begin{equation}
g_{\Lambda }^{AB}=-g_{\Lambda }^{BA}.  \label{gABlambda}
\end{equation}%
Looking at their expressions, it is easy to see that all the
quantities denoted by $Q$s or $R$s are invariant polynomials.
Putting together the above results we can state that
$H_{2}^{\text{\textrm{inv}}}\left( \delta \vert d\right) $ is
spanned by $\left( P_{\Lambda }\left( W\right) \right) ^{\mu \nu }$
listed in (\ref{3.15}), $\left( Q_{\Lambda }\left( f\right) \right)
^{\mu }$ expressed by (\ref{pm}), $R_{\Lambda }\left( g\right) $ given in (%
\ref{q}), and the undifferentiated antifields $\eta _{\mu \nu \rho }^{\ast
a} $ and $\eta _{a}^{\ast }$ (in agreement with the last formula from (\ref%
{bfa18}) and the first definition in (\ref{bfa18})).

In contrast to the spaces $\left( H_{J}(\delta \vert d)\right) _{J\geq 2}$
and $\left( H_{J}^{\mathrm{inv}}(\delta \vert d)\right) _{J\geq 2}$, which
are finite-dimensional, the cohomology $H_{1}(\delta \vert d)$ (known to be
related to global symmetries and ordinary conservation laws) is
infinite-dimensional since the theory is free. Fortunately, it will not be
needed in the sequel.

The previous results on $H(\delta \vert d)$ and $H^{\mathrm{inv}}(\delta
\vert d)$ in strictly positive antighost numbers are important because they
control the obstructions to removing the antifields from the first-order
deformation. More precisely, we can successively eliminate all the pieces of
antighost number strictly greater that $4$ from the nonintegrated density of
the first-order deformation by adding solely trivial terms, so we can take,
without loss of nontrivial objects, the condition $I\leq 4$ into (\ref{3.2}%
). In addition, the last representative is of the form (\ref{3.8}), where
the invariant polynomial is necessarily a nontrivial object from $H_{4}^{%
\mathrm{inv}}(\delta \vert d)$.

\subsection{First-order deformation\label{firstord}}

In the case $I=4$ the nonintegrated density of the first-order deformation
(see (\ref{3.2})) becomes
\begin{equation}
a=a_{0}+a_{1}+a_{2}+a_{3}+a_{4}.  \label{fo1}
\end{equation}%
We can further decompose $a$ in a natural manner as a sum between
two kinds of deformations
\begin{equation}
a=a^{\left( \mathrm{BF}\right) }+a^{\left( \mathrm{int}\right) },
\label{fo2}
\end{equation}%
where $a^{\left( \mathrm{BF}\right) }$ contains only
fields/ghosts/antifields from the BF sector and $a^{\left( \mathrm{int}%
\right) }$ describes the cross-interactions between the two theories.
Strictly speaking, we should have added to (\ref{fo2}) also a component $%
a^{\left( \mathrm{V}\right) }$ that involves only the two-form field sector.
As it will be seen at the end of this subsection, $a^{\left( \mathrm{V}%
\right) }$ will be automatically included into $a^{\left( \mathrm{int}%
\right) }$. The piece $a^{\left( \mathrm{BF}\right) }$ is completely known
(see~\cite{defBFijmpa,defBFepjc,defBFijmpajuvi06}) and (separately)
satisfies an equation of the type (\ref{3.1}). It admits a decomposition
similar to (\ref{fo1})
\begin{equation}
a^{\left( \mathrm{BF}\right) }=a_{0}^{\left( \mathrm{BF}\right)
}+a_{1}^{\left( \mathrm{BF}\right) }+a_{2}^{\left( \mathrm{BF}\right)
}+a_{3}^{\left( \mathrm{BF}\right) }+a_{4}^{\left( \mathrm{BF}\right) },
\label{descBF}
\end{equation}%
where
\begin{eqnarray}
a_{4}^{\left( \mathrm{BF}\right) } &=&\left( P_{ab}\left( W\right) \right)
^{\mu \nu \rho \lambda }\eta ^{a}C_{\mu \nu \rho \lambda }^{b}-\tfrac{1}{4}%
\left( P_{ab}^{c}\left( M\right) \right) _{\mu \nu \rho \lambda }\eta
^{a}\eta ^{b}\eta _{c}^{\mu \nu \rho \lambda }  \notag \\
&&+\tfrac{1}{2}\varepsilon _{\mu \nu \rho \lambda }\left( \left(
P^{ab}\left( M\right) \right) ^{\mu \nu \rho \lambda }\eta _{a\alpha \beta
\gamma \delta }\eta _{b}^{\alpha \beta \gamma \delta }\right.  \notag \\
&&\left. -\tfrac{1}{2\cdot \left( 4!\right) ^{2}}\left( P_{abcd}\left(
M\right) \right) ^{\mu \nu \rho \lambda }\eta ^{a}\eta ^{b}\eta ^{c}\eta
^{d}\right) ,  \label{a4}
\end{eqnarray}%
\begin{eqnarray}
a_{3}^{\left( \mathrm{BF}\right) } &=&\left( P_{ab}\left( W\right) \right)
^{\mu \nu \rho }\left( -\eta ^{a}C_{\mu \nu \rho }^{b}+4A^{a\lambda }C_{\mu
\nu \rho \lambda }^{b}\right)  \notag \\
&&+2\left( 6\left( P_{ab}\left( W\right) \right) ^{\mu \nu }B^{\ast a\rho
\lambda }+4\left( P_{ab}\left( W\right) \right) ^{\mu }\eta ^{\ast a\nu \rho
\lambda }+W_{ab}\eta ^{\ast a\mu \nu \rho \lambda }\right) C_{\mu \nu \rho
\lambda }^{b}  \notag \\
&&+\tfrac{1}{2}\left( P_{ab}^{c}\left( M\right) \right) _{\mu \nu \rho
}\left( \tfrac{1}{2}\eta ^{a}\eta ^{b}\eta _{c}^{\mu \nu \rho }-4A_{\lambda
}^{a}\eta ^{b}\eta _{c}^{\mu \nu \rho \lambda }\right)  \notag \\
&&-\left( 6\left( P_{ab}^{c}\left( M\right) \right) _{\mu \nu }B_{\rho
\lambda }^{\ast a}+4\left( P_{ab}^{c}\left( M\right) \right) _{\mu }\eta
_{\nu \rho \lambda }^{\ast a}+M_{ab}^{c}\eta _{\mu \nu \rho \lambda }^{\ast
a}\right) \eta ^{b}\eta _{c}^{\mu \nu \rho \lambda }  \notag \\
&&-\varepsilon _{\mu \nu \rho \lambda }\left( P^{ab}\left( M\right) \right)
_{\alpha \beta \gamma }\eta _{a}^{\alpha \beta \gamma }\eta _{b}^{\mu \nu
\rho \lambda }-\tfrac{1}{3!\cdot 4!}\varepsilon ^{\mu \nu \rho \lambda
}\left( \left( P_{abcd}\left( M\right) \right) _{\mu \nu \rho }A_{\lambda
}^{a}\right.  \notag \\
&&+3\left( P_{abcd}\left( M\right) \right) _{\mu \nu }B_{\rho \lambda
}^{\ast a}+2\left( P_{abcd}\left( M\right) \right) _{\mu }\eta _{\nu \rho
\lambda }^{\ast a}  \notag \\
&&\left. +M_{abcd}\eta _{\mu \nu \rho \lambda }^{\ast a}\right) \eta
^{b}\eta ^{c}\eta ^{d},  \label{a3}
\end{eqnarray}%
\begin{eqnarray}
a_{2}^{\left( \mathrm{BF}\right) } &=&\left( P_{ab}\left( W\right) \right)
^{\mu \nu }\left( \eta ^{a}C_{\mu \nu }^{b}-3A^{a\rho }C_{\mu \nu \rho
}^{b}\right) -2\left( 3\left( P_{ab}\left( W\right) \right) ^{\mu }B^{\ast
a\nu \rho }\right.  \notag \\
&&\left. +W_{ab}\eta ^{\ast a\mu \nu \rho }\right) C_{\mu \nu \rho }^{b}-%
\tfrac{1}{2}\left( P_{ab}^{c}\left( M\right) \right) ^{\mu \nu }\left(
\tfrac{1}{2}\eta ^{a}\eta ^{b}B_{c\mu \nu }-3A^{a\rho }\eta ^{b}\eta _{c\mu
\nu \rho }\right)  \notag \\
&&+\left( 3\left( P_{ab}^{c}\left( M\right) \right) _{\mu }B_{\nu \rho
}^{\ast a}+M_{ab}^{c}\eta _{\mu \nu \rho }^{\ast a}\right) \eta ^{b}\eta
_{c}^{\mu \nu \rho }+\tfrac{1}{2}\left( -\left( P_{ab}^{c}\left( M\right)
\right) _{\mu }A_{c}^{\ast \mu }\right.  \notag \\
&&\left. +M_{ab}^{c}\eta _{c}^{\ast }\right) \eta ^{a}\eta ^{b}+\left(
3\left( P_{ab}^{c}\left( M\right) \right) _{\mu \nu }A_{\rho }^{a}+12\left(
P_{ab}^{c}\left( M\right) \right) _{\mu }B_{\nu \rho }^{\ast a}\right.
\notag \\
&&\left. +4M_{ab}^{c}\eta _{\mu \nu \rho }^{\ast a}\right) A_{\lambda
}^{b}\eta _{c}^{\mu \nu \rho \lambda }+\tfrac{9}{2}\varepsilon ^{\mu \nu
\rho \lambda }\left( P^{ab}\left( M\right) \right) _{\mu \nu }\eta _{a\rho
\alpha \beta }\eta _{b\lambda }^{\;\;\;\alpha \beta }  \notag \\
&&-6M_{ab}^{c}B_{\mu \nu }^{\ast a}B_{\rho \lambda }^{\ast b}\eta _{c}^{\mu
\nu \rho \lambda }+\tfrac{1}{4\cdot 4!}\varepsilon ^{\mu \nu \rho \lambda
}\left( 3\left( P_{abcd}\left( M\right) \right) _{\mu \nu }A_{\rho
}^{a}A_{\lambda }^{b}\right.  \notag \\
&&\left. +12\left( P_{abcd}\left( M\right) \right) _{\mu }B_{\nu \rho
}^{\ast a}A_{\lambda }^{b}+4M_{abcd}\eta _{\mu \nu \rho }^{\ast a}A_{\lambda
}^{b}-6M_{abcd}B_{\mu \nu }^{\ast a}B_{\rho \lambda }^{\ast b}\right) \eta
^{c}\eta ^{d}  \notag \\
&&+\varepsilon _{\mu \nu \rho \lambda }\left( 2\left( P^{ab}\left( M\right)
\right) _{\alpha }A_{a}^{\ast \alpha }-2M^{ab}\eta _{a}^{\ast }\right.
\notag \\
&&\left. +\left( P^{ab}\left( M\right) \right) _{\alpha \beta }B_{a}^{\alpha
\beta }\right) \eta _{b}^{\mu \nu \rho \lambda },  \label{a2}
\end{eqnarray}%
\begin{eqnarray}
a_{1}^{\left( \mathrm{BF}\right) } &=&\left( P_{ab}\left( W\right) \right)
^{\mu }\left( -\eta ^{a}H_{\mu }^{b}+2A^{a\nu }C_{\mu \nu }^{b}\right)
+W_{ab}\left( 2B_{\mu \nu }^{\ast a}C^{b\mu \nu }-\varphi ^{\ast a}\eta
^{b}\right)  \notag \\
&&-\left( P_{ab}^{c}\left( M\right) \right) _{\mu }A_{\nu }^{a}\left( \eta
^{b}B_{c}^{\mu \nu }+\tfrac{3}{2}A_{\rho }^{b}\eta _{c}^{\mu \nu \rho
}\right) -M_{ab}^{c}\left( B_{\mu \nu }^{\ast a}\eta ^{b}B_{c}^{\mu \nu
}\right.  \notag \\
&&\left. +A_{\mu }^{a}\eta ^{b}A_{c}^{\ast \mu }+3B_{\mu \nu }^{\ast
a}A_{\rho }^{b}\eta _{c}^{\mu \nu \rho }\right)  \notag \\
&&+2\varepsilon _{\nu \rho \sigma \lambda }\left( \left( P^{ab}\left(
M\right) \right) _{\mu }B_{a}^{\mu \nu }-M^{ab}A_{a}^{\ast \nu }\right) \eta
_{b}^{\rho \sigma \lambda }  \notag \\
&&+\tfrac{1}{4!}\varepsilon ^{\mu \nu \rho \lambda }\left( \left(
P_{abcd}\left( M\right) \right) _{\mu }A_{\nu }^{a}+3M_{abcd}B_{\mu \nu
}^{\ast a}\right) A_{\rho }^{b}A_{\lambda }^{c}\eta ^{d},  \label{a1}
\end{eqnarray}%
\begin{eqnarray}
a_{0}^{\left( \mathrm{BF}\right) } &=&-W_{ab}A^{a\mu }H_{\mu }^{b}+\tfrac{1}{%
2}M_{ab}^{c}A_{\mu }^{a}A_{\nu }^{b}B_{c}^{\mu \nu }  \notag \\
&&+\tfrac{1}{2}\varepsilon ^{\mu \nu \rho \lambda }\left( M^{ab}B_{a\mu \nu
}B_{b\rho \lambda }-\tfrac{1}{2\cdot 4!}M_{abcd}A_{\mu }^{a}A_{\nu
}^{b}A_{\rho }^{c}A_{\lambda }^{d}\right) .  \label{a0}
\end{eqnarray}%
In (\ref{a4})--(\ref{a0}) the quantities denoted by $\left( P_{ab}\left(
W\right) \right) ^{\mu _{1}\ldots \mu _{k}}$, $\left( P_{ab}^{c}\left(
M\right) \right) ^{\mu _{1}\ldots \mu _{k}}$, $\left( P^{ab}\left( M\right)
\right) ^{\mu _{1}\ldots \mu _{k}}$, and $\left( P_{abcd}\left( M\right)
\right) ^{\mu _{1}\ldots \mu _{k}}$ read as in (\ref{3.13}), (\ref{3.14}), (%
\ref{3.15}), and (\ref{3.16}) for $k=4$, $k=3$, $k=2$, and $k=1$,
respectively, modulo the successive replacement of $W_{\Lambda }\left(
\varphi _{a}\right) $ with the functions $W_{ab}$, $M_{ab}^{c}$, $M^{ab}$,
and $M_{abcd}$, respectively. The last four kinds of functions depend only
on the undifferentiated scalar fields and satisfy various
symmetry/antisymmetry properties: $M_{ab}^{c}$ are antisymmetric in their
lower indices, $M^{ab}$ are symmetric, and $M_{abcd}$ are completely
antisymmetric.

Due to the fact that $a^{\left( \mathrm{BF}\right) }$ and $a^{\left( \mathrm{%
int}\right) }$ involve different types of fields and $a^{\left( \mathrm{BF}%
\right) }$ separately satisfies an equation of the type (\ref{3.1}), it
follows that $a^{\left( \mathrm{int}\right) }$ is subject to the equation
\begin{equation}
sa^{\left( \mathrm{int}\right) }=\partial _{\mu }m^{\left( \mathrm{int}%
\right) \mu },  \label{fo3}
\end{equation}%
for some local current $m^{\left( \mathrm{int}\right) \mu }$. In the sequel
we determine the general solution to (\ref{fo3}) that complies with all the
hypotheses mentioned in the beginning of the previous subsection.

In agreement with (\ref{fo1}), the solution to the equation $sa^{\left(
\mathrm{int}\right) }=\partial _{\mu }m^{\left( \mathrm{int}\right) \mu }$
can be decomposed as
\begin{equation}
a^{\left( \mathrm{int}\right) }=a_{0}^{\left( \mathrm{int}\right)
}+a_{1}^{\left( \mathrm{int}\right) }+a_{2}^{\left( \mathrm{int}\right)
}+a_{3}^{\left( \mathrm{int}\right) }+a_{4}^{\left( \mathrm{int}\right) },
\label{fo5}
\end{equation}%
where the components on the right-hand side of (\ref{fo5}) are subject to
the equations
\begin{eqnarray}
\gamma a_{4}^{\left( \mathrm{int}\right) } &=&0,  \label{fo6a} \\
\delta a_{k}^{\left( \mathrm{int}\right) }+\gamma a_{k-1}^{\left( \mathrm{int%
}\right) } &=&\partial _{\mu }\overset{(k-1)}{m}^{\left(
\mathrm{int}\right) \mu },\quad k=\overline{1,4}.  \label{fo6b}
\end{eqnarray}%
The piece $a_{4}^{\left( \mathrm{int}\right) }$ as solution to equation (\ref%
{fo6a}) has the general form expressed by (\ref{3.8}) for $I=4$, with $%
\alpha _{4}$ from $H_{4}^{\mathrm{inv}}(\delta \vert d)$ and $e^{4}$ spanned
by
\begin{equation}
\left( \eta ^{a}\eta ^{b}\eta ^{c}\eta ^{d},\eta ^{a}\eta ^{b}\eta _{c}^{\mu
\nu \rho \lambda },\eta ^{a}C_{\mu \nu \rho \lambda }^{b},\eta _{a}^{\mu \nu
\rho \lambda }\eta _{b}^{\alpha \beta \gamma \delta },\eta ^{a}\eta
^{b}C^{A},C^{A}C^{B},C^{A}\eta _{a}^{\mu \nu \rho \lambda }\right) .
\label{fo7}
\end{equation}%
Taking into account the result that the general representative of $H_{4}^{%
\mathrm{inv}}(\delta \vert d)$ is given by (\ref{3.13}) and recalling that $%
a_{4}^{\left( \mathrm{int}\right) }$ should mix the BF and the two-form
sectors (in order to provide cross-couplings), it follows that the eligible
representatives of $e^{4}$ from (\ref{fo7}) allowed to enter $a_{4}^{\left(
\mathrm{int}\right) }$ are those elements containing at least one ghost of
the type $C^{A}$. Therefore, up to trivial, $\gamma $-exact terms, we can
write
\begin{eqnarray}
a_{4}^{\left( \mathrm{int}\right) }&=&\tfrac{1}{2\cdot 4!}\varepsilon _{\mu
\nu \rho \lambda }\left( \left( P_{abA}\left( N\right) \right) ^{\mu \nu
\rho \lambda }\eta ^{a}\eta ^{b}C^{A}+\left( P_{AB}\left( N\right) \right)
^{\mu \nu \rho \lambda }C^{A}C^{B}\right)  \notag \\
&&+\left( P_{A}^{a}\left( N\right) \right) _{\mu \nu \rho \lambda }C^{A}\eta
_{a}^{\mu \nu \rho \lambda },  \label{fo8}
\end{eqnarray}%
where the objects denoted by $\left( P_{abA}\left( N\right) \right) ^{\mu
\nu \rho \lambda }$, $\left( P_{AB}\left( N\right) \right) ^{\mu \nu \rho
\lambda }$, and respectively $\left( P_{A}^{a}\left( N\right) \right) _{\mu
\nu \rho \lambda }$ are expressed as in (\ref{3.13}), being generated by the
arbitrary, smooth functions of the undifferentiated scalar fields $%
N_{abA}\left( \varphi _{m}\right) $, $N_{AB}\left( \varphi _{m}\right) $,
and $N_{A}^{a}\left( \varphi _{m}\right) $, respectively. In addition, the
functions $N_{abA}\left( \varphi _{m}\right) $ and $N_{AB}\left( \varphi
_{m}\right) $ satisfy the symmetry/antisymmetry properties
\begin{equation}
N_{abA}\left( \varphi _{m}\right) =-N_{baA}\left( \varphi _{m}\right)
,\;N_{AB}\left( \varphi _{m}\right) =N_{BA}\left( \varphi _{m}\right) .
\label{fo9}
\end{equation}

Inserting (\ref{fo8}) into equation (\ref{fo6b}) for $k=4$ and using
definitions (\ref{bfa15})--(\ref{bfa24}), after some computation we obtain
the interacting piece of antighost number $3$ from the first-order
deformation in the form
\begin{eqnarray}
a_{3}^{\left( \mathrm{int}\right) }&=&-\left( P_{A}^{a}\left( N\right)
\right) _{\mu \nu \rho }\left( C^{A}\eta _{a}^{\mu \nu \rho }+4C_{\lambda
}^{A}\eta _{a}^{\mu \nu \rho \lambda }\right)  \notag \\
&&-\tfrac{1}{3!}\varepsilon ^{\mu \nu \rho \lambda }\left[ \left(
P_{abA}\left( N\right) \right) _{\mu \nu \rho }\eta ^{a}\left( A_{\lambda
}^{b}C^{A}+\tfrac{1}{2}\eta ^{b}C_{\lambda }^{A}\right) \right.  \notag \\
&&+\left( P_{AB}\left( N\right) \right) _{\mu \nu \rho }C^{A}C_{\lambda
}^{B}-\left( 3\left( P_{abA}\left( N\right) \right) _{\mu \nu }B_{\rho
\lambda }^{\ast a}\right.  \notag \\
&&\left. \left. +2\left( P_{abA}\left( N\right) \right) _{\mu }\eta _{\nu
\rho \lambda }^{\ast a}+\tfrac{1}{2}N_{abA}\eta _{\mu \nu \rho \lambda
}^{\ast a}\right) \eta ^{b}C^{A}\right]  \notag \\
&&+Q_{aA}\left( f\right) \eta ^{a}C^{A} +\tfrac{1}{3!}Q_{abc}\left( f\right)
\eta ^{a}\eta ^{b}\eta ^{c}  \notag \\
&&+\tfrac{1}{4!}\varepsilon _{\alpha \beta \gamma \delta }\left(
Q_{\;\;b}^{a}\left( f\right) \eta ^{b}\eta _{a}^{\alpha \beta \gamma \delta
}+Q_{a}\left( f\right) C^{a\alpha \beta \gamma \delta }\right) .
\label{fo10}
\end{eqnarray}%
(Solution (\ref{fo10}) embeds also the general solution to the homogeneous
equation $\gamma \bar{a}_{3}^{\left( \mathrm{int}\right) }=0$.) The elements
denoted by $Q_{aA}\left( f\right) $, $Q_{abc}\left( f\right) $, $%
Q_{\;\;b}^{a}\left( f\right) $, and $Q_{a}\left( f\right) $ are generated
via formula (\ref{p}) by the smooth functions (of the undifferentiated
scalar fields) $f_{aB}^{A}$, $f_{abc}^{A}$, $f_{\;\;b}^{Aa}$, and $f_{a}^{A}$%
, respectively. In addition, the functions $f_{abc}^{A}$ are completely
antisymmetric in their BF collection indices.

The interacting component of antighost number $2$ results as solution to
equation (\ref{fo6b}) for $k=3$ by relying on formula (\ref{fo10}) and
definitions (\ref{bfa15})--(\ref{bfa24}), and takes the form
\begin{eqnarray}
a_{2}^{\prime \left( \mathrm{int}\right) } &=&-\tfrac{1}{2}\left(
P_{AB}\left( N\right) \right) ^{\mu \nu }\left( C^{A}V_{\mu \nu }^{B}-\tfrac{%
1}{2}\varepsilon _{\mu \nu \rho \lambda }C^{A\rho }C^{B\lambda }\right)
\notag \\
&&-\tfrac{1}{4}\left( P_{abA}\left( N\right) \right) ^{\mu \nu }\left[ \eta
^{a}\eta ^{b}V_{\mu \nu }^{A}+\varepsilon _{\mu \nu \rho \lambda }\left(
2A^{a\rho }\eta ^{b}C^{A\lambda }+A^{a\rho }A^{b\lambda }C^{A}\right) \right]
\notag \\
&&+\left( P_{A}^{a}\left( N\right) \right) _{\mu \nu }\left( C^{A}B_{a}^{\mu
\nu }+3C_{\rho }^{A}\eta _{a}^{\mu \nu \rho }+\tfrac{1}{2}\varepsilon
_{\alpha \beta \gamma \delta }V^{A\mu \nu }\eta _{a}^{\alpha \beta \gamma
\delta }\right)  \notag \\
&&-\varepsilon ^{\mu \nu \rho \lambda }\left( \left( P_{abA}\left( N\right)
\right) _{\mu }B_{\nu \rho }^{\ast a}+\tfrac{1}{3}N_{abA}\eta _{\mu \nu \rho
}^{\ast a}\right) \left( A_{\lambda }^{b}C^{A}+\eta ^{b}C_{\lambda
}^{A}\right)  \notag \\
&&+\tfrac{1}{4!}\varepsilon ^{\mu \nu \rho \lambda }\left( Q_{a}\left(
f\right) \right) _{\mu }C_{\nu \rho \lambda }^{a}-\left( Q_{aA}\left(
f\right) \right) _{\mu }\left( A^{a\mu }C^{A}+\eta ^{a}C^{A\mu }\right)
\notag \\
&&-\tfrac{1}{4!}\left( Q_{\;\;b}^{a}\left( f\right) \right) ^{\mu }\left(
\varepsilon _{\alpha \beta \gamma \delta }A_{\mu }^{b}\eta _{a}^{\alpha
\beta \gamma \delta }-\varepsilon _{\mu \alpha \beta \gamma }\eta ^{b}\eta
_{a}^{\alpha \beta \gamma }\right)  \notag \\
&&-\tfrac{1}{2}\left( Q_{abc}\left( f\right) \right) ^{\mu }A_{\mu }^{a}\eta
^{b}\eta ^{c}.  \label{ai2}
\end{eqnarray}%
Using definitions (\ref{bfa15})--(\ref{bfa24}), we obtain
\begin{equation}
\delta a_{2}^{\prime \left( \mathrm{int}\right) }=\delta c_{2}+\gamma
e_{1}+\partial _{\mu }j_{1}^{\mu }+h_{1},  \label{eca2}
\end{equation}%
where%
\begin{eqnarray}
c_{2} &=&\left( \left( P_{AB}\left( N\right) \right) ^{\mu }C^{A}+\tfrac{1}{2%
}\left( P_{abB}\left( N\right) \right) ^{\mu }\eta ^{a}\eta ^{b}\right.
\notag \\
&&\left. -\varepsilon _{\alpha \beta \gamma \delta }\left( P_{B}^{a}\left(
N\right) \right) ^{\mu }\eta _{a}^{\alpha \beta \gamma \delta }\right)
V_{\mu }^{\ast B}+2\left( N_{A}^{a}\eta _{a}^{\ast }-\left( P_{A}^{a}\left(
N\right) \right) _{\mu }A_{a}^{\ast \mu }\right) C^{A}  \notag \\
&&+\left( \left( Q_{aA}\left( f\right) \right) ^{\mu \nu }C^{A}+\tfrac{1}{2}%
\left( Q_{abc}\left( f\right) \right) ^{\mu \nu }\eta ^{b}\eta ^{c}\right)
B_{\mu \nu }^{\ast a}  \notag \\
&&+\tfrac{1}{3}\varepsilon ^{\mu \nu \rho \lambda }\eta _{\mu \nu \rho
}^{\ast a}V_{B\lambda }\left( f_{aA}^{B}C^{A}+\tfrac{1}{2}f_{abc}^{B}\eta
^{b}\eta ^{c}\right)  \notag \\
&&-\tfrac{1}{2}\varepsilon ^{\mu \nu \rho \lambda }N_{abA}B_{\mu \nu }^{\ast
a}B_{\rho \lambda }^{\ast b}C^{A}+\tfrac{1}{4!}\varepsilon _{\alpha \beta
\gamma \delta }\left( Q_{\;\;b}^{a}\left( f\right) \right) ^{\mu \nu }B_{\mu
\nu }^{\ast b}\eta _{a}^{\alpha \beta \gamma \delta }  \notag \\
&&-\tfrac{1}{3}f_{\;\;b}^{Ba}\eta _{\mu \nu \rho }^{\ast b}V_{B\lambda }\eta
_{a}^{\mu \nu \rho \lambda },  \label{c2}
\end{eqnarray}%
\begin{eqnarray}
e_{1}&=&A_{\mu }^{a}\eta ^{b}\left( \left( P_{abB}\left( N\right) \right)
_{\nu }V^{B\mu \nu }+N_{abB}V^{\ast B\mu }\right) +2\left( P_{A}^{a}\left(
N\right) \right) _{\mu }C_{\nu }^{A}B_{a}^{\mu \nu }  \notag \\
&&-\varepsilon _{\mu \alpha \beta \gamma }\eta _{a}^{\alpha \beta \gamma
}\left( \left( P_{A}^{a}\left( N\right) \right) _{\nu }V^{A\mu \nu
}+N_{B}^{a}V^{\ast B\mu }\right) -2N_{A}^{a}A_{a}^{\ast \mu }C_{\mu }^{A}
\notag \\
&&+N_{abA}B_{\mu \nu }^{\ast a}\eta ^{b}V^{A\mu \nu }-\varepsilon ^{\mu \nu
\rho \lambda }\left( \tfrac{1}{2}\left( P_{abA}\left( N\right) \right) _{\mu
}A_{\nu }^{a}+N_{abA}B_{\mu \nu }^{\ast a}\right) A_{\rho }^{b}C_{\lambda
}^{A}  \notag \\
&&-C_{\mu }^{A}\left( \left( P_{AB}\left( N\right) \right) _{\nu }V^{B\mu
\nu }+N_{AB}V^{\ast B\mu }\right) -\varepsilon ^{\mu \nu \rho \lambda
}f_{aA}^{B}B_{\mu \nu }^{\ast a}V_{B\rho }C_{\lambda }^{A}  \notag \\
&&+\left( Q_{aA}\left( f\right) \right) ^{\mu \nu }\left( A_{\mu }^{a}C_{\nu
}^{A}+\tfrac{1}{4}\varepsilon _{\mu \nu \rho \lambda }\eta ^{a}V^{A\rho
\lambda }\right) -\tfrac{1}{2}\left( Q_{abc}\left( f\right) \right) ^{\mu
\nu }A_{\mu }^{a}A_{\nu }^{b}\eta ^{c}  \notag \\
&&+\varepsilon ^{\mu \nu \rho \lambda }f_{abc}^{B}B_{\mu \nu }^{\ast
a}V_{B\rho }A_{\lambda }^{b}\eta ^{c}+\tfrac{1}{2\cdot 4!}\varepsilon ^{\mu
\nu \rho \lambda }\left( Q_{a}\left( f\right) \right) _{\mu \nu }C_{\rho
\lambda }^{a}  \notag \\
&&+\tfrac{1}{4!}\left( Q_{\;\;b}^{a}\left( f\right) \right) ^{\mu \nu
}\left( \tfrac{1}{2}\varepsilon _{\mu \nu \rho \lambda }\eta ^{b}B_{a}^{\rho
\lambda }-\varepsilon _{\nu \alpha \beta \gamma }A_{\mu }^{b}\eta
_{a}^{\alpha \beta \gamma }\right)  \notag \\
&&+\tfrac{1}{4}f_{\;\;b}^{Ba}B_{\mu \nu }^{\ast b}V_{B\rho }\eta _{a}^{\mu
\nu \rho },  \label{e1}
\end{eqnarray}%
\begin{eqnarray}
j_{1}^{\mu } &=&-\left( N_{AB}C^{A}+\tfrac{1}{2}N_{abB}\eta ^{a}\eta
^{b}-\varepsilon _{\alpha \beta \gamma \delta }N_{B}^{a}\eta _{a}^{\alpha
\beta \gamma \delta }\right) V^{\ast B\mu }+2\left( N_{A}^{a}A_{a}^{\ast \mu
}\right.  \notag \\
&&\left. +\left( P_{A}^{a}\left( N\right) \right) _{\nu }B_{a}^{\mu \nu
}\right) C^{A}+\left( P_{A}^{a}\left( N\right) \right) _{\nu }\left(
6C_{\rho }^{A}\eta _{a}^{\mu \nu \rho }+\varepsilon _{\alpha \beta \gamma
\delta }V^{A\mu \nu }\eta _{a}^{\alpha \beta \gamma \delta }\right)  \notag
\\
&&-\left( P_{AB}\left( N\right) \right) _{\nu }\left( C^{A}V^{B\mu \nu }-%
\tfrac{1}{2}\varepsilon ^{\mu \nu \rho \lambda }C_{\rho }^{A}C_{\lambda
}^{B}\right)  \notag \\
&&-\varepsilon ^{\mu \nu \rho \lambda }N_{abA}B_{\nu \rho }^{\ast a}\left(
\eta ^{b}C_{\lambda }^{A}+A_{\lambda }^{b}C^{A}\right) -\tfrac{1}{2}\left(
P_{abA}\left( N\right) \right) _{\nu }\eta ^{a}\eta ^{b}V^{A\mu \nu }  \notag
\\
&&-\varepsilon ^{\mu \nu \rho \lambda }\left( P_{abA}\left( N\right) \right)
_{\nu }A_{\rho }^{a}\left( \eta ^{b}C_{\lambda }^{A}+\tfrac{1}{2}A_{\lambda
}^{b}C^{A}\right) +f_{\;\;b}^{Ba}B_{\nu \rho }^{\ast b}V_{B\lambda }\eta
_{a}^{\mu \nu \rho \lambda }  \notag \\
&&+\left( Q_{aA}\left( f\right) \right) ^{\mu \nu }\left( A_{\nu
}^{a}C^{A}+\eta ^{a}C_{\nu }^{A}\right) +\tfrac{1}{2}\left( Q_{abc}\left(
f\right) \right) ^{\mu \nu }A_{\nu }^{a}\eta ^{b}\eta ^{c}  \notag \\
&&-\varepsilon ^{\mu \nu \rho \lambda }B_{\nu \rho }^{\ast a}V_{B\lambda
}\left( f_{aA}^{B}C^{A}+\tfrac{1}{2}f_{abc}^{B}\eta ^{b}\eta ^{c}\right) -%
\tfrac{1}{4!}\varepsilon _{\nu \alpha \beta \gamma }\left( Q_{a}\left(
f\right) \right) ^{\mu \nu }C^{a\alpha \beta \gamma }  \notag \\
&&+\tfrac{1}{4!}\left( Q_{\;\;b}^{a}\left( f\right) \right) ^{\mu \nu
}\left( \varepsilon _{\alpha \beta \gamma \delta }A_{\nu }^{b}\eta
_{a}^{\alpha \beta \gamma \delta }-\varepsilon _{\nu \alpha \beta \gamma
}\eta ^{b}\eta _{a}^{\alpha \beta \gamma }\right) ,  \label{j1m}
\end{eqnarray}%
\begin{eqnarray}
h_{1} &=&\left( \left( P_{AB}\left( N\right) \right) ^{\mu }C^{A}+\tfrac{1}{2%
}\left( P_{abB}\left( N\right) \right) ^{\mu }\eta ^{a}\eta ^{b}-\varepsilon
_{\alpha \beta \gamma \delta }\left( P_{B}^{a}\left( N\right) \right) ^{\mu
}\eta _{a}^{\alpha \beta \gamma \delta }\right) V_{\mu }^{B}  \notag \\
&&+\left( N_{AB}C^{A}+\tfrac{1}{2}N_{abB}\eta ^{a}\eta ^{b}-\varepsilon
_{\alpha \beta \gamma \delta }N_{B}^{a}\eta _{a}^{\alpha \beta \gamma \delta
}\right) \partial ^{\mu }V_{\mu }^{\ast B}.  \label{h1}
\end{eqnarray}%
If we make the notation
\begin{equation}
a_{2}^{\left( \mathrm{int}\right) }\equiv a_{2}^{\prime \left( \mathrm{int}%
\right) }-c_{2},  \label{notata2}
\end{equation}%
then (\ref{eca2}) is equivalent with the equation%
\begin{equation}
\delta a_{2}^{\left( \mathrm{int}\right) }=\gamma e_{1}+\partial _{\mu
}j_{1}^{\mu }+h_{1}.  \label{eca3}
\end{equation}%
Comparing (\ref{eca3}) with equation (\ref{fo6b}) for $k=2$, we obtain that
a necessary condition for the existence of a local $a_{1}^{\left( \mathrm{int%
}\right) }$ is
\begin{equation}
h_{1}=\delta g_{2}+\gamma f_{1}+\partial _{\mu }l_{1}^{\mu },  \label{eca3a}
\end{equation}%
with $g_{2}$, $f_{1}$, and $l_{1}^{\mu }$ local functions. We show that
equation (\ref{eca3a}) cannot hold (locally) unless $h_{1}=0$. Indeed, assuming (\ref%
{eca3a}) is satisfied, we act with $\delta $ on it and use its nilpotency
and anticommutation with $\gamma $, which yields the necessary condition%
\begin{equation}
\delta h_{1}=\gamma (-\delta f_{1})+\partial _{\mu }\left( \delta l_{1}^{\mu
}\right) .  \label{eca3b}
\end{equation}%
On the other hand, direct computation provides
\begin{eqnarray}
\delta h_{1} &=&\gamma \left[ \left( N_{AB}C_{\mu }^{A}-N_{abB}A_{\mu
}^{a}\eta ^{b}+\varepsilon _{\mu \alpha \beta \gamma }N_{B}^{a}\eta
_{a}^{\alpha \beta \gamma }\right) V^{B\mu }\right]  \notag \\
&&+\partial _{\mu }\left[ -\left( N_{AB}C^{A}+\tfrac{1}{2}N_{abB}\eta
^{a}\eta ^{b}-\varepsilon _{\alpha \beta \gamma \delta }N_{B}^{a}\eta
_{a}^{\alpha \beta \gamma \delta }\right) V^{B\mu }\right] .  \label{eca3c}
\end{eqnarray}%
Juxtaposing (\ref{eca3b}) and (\ref{eca3c}) and looking at definitions (\ref%
{bfa15})--(\ref{bfa24}), it follows that $V^{B\mu }$ must necessarily be $%
\delta $-exact modulo $d$ in the space of local functions. Since this is
obviously not true, we find that (\ref{eca3b}) cannot be satisfied and
consequently neither does equation (\ref{eca3a}). Thus, the consistency of $%
a_{2}^{\left( \mathrm{int}\right) }$ leads to the equation
\begin{equation}
h_{1}=0,  \label{ecca2}
\end{equation}%
which further implies that the functions $N_{abA}$, $N_{AB}$, and $N_{A}^{a}$
must vanish
\begin{equation}
N_{abA}=N_{AB}=N_{A}^{a}=0.  \label{negal0}
\end{equation}%
Based on (\ref{negal0}), from (\ref{fo8}), (\ref{fo10}), (\ref{ai2}), (\ref%
{c2}), (\ref{e1}), (\ref{notata2}), and (\ref{eca3}) we get the components
of antighost number $4$, $3$, and $2$ from the nonintegrated density of the
first-order deformation as
\begin{equation}
a_{4}^{\left( \mathrm{int}\right) }=0,  \label{fo12}
\end{equation}%
\begin{eqnarray}
a_{3}^{\left( \mathrm{int}\right) } &=&Q_{aA}\left( f\right) \eta ^{a}C^{A}+%
\tfrac{1}{3!}Q_{abc}\left( f\right) \eta ^{a}\eta ^{b}\eta ^{c}  \notag \\
&&+\tfrac{1}{4!}\varepsilon _{\alpha \beta \gamma \delta }\left(
Q_{\;\;b}^{a}\left( f\right) \eta ^{b}\eta _{a}^{\alpha \beta \gamma \delta
}+Q_{a}\left( f\right) C^{a\alpha \beta \gamma \delta }\right) ,
\label{fo13}
\end{eqnarray}%
\begin{eqnarray}
a_{2}^{\left( \mathrm{int}\right) } &=&\tfrac{1}{4!}\varepsilon ^{\mu \nu
\rho \lambda }\left( Q_{a}\left( f\right) \right) _{\mu }C_{\nu \rho \lambda
}^{a}-\left( Q_{aA}\left( f\right) \right) ^{\mu }\left( A_{\mu
}^{a}C^{A}+\eta ^{a}C_{\mu }^{A}\right)  \notag \\
&&-\tfrac{1}{2}\left( Q_{abc}\left( f\right) \right) ^{\mu }A_{\mu }^{a}\eta
^{b}\eta ^{c}-\tfrac{1}{4!}\left( Q_{\;\;b}^{a}\left( f\right) \right) ^{\mu
}\left( \varepsilon _{\alpha \beta \gamma \delta }A_{\mu }^{b}\eta
_{a}^{\alpha \beta \gamma \delta }\right.  \notag \\
&&\left. -\varepsilon _{\mu \alpha \beta \gamma }\eta ^{b}\eta _{a}^{\alpha
\beta \gamma }\right) -\left( \left( Q_{aA}\left( f\right) \right) ^{\mu \nu
}C^{A}+\tfrac{1}{2}\left( Q_{abc}\left( f\right) \right) ^{\mu \nu }\eta
^{b}\eta ^{c}\right) B_{\mu \nu }^{\ast a}  \notag \\
&&-\tfrac{1}{3}\varepsilon ^{\mu \nu \rho \lambda }\eta _{\mu \nu \rho
}^{\ast a}V_{B\lambda }\left( f_{aA}^{B}C^{A}+\tfrac{1}{2}f_{abc}^{B}\eta
^{b}\eta ^{c}\right) +\tfrac{1}{3}f_{\;\;b}^{Ba}\eta _{\mu \nu \rho }^{\ast
b}V_{B\lambda }\eta _{a}^{\mu \nu \rho \lambda }  \notag \\
&&-\tfrac{1}{4!}\varepsilon _{\alpha \beta \gamma \delta }\left(
Q_{\;\;b}^{a}\left( f\right) \right) ^{\mu \nu }B_{\mu \nu }^{\ast b}\eta
_{a}^{\alpha \beta \gamma \delta }+\tfrac{1}{2}R_{ab}\left( g\right) \eta
^{a}\eta ^{b}  \notag \\
&&+R_{A}\left( g\right) C^{A}+\tfrac{1}{4!}\varepsilon _{\mu \nu \rho
\lambda }R^{a}\left( g\right) \eta _{a}^{\mu \nu \rho \lambda }.
\label{fo14}
\end{eqnarray}%
The objects $R_{ab}\left( g\right) $, $R_{A}\left( g\right) $, and $%
R^{a}\left( g\right) $ are generated by formula (\ref{q}) via the smooth
functions of the undifferentiated scalar fields $g_{ab}^{AB}$, $g_{\quad
C}^{AB}$, and $g^{aAB}$, respectively. All these functions are antisymmetric
in $A$ and $B$ and in addition $g_{ab}^{AB}$ are antisymmetric also in their
(lower) BF collection indices.

Replacing now expression (\ref{fo14}) into equation (\ref{fo6b}) for
$k=2$, we obtain that the interacting piece of antighost number $1$
from the first-order deformation is written as
\begin{eqnarray}
a_{1}^{\prime \left( \mathrm{int}\right) } &=&-\tfrac{1}{2\cdot 4!}%
\varepsilon ^{\mu \nu \rho \lambda }\left( Q_{a}\left( f\right) \right)
_{\mu \nu }C_{\rho \lambda }^{a}-\left( Q_{aA}\left( f\right) \right) ^{\mu
\nu }\left( A_{\mu }^{a}C_{\nu }^{A}\right.  \notag \\
&&\left. +\tfrac{1}{4}\varepsilon _{\mu \nu \rho \lambda }\eta ^{a}V^{A\rho
\lambda }\right) +\tfrac{1}{4!}\left( Q_{\;\;b}^{a}\left( f\right) \right)
^{\mu \nu }\left( \varepsilon _{\nu \alpha \beta \gamma }A_{\mu }^{b}\eta
_{a}^{\alpha \beta \gamma }-\tfrac{1}{2}\varepsilon _{\mu \nu \alpha \beta
}\eta ^{b}B_{a}^{\alpha \beta }\right)  \notag \\
&&+\left( R_{A}\left( g\right) \right) ^{\mu }C_{\mu }^{A}-\left(
R_{ab}\left( g\right) \right) ^{\mu }A_{\mu }^{a}\eta ^{b}-\tfrac{1}{4!}%
\varepsilon _{\mu \nu \rho \lambda }\left( R^{a}\left( g\right) \right)
^{\mu }\eta _{a}^{\nu \rho \lambda }  \notag \\
&&+\varepsilon ^{\mu \nu \rho \lambda }B_{\mu \nu }^{\ast a}V_{B\rho }\left(
f_{aA}^{B}C_{\lambda }^{A}-f_{abc}^{B}A_{\lambda }^{b}\eta ^{c}-\tfrac{1}{4!}%
\varepsilon _{\lambda \alpha \beta \gamma }f_{\;\;a}^{Bb}\eta _{b}^{\alpha
\beta \gamma }\right)  \notag \\
&&+\tfrac{1}{2}\left( Q_{abc}\left( f\right) \right) ^{\mu \nu }A_{\mu
}^{a}A_{\nu }^{b}\eta ^{c}.  \label{fo15}
\end{eqnarray}%
Using definitions (\ref{bfa15})--(\ref{bfa24}), by direct computation we
obtain that
\begin{equation}
\delta a_{1}^{\prime \left( \mathrm{int}\right) }=\delta c_{1}+\gamma
e_{0}+\partial _{\mu }j_{0}^{\mu }+h_{0},  \label{cond00}
\end{equation}%
with
\begin{equation}
c_{1}=-\eta ^{a}V_{B\mu }\left( f_{aA}^{B}V^{\ast A\mu }+\tfrac{1}{12}%
f_{\;\;a}^{Bb}A_{b}^{\ast \mu }+\tfrac{1}{2}\varepsilon ^{\mu \nu \rho
\lambda }g_{ab}^{AB}V_{A\nu }B_{\rho \lambda }^{\ast b}\right) ,  \label{c1}
\end{equation}%
\begin{eqnarray}
e_{0} &=&-\tfrac{1}{2}\varepsilon ^{\mu \nu \rho \lambda }V_{A\mu }\left( -%
\tfrac{1}{3}f_{abc}^{A}A_{\nu }^{c}+\tfrac{1}{2}g_{ab}^{AB}V_{B\nu }\right)
A_{\rho }^{a}A_{\lambda }^{b}  \notag \\
&&+\tfrac{1}{4!}f_{a}^{A}V_{A}^{\mu }H_{\mu }^{a}-A_{\mu }^{a}V_{A\nu
}\left( f_{aB}^{A}V^{B\mu \nu }+\tfrac{1}{12}f_{\;\;a}^{Ab}B_{b}^{\mu \nu
}\right)  \notag \\
&&-\tfrac{1}{2}\left( g_{\quad C}^{AB}V_{\mu \nu }^{C}+\tfrac{1}{12}%
g^{aAB}B_{a\mu \nu }\right) V_{A}^{\mu }V_{B}^{\nu },  \label{e0}
\end{eqnarray}%
\begin{eqnarray}
j_{0}^{\mu } &=&V_{A\nu }\left( \tfrac{1}{12}f_{a}^{A}C^{a\mu \nu
}+f_{aB}^{A}\eta ^{a}V^{B\mu \nu }\right) +\tfrac{1}{4}f_{\;\;b}^{Aa}V_{A\nu
}\left( A_{\rho }^{b}\eta _{a}^{\mu \nu \rho }\right.  \notag \\
&&\left. +\tfrac{1}{3}\eta ^{b}B_{a}^{\mu \nu }\right) -\tfrac{1}{8}%
g^{aAB}V_{A\nu }V_{B\rho }\eta _{a}^{\mu \nu \rho }-\varepsilon ^{\mu \nu
\rho \lambda }\left[ f_{aB}^{A}A_{\nu }^{a}V_{A\lambda }C_{\rho }^{B}\right.
\notag \\
&&\left. -\tfrac{1}{2}f_{abc}^{A}A_{\nu }^{a}A_{\rho }^{b}\eta
^{c}V_{A\lambda }-\tfrac{1}{2}V_{A\nu }V_{B\rho }\left( g_{\quad
C}^{AB}C_{\lambda }^{C}-g_{ab}^{AB}A_{\lambda }^{a}\eta ^{b}\right) \right] ,
\label{j0}
\end{eqnarray}%
\begin{equation}
h_{0}=-f_{aB}^{A}\eta ^{a}V_{A}^{\mu }V_{\mu }^{B}.  \label{h0}
\end{equation}%
At this stage we act like between formulas (\ref{notata2}) and (\ref{negal0}%
). If we make the notation
\begin{equation}
a_{1}^{\left( \mathrm{int}\right) }=a_{1}^{\prime \left( \mathrm{int}\right)
}-c_{1},  \label{notata1}
\end{equation}%
then (\ref{cond00}) becomes%
\begin{equation}
\delta a_{1}^{\left( \mathrm{int}\right) }=\gamma e_{0}+\partial _{\mu
}j_{0}^{\mu }+h_{0},  \label{eca4}
\end{equation}%
which, compared with equation (\ref{fo6b}) for $k=1$, reveals that the
existence of $a_{0}^{(\mathrm{int})}$ demands%
\begin{equation}
h_{0}=\delta g_{1}+\gamma f_{0}+\partial _{\mu }l_{0}^{\mu },  \label{eca4a}
\end{equation}%
with $g_{1}$, $f_{0}$, and $l_{0}^{\mu }$ some local elements. Using (\ref%
{h0}) and definitions (\ref{bfa15})--(\ref{bfa24}), straightforward
calculation shows that (\ref{eca4a}) cannot be valid, and hence the
consistency of $a_{1}^{\left( \mathrm{int}\right) }$ leads to the equation
\begin{equation}
h_{0}=0,  \label{ph14}
\end{equation}%
which requires the antisymmetry of the functions $f_{aAB}$ ($\equiv
k_{AM}f_{aB}^{M}$) with respect to their collection indices from the
two-form sector
\begin{equation}
f_{aAB}=-f_{aBA}.  \label{ant}
\end{equation}%
With the help of (\ref{fo15}), (\ref{c1}), (\ref{e0}), (\ref{notata1}), (\ref%
{eca4}), and (\ref{ant}) we completely determine $a_{1}^{\left( \mathrm{int}%
\right) }$ and then $a_{0}^{\left( \mathrm{int}\right) }$ as solution to (%
\ref{fo6b}) for $k=1$ in the form
\begin{eqnarray}
a_{1}^{\left( \mathrm{int}\right) } &=&-\tfrac{1}{2\cdot 4!}\varepsilon
^{\mu \nu \rho \lambda }\left( Q_{a}\left( f\right) \right) _{\mu \nu
}C_{\rho \lambda }^{a}-\left( Q_{aA}\left( f\right) \right) ^{\mu \nu
}\left( A_{\mu }^{a}C_{\nu }^{A}\right.  \notag \\
&&\left. +\tfrac{1}{4}\varepsilon _{\mu \nu \rho \lambda }\eta ^{a}V^{A\rho
\lambda }\right) +\tfrac{1}{4!}\left( Q_{\;\;b}^{a}\left( f\right) \right)
^{\mu \nu }\left( \varepsilon _{\nu \alpha \beta \gamma }A_{\mu }^{b}\eta
_{a}^{\alpha \beta \gamma }-\tfrac{1}{2}\varepsilon _{\mu \nu \alpha \beta
}\eta ^{b}B_{a}^{\alpha \beta }\right)  \notag \\
&&+\left( R_{A}\left( g\right) \right) ^{\mu }C_{\mu }^{A}-\left(
R_{ab}\left( g\right) \right) ^{\mu }A_{\mu }^{a}\eta ^{b}-\tfrac{1}{4!}%
\varepsilon _{\mu \nu \rho \lambda }\left( R^{a}\left( g\right) \right)
^{\mu }\eta _{a}^{\nu \rho \lambda }  \notag \\
&&+\varepsilon ^{\mu \nu \rho \lambda }B_{\mu \nu }^{\ast a}V_{B\rho }\left(
f_{aA}^{B}C_{\lambda }^{A}-f_{abc}^{B}A_{\lambda }^{b}\eta ^{c}-\tfrac{1}{4!}%
\varepsilon _{\lambda \alpha \beta \gamma }f_{\;\;a}^{Bb}\eta _{b}^{\alpha
\beta \gamma }\right)  \notag \\
&&+\tfrac{1}{2}\left( Q_{abc}\left( f\right) \right) ^{\mu \nu }A_{\mu
}^{a}A_{\nu }^{b}\eta ^{c}+\eta ^{a}V_{B\mu }\left( f_{aA}^{B}V^{\ast A\mu }+%
\tfrac{1}{12}f_{\;\;a}^{Bb}A_{b}^{\ast \mu }\right.  \notag \\
&&\left. +\tfrac{1}{2}\varepsilon ^{\mu \nu \rho \lambda }g_{ab}^{AB}V_{A\nu
}B_{\rho \lambda }^{\ast b}\right) ,  \label{fo16}
\end{eqnarray}%
\begin{eqnarray}
a_{0}^{\left( \mathrm{int}\right) } &=&\tfrac{1}{2}\varepsilon ^{\mu \nu
\rho \lambda }V_{A\mu }\left( -\tfrac{1}{3}f_{abc}^{A}A_{\nu }^{c}+\tfrac{1}{%
2}g_{ab}^{AB}V_{B\nu }\right) A_{\rho }^{a}A_{\lambda }^{b}  \notag \\
&&-\tfrac{1}{4!}f_{a}^{A}V_{A}^{\mu }H_{\mu }^{a}+f_{aB}^{A}A_{\mu
}^{a}V_{A\nu }V^{B\mu \nu }+\tfrac{1}{12}f_{\;\;a}^{Ab}A_{\mu }^{a}V_{A\nu
}B_{b}^{\mu \nu }  \notag \\
&&+\tfrac{1}{2}\left( g_{\quad C}^{AB}V_{\mu \nu }^{C}+\tfrac{1}{12}%
g^{aAB}B_{a\mu \nu }\right) V_{A}^{\mu }V_{B}^{\nu }.  \label{fo17}
\end{eqnarray}

Thus, we can write the final form of the interacting part from the
first-order deformation of the solution to the master equation for a
collection of BF models and a set of two-form gauge fields as%
\begin{equation}
S_{1}^{\left( \mathrm{int}\right) }\equiv \int d^{4}x\,a^{(\mathrm{int}%
)}=\int d^{4}x\left( a_{3}^{\left( \mathrm{int}\right) }+a_{2}^{\left(
\mathrm{int}\right) }+a_{1}^{\left( \mathrm{int}\right) }+a_{0}^{\left(
\mathrm{int}\right) }\right) ,  \label{s1int}
\end{equation}%
where the $4$ components from (\ref{s1int}) read as in formulas (\ref{fo13}%
)--(\ref{fo14}) and (\ref{fo16})--(\ref{fo17}), respectively. The previous
first-order deformation is parameterized by $7$ functions, $f_{abc}^{A}$, $%
g_{ab}^{AB}$, $f_{a}^{A}$, $f_{aB}^{A}$, $f_{\;\;a}^{Ab}$, $g_{\quad C}^{AB}$%
, and $g^{aAB}$, which depend smoothly on the undifferentiated scalar fields
$\varphi _{d}$ and are antisymmetric as follows: $f_{abc}^{A}$ in the
indices $\left\{ a,b,c\right\} $, $g_{ab}^{AB}$ with respect to $\left\{
A,B\right\} $ and $\left\{ a,b\right\} $, and $f_{aAB}\equiv
k_{AM}f_{aB}^{M} $ together with $g_{\quad C}^{AB}$ and $g^{aAB}$ in $%
\left\{ A,B\right\} $. It is easy to see that (\ref{s1int}) also
includes the general solution that describes the self-interactions
among the two-form gauge fields. Indeed, if we isolate from
$S_{1}^{\left( \mathrm{int}\right) } $ the part containing the
functions $g_{\quad C}^{AB}$, represent these functions as some
series in the undifferentiated scalar fields, $g_{\quad
C}^{AB}\left( \varphi _{a}\right) =k_{\quad C}^{AB}+k_{\quad
C}^{ABa}\varphi _{a}+\cdots $, where $k_{\quad C}^{AB}$ and
$k_{\quad C}^{ABa}$ are some real constants, antisymmetric in their
upper, capital indices, and retain only the terms including
$k_{\quad C}^{AB}$, then we obtain
\begin{eqnarray}
S_{1}^{\left( \mathrm{int}\right) }(k) &\equiv &\int d^{4}x\,a^{(\mathrm{V}%
)}=\int d^{4}x\left( a_{2}^{\left( \mathrm{V}\right) }+a_{1}^{\left( \mathrm{%
V}\right) }+a_{0}^{\left( \mathrm{V}\right) }\right)  \notag \\
&=&k_{\quad C}^{AB}\int d^{4}x\left[ \left( C_{A}^{\ast \mu }V_{B\mu }+%
\tfrac{1}{2}\varepsilon _{\mu \nu \rho \lambda }V_{A}^{\ast \mu \nu
}V_{B}^{\ast \rho \lambda }\right) C^{C}\right.  \notag \\
&&\left. +\varepsilon _{\mu \nu \rho \lambda }V_{A}^{\ast \mu \nu
}V_{B}^{\rho }C^{C\lambda }+\tfrac{1}{2}V_{\mu \nu }^{C}V_{A}^{\mu
}V_{B}^{\nu }\right] ,  \label{s1V}
\end{eqnarray}%
which has been shown in~\cite{henfreedman} to be the most general form of
the first-order deformation for a set of two-form gauge fields in four
spacetime dimensions with the Lagrangian action written in first-order form.
In conclusion, the overall first-order deformation of the solution to the
master equation for the model under study is expressed like the sum between (%
\ref{s1int}) and the piece responsible for the interactions from the BF
sector%
\begin{equation}
S_{1}=S_{1}^{\left( \mathrm{BF}\right) }+S_{1}^{\left( \mathrm{int}\right) },
\label{s1final}
\end{equation}%
where%
\begin{equation}
S_{1}^{\left( \mathrm{BF}\right) }=\int d^{4}x\,a^{(\mathrm{BF})},
\label{s1BF}
\end{equation}%
with $a^{(\mathrm{BF})}$ provided by (\ref{descBF}) and (\ref{a4})--(\ref{a0}%
). We recall that $S_{1}^{\left( \mathrm{BF}\right) }$ is parameterized by $%
4 $ kinds of smooth functions of the undifferentiated scalar fields: $W_{ab}$%
, $M_{ab}^{c}$, $M^{ab}$, and $M_{abcd}$, where $M_{ab}^{c}$ are
antisymmetric in their lower indices, $M^{ab}$ are symmetric, and $M_{abcd}$
are completely antisymmetric.

\subsection{Second-order deformation\label{highord}}

Next, we investigate the equations responsible for higher-order
deformations. The second-order deformation is governed by equation (\ref%
{bff3.5}). Making use of the first-order deformation derived in the previous
subsection, after some computation we organize the second term on the
left-hand side of (\ref{bff3.5}) like
\begin{equation}
\left( S_{1},S_{1}\right) =\int d^{4}x\left( \Delta +\bar{\Delta}\right) ,
\label{so1}
\end{equation}%
where
\begin{eqnarray}
\Delta &=&\sum\limits_{p=0}^{4}\left( K_{,m_{1}\ldots m_{p}}^{abc}\frac{%
\partial ^{p}t_{abc}}{\partial \varphi _{m_{1}}\ldots \partial \varphi
_{m_{p}}}+K_{d,m_{1}\ldots m_{p}}^{abc}\frac{\partial ^{p}t_{abc}^{d}}{%
\partial \varphi _{m_{1}}\ldots \partial \varphi _{m_{p}}}\right.  \notag \\
&&+K_{m_{1}\ldots m_{p}}^{abcdf}\frac{\partial ^{p}t_{abcdf}}{\partial
\varphi _{m_{1}}\ldots \partial \varphi _{m_{p}}}+K_{b,m_{1}\ldots m_{p}}^{a}%
\frac{\partial ^{p}t_{a}^{b}}{\partial \varphi _{m_{1}}\ldots \partial
\varphi _{m_{p}}}  \notag \\
&&\left. +K_{ab,m_{1}\ldots m_{p}}^{c}\frac{\partial ^{p}t_{c}^{ab}}{%
\partial \varphi _{m_{1}}\ldots \partial \varphi _{m_{p}}}\right)
\label{so2}
\end{eqnarray}%
and
\begin{eqnarray}
\bar{\Delta} &=&\sum\limits_{p=0}^{3}\left( X_{A,m_{1}\ldots m_{p}}^{abB}%
\frac{\partial ^{p}T_{abB}^{A}}{\partial \varphi _{m_{1}}\ldots \partial
\varphi _{m_{p}}}+X_{A,m_{1}\ldots m_{p}}^{abcd}\frac{\partial
^{p}T_{abcd}^{A}}{\partial \varphi _{m_{1}}\ldots \partial \varphi _{m_{p}}}%
\right.  \notag \\
&&+X_{A,m_{1}\ldots m_{p}}^{ab}\frac{\partial ^{p}T_{ab}^{A}}{\partial
\varphi _{m_{1}}\ldots \partial \varphi _{m_{p}}}+X_{Ac,m_{1}\ldots
m_{p}}^{ab}\frac{\partial ^{p}T_{ab}^{Ac}}{\partial \varphi _{m_{1}}\ldots
\partial \varphi _{m_{p}}}  \notag \\
&&\left. +X_{Aab,m_{1}\ldots m_{p}}\frac{\partial ^{p}T^{Aab}}{\partial
\varphi _{m_{1}}\ldots \partial \varphi _{m_{p}}}+X_{a,m_{1}\ldots
m_{p}}^{AB}\frac{\partial ^{p}T_{AB}^{a}}{\partial \varphi _{m_{1}}\ldots
\partial \varphi _{m_{p}}}\right)  \notag \\
&&+\sum\limits_{p=0}^{2}\left( X_{m_{1}\ldots m_{p}}^{aABC}\frac{\partial
^{p}T_{aABC}}{\partial \varphi _{m_{1}}\ldots \partial \varphi _{m_{p}}}%
+X_{AB,m_{1}\ldots m_{p}}^{abc}\frac{\partial ^{p}T_{abc}^{AB}}{\partial
\varphi _{m_{1}}\ldots \partial \varphi _{m_{p}}}\right.  \notag \\
&&\left. +X_{AB,m_{1}\ldots m_{p}}^{a}\frac{\partial ^{p}T_{a}^{AB}}{%
\partial \varphi _{m_{1}}\ldots \partial \varphi _{m_{p}}}%
+X_{ABa,m_{1}\ldots m_{p}}^{b}\frac{\partial ^{p}T_{b}^{ABa}}{\partial
\varphi _{m_{1}}\ldots \partial \varphi _{m_{p}}}\right)  \notag \\
&&+\sum\limits_{p=0}^{1}\left( X_{ABCD,m_{1}\ldots m_{p}}\frac{\partial
^{p}T^{ABCD}}{\partial \varphi _{m_{1}}\ldots \partial \varphi _{m_{p}}}%
+X_{ABC,m_{1}\ldots m_{p}}^{ab}\frac{\partial ^{p}T_{ab}^{ABC}}{\partial
\varphi _{m_{1}}\ldots \partial \varphi _{m_{p}}}\right.  \notag \\
&&\left. +X_{a,m_{1}\ldots m_{p}}^{ABC}\frac{\partial ^{p}T_{ABC}^{a}}{%
\partial \varphi _{m_{1}}\ldots \partial \varphi _{m_{p}}}\right)
+X_{ABCD}^{a}T_{a}^{ABCD}.  \label{so3}
\end{eqnarray}%
In formulas (\ref{so2}) and (\ref{so3}) we used the notations
\begin{eqnarray}
t_{abc} &=&W_{ec}M_{ab}^{e}+W_{ea}\frac{\partial W_{bc}}{\partial \varphi
_{e}}+W_{eb}\frac{\partial W_{ca}}{\partial \varphi _{e}},  \label{so4a} \\
t_{abc}^{d} &=&W_{e[a}\frac{\partial M_{bc]}^{d}}{\partial \varphi _{e}}%
+M_{e[a}^{d}M_{bc]}^{e}+M^{de}M_{eabc},  \label{so4b} \\
t_{abcdf} &=&W_{e[a}\frac{\partial M_{bcdf]}}{\partial \varphi _{e}}%
+M_{e[abc}M_{df]}^{e},  \label{so4c} \\
t_{a}^{b} &=&M^{be}W_{ea},  \label{so4d} \\
t_{a}^{bc} &=&W_{ea}\frac{\partial M^{bc}}{\partial \varphi _{e}}%
+M_{ea}^{(b}M_{\left. {}\right. }^{c)e},  \label{so4f}
\end{eqnarray}%
\begin{equation}
T_{ab}^{A}=f_{aM}^{A}f_{b}^{M}+f_{e}^{A}\frac{\partial W_{ab}}{\partial
\varphi _{e}}+W_{ea}\frac{\partial f_{b}^{A}}{\partial \varphi _{e}}%
+2W_{eb}f_{\;\;a}^{Ae},  \label{so5}
\end{equation}%
\begin{equation}
T_{a}^{AB}=f_{e}^{A}\frac{\partial f_{a}^{B}}{\partial \varphi _{e}}%
-f_{e}^{B}\frac{\partial f_{a}^{A}}{\partial \varphi _{e}}-4!\left( g_{\quad
M}^{AB}f_{a}^{M}+2W_{ea}g^{eAB}\right) ,  \label{so6}
\end{equation}%
\begin{eqnarray}
T_{ab}^{Ac} &=&f_{aM}^{A}f_{\;\;b}^{Mc}-f_{bM}^{A}f_{\;\;a}^{Mc}-\tfrac{1}{2}%
f_{e}^{A}\frac{\partial M_{ab}^{c}}{\partial \varphi _{e}}%
+f_{\;\;e}^{Ac}M_{ab}^{e}  \notag \\
&&+f_{\;\;[a}^{Ae}M_{b]e}^{c}-2\cdot 4!f_{eab}^{A}M^{ec}+W_{e[a}\frac{%
\partial f_{\;\;b]}^{Ac}}{\partial \varphi _{e}},  \label{so7}
\end{eqnarray}%
\begin{eqnarray}
T_{abcd}^{A} &=&W_{e[a}\frac{\partial f_{bcd]}^{A}}{\partial \varphi _{e}}%
+f_{e[ab}^{A}M_{cd]}^{e}+f_{M[a}^{A}f_{bcd]}^{M}  \notag \\
&&+\tfrac{1}{2\cdot 4!}\left( \tfrac{1}{2}f_{e}^{A}\frac{\partial M_{abcd}}{%
\partial \varphi _{e}}-f_{\;\;[a}^{Ae}M_{bcd]e}^{\left. {}\right. }\right) ,
\label{so8}
\end{eqnarray}%
\begin{equation}
T^{Aab}=f_{e}^{A}\frac{\partial M^{ab}}{\partial \varphi _{e}}%
-2f_{\;\;e}^{Aa}M^{be}-2f_{\;\;e}^{Ab}M^{ae},  \label{so9}
\end{equation}%
\begin{equation}
T_{abB}^{A}=f_{M[a}^{A}f_{b]B}^{M}+f_{eB}^{A}M_{ab}^{e}+W_{e[a}\frac{%
\partial f_{b]B}^{A}}{\partial \varphi _{e}},  \label{so10}
\end{equation}%
\begin{eqnarray}
T_{aABC} &=&f_{Ae}\frac{\partial f_{aBC}}{\partial \varphi _{e}}-f_{Be}\frac{%
\partial f_{aAC}}{\partial \varphi _{e}}+2f_{\;\;Aa}^{e}f_{eBC}-2f_{\;%
\;Ba}^{e}f_{eAC}  \notag \\
&&+4!\left( -g_{ABM}f_{aC}^{M}+W_{ea}\frac{\partial g_{ABC}}{\partial
\varphi _{e}}+f_{a[A}^{M}g_{B]MC}^{\left. {}\right. }\right) ,  \label{so11}
\end{eqnarray}%
\begin{equation}
T_{AB}^{a}=f_{eAB}M^{ea},  \label{so12}
\end{equation}%
\begin{eqnarray}
T_{abc}^{AB} &=&f_{e}^{A}\frac{\partial f_{abc}^{B}}{\partial \varphi _{e}}%
-f_{e}^{B}\frac{\partial f_{abc}^{A}}{\partial \varphi _{e}}%
+2f_{\;\;[a}^{Ae}f_{bc]e}^{B}-2f_{\;\;[a}^{Be}f_{bc]e}^{A}  \notag \\
&&+\tfrac{1}{2}g^{eAB}M_{abce}+4!\left( g_{e[a}^{AB}M_{bc]}^{e}+W_{e[a}\frac{%
\partial g_{bc]}^{AB}}{\partial \varphi _{e}}\right)  \notag \\
&&-4!\left( g_{\quad M}^{AB}f_{abc}^{M}+f_{M[a}^{[A}g_{bc]}^{B]M}\right) ,
\label{so13}
\end{eqnarray}%
\begin{eqnarray}
T_{b}^{ABa} &=&f_{e}^{A}\frac{\partial f_{\;\;b}^{Ba}}{\partial \varphi _{e}}%
-f_{e}^{B}\frac{\partial f_{\;\;b}^{Aa}}{\partial \varphi _{e}}%
-2f_{\;\;e}^{Aa}f_{\;\;b}^{Be}+2f_{\;\;e}^{Ba}f_{\;\;b}^{Ae}  \notag \\
&&+4!\left( g^{eAB}M_{eb}^{a}+W_{eb}\frac{\partial g^{aAB}}{\partial \varphi
_{e}}\right) -4!\left( g_{\quad M}^{AB}f_{\;\;b}^{Ma}\right.  \notag \\
&&\left. +2\cdot
4!g_{eb}^{AB}M^{ea}+f_{bM}^{A}g^{aBM}-f_{bM}^{B}g^{aAM}\right) ,
\label{so14}
\end{eqnarray}%
\begin{equation}
T^{ABCD}=g_{\left. {}\right. }^{e[AB}f_{e}^{C]D}-\tfrac{1}{2}f_{e}^{[A}\frac{%
\partial g^{BC]D}}{\partial \varphi _{e}}-12g_{\quad M}^{[AB}g_{\left.
{}\right. }^{C]MD},  \label{so15}
\end{equation}%
\begin{equation}
T_{ab}^{ABC}=g_{\left. {}\right. }^{e[AB}f_{eab}^{C]}-\tfrac{1}{2}f_{e}^{[A}%
\frac{\partial g_{ab}^{BC]}}{\partial \varphi _{e}}-12g_{\quad
M}^{[AB}g_{ab}^{C]M}+g_{e[a}^{[AB}f_{\;\;b]}^{C]e},  \label{so16}
\end{equation}%
\begin{equation}
T_{ABC}^{a}=g_{[AB}^{e}f_{\;\;C]e}^{a}-\tfrac{1}{2}f_{e[A}\frac{\partial
g_{BC]}^{a}}{\partial \varphi _{e}}-12g_{[AB}^{\quad M}g_{C]M}^{a},
\label{so17}
\end{equation}%
\begin{equation}
T_{a}^{ABCD}=g_{\left. {}\right. }^{e[AB}g_{ea}^{CD]},  \label{so18}
\end{equation}%
where the functions $g_{ABC}$, $g^{CMD}$, and $g_{AB}^{\quad M}$ result from
$g_{\quad M}^{AB}$ by appropriately lowering or raising the two-form
collection indices with the help of the metric $k_{AB}$ or its inverse $%
k^{AB}$: $g_{ABC}=k_{AM}k_{BN}g_{\quad C}^{MN}$, $g^{CMD}=g_{\quad
E}^{CM}k^{ED}$, $g_{AB}^{\quad M}=k_{AE}k_{BF}g_{\quad N}^{EF}k^{NM}$.\ The
remaining objects, of the type $K$ or $X$, are listed in Appendix \ref%
{appendixA}. Each of them is a polynomial of ghost number $1$ involving only
the \emph{undifferentiated} fields/ghosts and antifields. Comparing equation
(\ref{bff3.5}) with (\ref{so1}), we obtain that the existence of $S_{2}$
requires that $\int d^{4}x\left( \Delta +\bar{\Delta}\right) $ is $s$-exact.
This is not possible since all the objects denoted by $K$ or $X$ are
polynomials comprising only undifferentiated fields/ghosts/antifields, so (%
\ref{bff3.5}) takes place if and only if the following equations are
simultaneously obeyed
\begin{gather}
t_{abc}=0,\quad t_{abc}^{d}=0,\quad t_{abcdf}=0,\quad
t_{a}^{b}=0,\quad t_{a}^{bc}=0,
\label{eqs1} \\
T_{ab}^{A}=0,\quad T_{a}^{AB}=0,\quad T_{ab}^{Ac}=0,\quad
T_{abcd}^{A}=0,\quad T^{Aab}=0,
\label{eqs2} \\
T_{abB}^{A}=0,\quad T_{aABC}=0,\quad T_{AB}^{a}=0,\quad
T_{abc}^{AB}=0,\quad T_{b}^{ABa}=0,
\label{eqs3} \\
T^{ABCD}=0,\quad T_{ab}^{ABC}=0,\quad T_{ABC}^{a}=0,\quad
T_{a}^{ABCD}=0. \label{eqs4}
\end{gather}%
Based on the last equations, which enforce $\Delta =0=\bar{\Delta}$, from (%
\ref{so1}) compared with (\ref{bff3.5}) it follows that we can take
\begin{equation}
S_{2}=0.  \label{s2int}
\end{equation}%
On behalf of (\ref{s2int}) it is easy to show that one can safely set zero
the solutions to the higher-order deformation equations, (\ref{bff3.6}),
etc.
\begin{equation}
S_{k}=0,\quad k>2.  \label{skint}
\end{equation}

Collecting formulas (\ref{s2int}) and (\ref{skint}), we can state that the
complete deformed solution to the master equation for the model under study,
which is consistent to all orders in the coupling constant, reads as
\begin{equation}
S=\bar{S}+\lambda S_{1},  \label{defsolmast}
\end{equation}%
where $\bar{S}$ is given in (\ref{solfree}) and $S_{1}$ is expressed by (\ref%
{s1final}). The full deformed solution to the master equation
comprises $11$
types of smooth functions of the undifferentiated scalar fields: $W_{ab}$, $%
M_{bc}^{a}$, $M_{abcd}$, $M^{ab}$, $f_{abc}^{A}$, $g_{ab}^{AB}$, $f_{a}^{A}$%
, $f_{aB}^{A}$, $f_{\;\;a}^{Ab}$, $g_{\quad C}^{AB}$, and $g^{aAB}$. They
are subject to equations (\ref{eqs1})--(\ref{eqs4}), imposed by the
consistency of the first-order deformation.

\section{Lagrangian formulation of the interacting model\label{lagint}}

\setcounter{equation}{0}

The piece of antighost number $0$ from the full deformed solution to the
master equation, of the form (\ref{defsolmast}), furnishes us with the
Lagrangian action of the interacting theory
\begin{eqnarray}
S^{\mathrm{L}}[A_{\mu }^{a},H_{\mu }^{a},\varphi _{a},B_{a}^{\mu \nu
},V_{\mu \nu }^{A},V_{\mu }^{A}] &=&\int d^{4}x\left[ H_{\mu }^{a}D^{\mu
}\varphi _{a}+\tfrac{1}{2}B_{a}^{\mu \nu }\bar{F}_{\mu \nu }^{a}\right.
\notag \\
&&+\tfrac{1}{2}\left( V_{A}^{\mu \nu }\bar{F}_{\mu \nu }^{A}+V_{\mu
}^{A}V_{A}^{\mu }\right)  \notag \\
&&-\tfrac{\lambda }{4}\varepsilon ^{\mu \nu \rho \lambda }\left( \tfrac{1}{4!%
}M_{abcd}A_{\mu }^{a}A_{\nu }^{b}+\tfrac{2}{3}f_{Aacd}V_{\mu }^{A}A_{\nu
}^{a}\right.  \notag \\
&&\left. \left. -g_{ABcd}V_{\mu }^{A}V_{\nu }^{B}\right) A_{\rho
}^{c}A_{\lambda }^{d}\right] ,  \label{ldef}
\end{eqnarray}%
where we used the notations
\begin{equation}
D^{\mu }\varphi _{a}=\partial ^{\mu }\varphi _{a}+\lambda W_{ab}A^{b\mu }-%
\tfrac{\lambda }{4!}f_{Aa}V^{A\mu },  \label{n1}
\end{equation}%
\begin{eqnarray}
\bar{F}_{\mu \nu }^{a} &=&\partial _{\lbrack \mu }^{\left. {}\right. }A_{\nu
]}^{a}+\lambda M_{bc}^{a}A_{\mu }^{b}A_{\nu }^{c}+\lambda \varepsilon _{\mu
\nu \rho \lambda }M^{ab}B_{b}^{\rho \lambda }  \notag \\
&&+\tfrac{\lambda }{12}\left( f_{Ab}^{a}A_{[\mu }^{b}V_{\nu
]}^{A}+g_{AB}^{a}V_{\mu }^{A}V_{\nu }^{B}\right) ,  \label{n2}
\end{eqnarray}%
\begin{equation}
\bar{F}_{\mu \nu }^{A}=\partial _{\lbrack \mu }^{\left. {}\right. }V_{\nu
]}^{A}-\lambda f_{aB}^{A}A_{[\mu }^{a}V_{\nu ]}^{B}+\lambda g_{BC}^{\quad
A}V_{\mu }^{B}V_{\nu }^{C}.  \label{n3}
\end{equation}%
Formula (\ref{ldef}) expresses the most general form of the Lagrangian
action describing the interactions between a finite collection of BF models
and a finite set of two-form gauge fields that complies with our working
hypotheses and whose free limit is precisely action (\ref{bfa1}). We note
that the deformed Lagrangian action is of maximum order $1$ in the coupling
constant and includes two main types of vertices: one generates
self-interactions among the BF fields and the other couples the two-form
field spectrum to the BF field spectrum. The first type is already known
from the literature and we will not comment on it. The second is yielded by
the expression%
\begin{eqnarray}
&&-\tfrac{\lambda }{4!}f_{Aa}V^{A\mu }H_{\mu }^{a}+\tfrac{\lambda }{24}%
B_{a}^{\mu \nu }\left( f_{Ab}^{a}A_{[\mu }^{b}V_{\nu ]}^{A}+g_{AB}^{a}V_{\mu
}^{A}V_{\nu }^{B}\right)  \notag \\
&&-\tfrac{\lambda }{2}V_{A}^{\mu \nu }\left( f_{aB}^{A}A_{[\mu }^{a}V_{\nu
]}^{B}-g_{BC}^{\quad A}V_{\mu }^{B}V_{\nu }^{C}\right)  \notag \\
&&-\tfrac{\lambda }{4}\varepsilon ^{\mu \nu \rho \lambda }\left( \tfrac{2}{3}%
f_{Aacd}V_{\mu }^{A}A_{\nu }^{a}-g_{ABcd}V_{\mu }^{A}V_{\nu }^{B}\right)
A_{\rho }^{c}A_{\lambda }^{d}.  \label{intvert}
\end{eqnarray}%
We observe that the vector fields $V^{A\mu }$ couple to all the BF fields
from the collection, while the two-form gauge fields $V_{A}^{\mu \nu }$
interact only with the one-forms $A_{\mu }^{a}$ from the BF sector. Also,
all the interaction vertices are derivative-free (we recall that the various
functions that parameterize (\ref{ldef}) depend only on the \textit{%
undifferentiated} scalar fields). One of this couplings, $\tfrac{\lambda }{2}%
g_{BC}^{\quad A}V_{A}^{\mu \nu }V_{\mu }^{B}V_{\nu }^{C}$, is nothing but
the generalized version of non-Abelian Freedman-Townsend vertex. (By
`generalized' we mean that its form is identical with the standard
non-Abelian Freedman-Townsend vertex up to the point that $g_{BC}^{\quad A}$
are \textit{not} the structure constants of a Lie algebra, but depend on the
undifferentiated scalar fields.) Thus, action (\ref{ldef}) contains the
generalized version of non-Abelian Freedman-Townsend action
\begin{equation}
S^{\mathrm{FT}}_{\mathrm{gen}}[V_{\mu \nu }^{A},V_{\mu }^{A},\varphi _{a}]=%
\tfrac{1}{2}\int d^{4}x\left[ V_{A}^{\mu \nu }\left( \partial _{\lbrack \mu
}^{\left. {}\right. }V_{\nu ]}^{A}+\lambda g_{BC}^{\quad A}V_{\mu
}^{B}V_{\nu }^{C}\right) +V_{\mu }^{A}V_{A}^{\mu }\right] .  \label{lFT}
\end{equation}

From the terms of antighost number $1$ present in (\ref{defsolmast}) we read
the deformed gauge transformations (which leave invariant action (\ref{ldef}%
)), namely%
\begin{equation}
\bar{\delta}_{\epsilon }A_{\mu }^{a}=\left( D_{\mu }\right)
_{\;\;b}^{a}\epsilon ^{b}-2\lambda M^{ab}\varepsilon _{\mu \nu \rho \lambda
}\epsilon _{b}^{\nu \rho \lambda },  \label{gaugeA}
\end{equation}%
\begin{eqnarray}
\bar{\delta}_{\epsilon }H_{\mu }^{a} &=&2\left( \bar{D}^{\nu }\right)
_{\;\;b}^{a}\epsilon _{\mu \nu }^{b}+\tfrac{\lambda }{2}\varepsilon _{\mu
\nu \rho \lambda }\left[ \left( -\tfrac{1}{12}\frac{\partial M_{bcde}}{%
\partial \varphi _{a}}A^{c\nu }+\frac{\partial f_{bde}^{A}}{\partial \varphi
_{a}}V_{A}^{\nu }\right) A^{d\rho }\right.  \notag \\
&&\left. +\frac{\partial g_{be}^{AB}}{\partial \varphi _{a}}V_{A}^{\nu
}V_{B}^{\rho }\right] A^{e\lambda }\epsilon ^{b}+\lambda \left( -\frac{%
\partial W_{bc}}{\partial \varphi _{a}}H_{\mu }^{c}+\frac{\partial f_{bB}^{A}%
}{\partial \varphi _{a}}V_{A}^{\nu }V_{\mu \nu }^{B}\right) \epsilon ^{b}
\notag \\
&&-\frac{\partial \left( D^{\nu }\right) _{\;\;b}^{d}}{\partial \varphi _{a}}%
B_{d\mu \nu }\epsilon ^{b}-\tfrac{3\lambda }{2}\frac{\partial M_{cd}^{b}}{%
\partial \varphi _{a}}A^{c\nu }A^{d\rho }\epsilon _{b\mu \nu \rho }+2\lambda
\frac{\partial M^{bc}}{\partial \varphi _{a}}B_{c\mu \nu }\varepsilon ^{\nu
\alpha \beta \gamma }\epsilon _{b\alpha \beta \gamma }  \notag \\
&&+\tfrac{\lambda }{4}\left( \frac{\partial f_{Ac}^{b}}{\partial \varphi _{a}%
}V^{A\nu }A^{c\rho }-\tfrac{1}{2}\frac{\partial g_{AB}^{b}}{\partial \varphi
_{a}}V^{A\nu }V^{B\rho }\right) \epsilon _{b\mu \nu \rho }  \notag \\
&&+\lambda \varepsilon _{\mu \nu \rho \lambda }\left( \frac{\partial f_{bAB}%
}{\partial \varphi _{a}}V^{B\nu }A^{b\rho }+\tfrac{1}{2}\frac{\partial
g_{\quad A}^{BC}}{\partial \varphi _{a}}V_{B}^{\nu }V_{C}^{\rho }\right)
\epsilon ^{A\lambda },  \label{gaugeH}
\end{eqnarray}%
\begin{equation}
\bar{\delta}_{\epsilon }\varphi _{a}=-\lambda W_{ab}\epsilon ^{b},
\label{gaugefi}
\end{equation}%
\begin{eqnarray}
\bar{\delta}_{\epsilon }B_{a}^{\mu \nu } &=&-3\left( D_{\rho }\right)
_{a}^{\;\;b}\epsilon _{b}^{\mu \nu \rho }+2\lambda W_{ab}\epsilon ^{b\mu \nu
}-\lambda \varepsilon ^{\mu \nu \rho \lambda }f_{aAB}V_{\rho }^{B}\epsilon
_{\lambda }^{A}-\lambda M_{ab}^{c}B_{c}^{\mu \nu }\epsilon ^{b}  \notag \\
&&+\lambda \varepsilon ^{\mu \nu \rho \lambda }\left( \tfrac{1}{8}%
M_{abcd}A_{\rho }^{c}A_{\lambda }^{d}+f_{Aabc}V_{\rho }^{A}A_{\lambda }^{c}-%
\tfrac{1}{2}g_{ABab}V_{\rho }^{A}V_{\lambda }^{B}\right) \epsilon ^{b},
\label{gaugeB}
\end{eqnarray}%
\begin{eqnarray}
\bar{\delta}_{\epsilon }V_{\mu \nu }^{A} &=&\varepsilon _{\mu \nu \rho
\lambda }\left( D^{\rho }\right) _{\;\;B}^{A}\epsilon ^{B\lambda }+\tfrac{%
\lambda }{12}f_{a}^{A}\epsilon _{\mu \nu }^{a}+\tfrac{\lambda }{4}\left(
f_{\;\;b}^{Aa}A^{b\rho }-g^{aAB}V_{B}^{\rho }\right) \epsilon _{a\mu \nu
\rho }  \notag \\
&&+\lambda \left[ \varepsilon _{\mu \nu \rho \lambda }\left( \tfrac{1}{2}%
f_{abc}^{A}A^{b\rho }+g_{ac}^{AB}V_{B}^{\rho }\right) A^{c\lambda }\right.
\notag \\
&&\left. +f_{aB}^{A}V_{\mu \nu }^{B}+\tfrac{1}{12}f_{\;\;a}^{Ab}B_{b\mu \nu }%
\right] \epsilon ^{a},  \label{gaugeV2}
\end{eqnarray}%
\begin{equation}
\bar{\delta}_{\epsilon }V_{\mu }^{A}=\lambda f_{aB}^{A}V_{\mu }^{B}\epsilon
^{a}.  \label{gaugeV1}
\end{equation}%
In (\ref{gaugeA})--(\ref{gaugeV1}) we employed the following notations for
the various types of (generalized) covariant derivatives:
\begin{eqnarray}
\left( \bar{D}^{\mu }\right) _{\;\;b}^{a} &=&\delta _{b}^{a}\partial ^{\mu
}-\lambda \left( \frac{\partial W_{bc}}{\partial \varphi _{a}}A^{c\mu }-%
\tfrac{1}{12}\frac{\partial f_{Ab}}{\partial \varphi _{a}}V^{A\mu }\right) ,
\label{dbarab} \\
\left( D_{\mu }\right) _{\;\;b}^{a} &=&\delta _{b}^{a}\partial _{\mu
}-\lambda M_{bc}^{a}A_{\mu }^{c}-\tfrac{\lambda }{12}f_{Ab}^{a}V_{\mu }^{A},
\label{dabdir} \\
\left( D_{\mu }\right) _{a}^{\;\;b} &=&\delta _{a}^{b}\partial _{\mu
}+\lambda \left( M_{ac}^{b}A_{\mu }^{c}+\tfrac{1}{12}f_{Aa}^{b}V_{\mu
}^{A}\right) ,  \label{dabinv} \\
\left( D^{\mu }\right) _{\;\;B}^{A} &=&\delta _{B}^{A}\partial ^{\mu
}-\lambda f_{aB}^{A}A^{a\mu }+\lambda g_{\quad B}^{AC}V_{C}^{\mu }.
\label{dAB}
\end{eqnarray}%
It is interesting to see that the gauge transformations of all fields get
modified by the deformation procedure. Also, the gauge transformations of
the BF fields $H_{\mu }^{a}$ and $B_{a}^{\mu \nu }$ involve the gauge
parameters $\epsilon ^{A\lambda }$, which are specific to the two-form
sector. Similarly, the gauge transformations of $V_{\mu \nu }^{A}$ and $%
V_{\mu }^{A}$ include pure BF gauge parameters. By contrast to the standard
non-Abelian Freedman-Townsend model, where the vector fields $V_{\mu }^{A}$
are gauge-invariant, here these fields gain nonvanishing gauge
transformations, proportional with the BF gauge parameters $\epsilon ^{a}$.
The nonvanishing commutators among the deformed gauge transformations result
from the terms quadratic in the ghosts with pure ghost number $1$ present in
(\ref{defsolmast}). The concrete form of the gauge generators and of the
corresponding nonvanishing commutators is included in Appendix \ref%
{appendixB} and \ref{appendixD}, respectively (see relations (\ref{g1})--(%
\ref{g6d}) and (\ref{co1})--(\ref{co19}), respectively). With the help of
these relations we observe that the original Abelian gauge algebra is
deformed into an open one, meaning that the commutators among the gauge
transformations only close on-shell, i.e. on the field equations resulting
from the deformed Lagrangian action (\ref{ldef}). The deformed gauge
generators remain reducible of order two, just like the original ones, but
the reducibility relations of order one and two hold now only on the field
equations resulting from the deformed Lagrangian action (on-shell
reducibility). The expressions of the reducibility functions and relations
are given in detail in Appendix \ref{appendixC} (see formulas (\ref{r1})--(%
\ref{r25})). They are deduced from certain elements in (\ref{defsolmast})
that are linear in the ghosts with the pure ghost number greater or equal to
$2$.

We recall that the entire gauge structure of the interacting model is
controlled by the functions $W_{ab}$, $M_{bc}^{a}$, $M_{abcd}$, $M^{ab}$, $%
f_{abc}^{A}$, $g_{ab}^{AB}$, $f_{a}^{A}$, $f_{aB}^{A}$, $f_{\;\;a}^{Ab}$, $%
g_{\quad C}^{AB}$, and $g^{aAB}$, which are restricted to satisfy equations (%
\ref{eqs1})--(\ref{eqs4}). Thus, our procedure is consistent provided these
equations are shown to possess solutions. We give below some classes of
solutions to (\ref{eqs1})--(\ref{eqs4}), without pretending to exhaust all
possibilities.

\begin{itemize}
\item \textbf{Type I solutions}

A first class of solutions to equations (\ref{eqs1}) is given by
\begin{equation}
M_{ab}^{c}=\frac{\partial W_{ab}}{\partial \varphi _{c}},\quad
M_{abcd}=f_{e[ab}\frac{\partial W_{cd]}}{\partial \varphi _{e}},\quad
M^{ab}=0,  \label{z3}
\end{equation}%
where $f_{eab}$ are arbitrary, antisymmetric constants and the functions $%
W_{ab}$\ are required to fulfill the equations
\begin{equation}
W_{e[a}\frac{\partial W_{bc]}}{\partial \varphi _{e}}=0.  \label{z5}
\end{equation}%
We remark that all the nonvanishing solutions are parameterized by the
antisymmetric functions $W_{ab}$. Like in the pure BF case \cite{defBFjhep},
we can interpret the functions $W_{ab}$ like the components of a two-tensor
on a Poisson manifold with the target space locally parameterized by the
scalar fields $\varphi _{e}$. Consequently, the first and third equations
among (\ref{eqs2}) are verified if we take
\begin{equation}
f_{aB}^{A}=\lambda _{\;\;B}^{A}f_{a},\quad f_{a}^{A}=\tau
^{A}k^{c}W_{ac},\quad f_{\;\;b}^{Aa}=-\tfrac{1}{2}\tau ^{A}k^{c}\frac{%
\partial W_{bc}}{\partial \varphi _{a}},  \label{qq1}
\end{equation}%
where $f_{a}$ are arbitrary functions of $\varphi _{b}$, $k^{c}$\ stand for
some arbitrary constants, and $\tau ^{A}$\ and $\lambda _{\;\;B}^{A}$\ ($%
\lambda ^{AB}=-\lambda ^{BA}$, $\lambda ^{AB}=k^{AC}\lambda _{\;\;C}^{B}$)
represent some constants subject to the conditions
\begin{equation}
\lambda _{\;\;B}^{A}\tau ^{B}=0.  \label{qq2}
\end{equation}%
Inserting (\ref{qq1}) into the second equation from (\ref{eqs2}), we obtain
\begin{equation}
g_{AB}^{a}=\tfrac{1}{2}g_{ABC}\tau ^{C}k^{a}+\mu _{AB}\nu ^{a},  \label{qq3}
\end{equation}%
where $\mu _{AB}$ are some arbitrary, antisymmetric constants and $\nu
^{a}\left( \varphi \right) $ are null vectors of $W_{ab}$ (if the matrix of
elements $W_{ab}$ is degenerate), i.e.
\begin{equation}
W_{ab}\nu ^{a}=0.  \label{qq4}
\end{equation}%
In the presence of the previous solutions the fourth equation from (\ref%
{eqs2}) is solved for
\begin{equation}
f_{abc}^{A}=\tfrac{1}{4\cdot 4!}\tau ^{A}k^{d}f_{e[ab}\frac{\partial W_{cd]}%
}{\partial \varphi _{e}}.  \label{qq5}
\end{equation}%
Due to the last relation in (\ref{z3}), it is easy to see that the
fifth equation from (\ref{eqs2}) is now automatically satisfied.
Next, we investigate equations (\ref{eqs3}). The former equation is
checked if we make the choice
\begin{equation}
f_{a}=\bar{k}^{b}W_{ab},  \label{qq6}
\end{equation}%
with $\bar{k}^{b}$ some arbitrary constants. The next equation from (\ref%
{eqs3}) is fulfilled for
\begin{equation}
g_{ABC}=C_{ABC}(1+\chi ),\quad \lambda _{\;\;B}^{A}=C_{CB}^{\quad A}\tau
^{C},\quad k^{a}=\bar{k}^{a},  \label{qq7}
\end{equation}%
where $\chi \left( \varphi \right) $ has the property
\begin{equation}
W_{ab}\frac{\partial \chi }{\partial \varphi _{b}}=0  \label{qq8}
\end{equation}%
(if $W_{ab}$ allows for nontrivial null vectors) and the completely
antisymmetric constants $C_{ABC}$ are imposed to satisfy the Jacobi identity
\begin{equation}
C_{EA[B}C_{DC]}^{\quad E}=0.  \label{qq9}
\end{equation}%
Now, the third equation from (\ref{eqs3}) is automatically verified
by the last relation in (\ref{z3}). The solution to the fourth
equation reads as
\begin{equation}
g_{ab}^{AB}=C^{ABC}\tau _{C}W_{ab},\quad \mu _{AB}=0.  \label{qq10}
\end{equation}%
So far we have determined all the unknown functions. The above solutions
also fulfill the remaining equations from (\ref{eqs3}) and the first three
ones in (\ref{eqs4}). However, the last equation present in (\ref{eqs4})
produces the restriction
\begin{equation}
C^{E[AB}C^{CD]F}\tau _{E}\tau _{F}=0.  \label{qq11}
\end{equation}%
The last equation possesses at least two different types of solutions,
namely
\begin{equation}
C^{ABC}=\varepsilon ^{ijk}e_{i}^{A}e_{j}^{B}e_{k}^{C},\quad i,j,k=1,2,3
\label{qq12}
\end{equation}%
and
\begin{equation}
C^{ABC}=\varepsilon ^{\bar{A}\bar{B}\bar{C}}l_{\bar{A}}^{A}l_{\bar{B}}^{B}l_{%
\bar{C}}^{C},\quad \bar{A},\bar{B},\bar{C}=1,2,3,4,  \label{qq13}
\end{equation}%
respectively, where $e_{i}^{A}$ and $l_{\bar{A}}^{A}$\ are all constants and
$\varepsilon ^{ijk}$\ together with $\varepsilon ^{\bar{A}\bar{B}\bar{C}}$\
are completely antisymmetric symbols. These symbols are defined via the
conventions $\varepsilon ^{123}=+1$ and $\varepsilon ^{124}=\varepsilon
^{134}=\varepsilon ^{234}=+1$, respectively. It is straightforward to see
that the quantities $C^{ABC}$ given by either of the relations (\ref{qq12})
or (\ref{qq13}) indeed check (\ref{qq9}). By assembling the previous
results, we find the type I solutions to equations (\ref{eqs1})--(\ref{eqs4}%
) being expressed via relations (\ref{z3}), (\ref{qq5}), and
\begin{gather}
f_{aB}^{A}=C_{DB}^{\quad A}\tau ^{D}k^{b}W_{ab},\quad f_{a}^{A}=\tau
^{A}k^{c}W_{ac},  \label{qq14} \\
\,f_{\;\;b}^{Aa}=-\tfrac{1}{2}\tau ^{A}k^{c}\frac{\partial W_{bc}}{\partial
\varphi _{a}},\quad g_{ABC}=C_{ABC}(1+\chi ),  \label{qq14a} \\
g_{AB}^{a}=\tfrac{1}{2}C_{ABC}(1+\chi )\tau ^{C}k^{a},\quad
g_{ab}^{AB}=C^{ABC}\tau _{C}W_{ab},  \label{qq15}
\end{gather}%
where $\tau ^{A}$ and $k^{a}$ represent some arbitrary constants, $W_{ab}$
are assumed to satisfy equations (\ref{z5}), and $\chi $ is subject to (\ref%
{qq8}) (if the matrix of elements $W_{ab}$ is degenerate). The antisymmetric
constants $C^{ABC}$ are imposed to verify relations (\ref{qq11}) (which
ensure that (\ref{qq9}) are automatically checked). Two sets of solutions to
(\ref{qq11}) (and hence also to (\ref{qq9})) are provided by formulas (\ref%
{qq12}) and (\ref{qq13})).

\item \textbf{Type II solutions}

Another set of solutions to equations (\ref{eqs1}) can be written as
\begin{equation}
W_{ab}=0,\quad M_{ab}^{c}=C_{\;\;ab}^{c}\hat{M},\quad M_{abcd}=0,\quad
M^{ab}=\mu ^{ab}M,  \label{z7}
\end{equation}%
with $\hat{M}$\ and $M$\ arbitrary functions of the undifferentiated scalar
fields. The coefficients $\mu ^{ab}$\ represent the elements of the inverse
of the Killing metric $\bar{\mu}_{ad}$ of a semi-simple Lie algebra with the
structure constants $C_{\;\;ab}^{c}$ ($\bar{\mu}_{ad}\mu ^{de}=\delta
_{a}^{e}$), where, in addition $C_{abc}=\bar{\mu}_{ad}C_{\;\;bc}^{d}$ must
be completely antisymmetric. Under these circumstances, the first equation
from (\ref{eqs2}) is solved if we take
\begin{equation}
f_{aB}^{A}=\tilde{\lambda}_{\;\;B}^{A}\hat{f}_{a},\quad f_{a}^{A}=\sigma ^{A}%
\bar{f}_{a},  \label{qq16}
\end{equation}%
where $\hat{f}_{a}$ and $\bar{f}_{a}$ are arbitrary functions of the
undifferentiated scalar fields, and $\tilde{\lambda}_{\;\;B}^{A}$ as well as
$\sigma ^{A}$ are some constants that must satisfy the relations
\begin{equation}
\tilde{\lambda}_{\;\;B}^{A}\sigma ^{B}=0.  \label{qq17}
\end{equation}%
Then, the second equation from (\ref{eqs2}) implies the fact that $%
g_{AB}^{\quad C}$ is restricted to fulfill the condition
\begin{equation}
g_{AB}^{\quad C}\sigma _{C}=0.  \label{qq18}
\end{equation}%
Replacing the above solutions into the third equation from (\ref{eqs2}), we
get the relation
\begin{equation}
f_{\;\;b}^{Aa}=\sigma ^{A}C_{\;\;bc}^{a}\frac{\partial P}{\partial \varphi
_{c}},\quad f_{abc}^{A}=\sigma ^{A}C_{abc}N,  \label{qq19}
\end{equation}%
where $P$ and $N$ are functions of the undifferentiated scalar fields, with $%
N$ restricted to verify the equation
\begin{equation}
\bar{f}_{a}\frac{\partial \hat{M}}{\partial \varphi _{a}}+4\cdot 4!NM=0.
\label{qq20}
\end{equation}%
Having in mind the solutions deduced until now, we find that the fourth
equation from (\ref{eqs2}) is automatically checked and the last equation in
(\ref{eqs2}) constrains the function $M$ to be constant (for the sake of
simplicity, we take this constant to be equal to unity)
\begin{equation}
M=1.  \label{qq21}
\end{equation}%
The first and the third equations from (\ref{eqs3}) immediately yield $\hat{f%
}_{a}=0$, which further leads to $f_{aB}^{A}=0$. Under these circumstances,
the second equation entering (\ref{eqs3}) is identically satisfied and the
fourth equation from the same formula possesses the solution
\begin{equation}
g_{ab}^{AB}=C_{abc}\bar{\lambda}^{AB}\frac{\partial Q}{\partial \varphi _{c}}%
,  \label{qq22}
\end{equation}%
where $Q$ is an arbitrary function of the undifferentiated scalar fields and
$\bar{\lambda}^{AB}$ denote some arbitrary, completely antisymmetric
constants. Substituting the solutions deduced so far into the last equation
from (\ref{eqs3}), we get
\begin{equation}
g_{AB}^{a}=\bar{\lambda}_{AB}\frac{\partial g}{\partial \varphi _{a}},
\label{qq23}
\end{equation}%
where $g$ is a function of the undifferentiated scalar fields that is
restricted to fulfill the equation
\begin{equation}
\frac{\partial Q}{\partial \varphi _{a}}=\tfrac{1}{2\cdot 4!}\hat{M}\frac{%
\partial g}{\partial \varphi _{a}}.  \label{qq24}
\end{equation}%
The first equation from (\ref{eqs4}) exhibits the solution
\begin{equation}
g_{ABC}=\sigma _{\lbrack A}\hat{\lambda}_{B]C}\hat{\Phi},  \label{qq25}
\end{equation}%
with $\hat{\Phi}$ an arbitrary function of the undifferentiated scalar
fields and $\hat{\lambda}_{BC}$ some arbitrary, completely antisymmetric
constants, which check the relations
\begin{equation}
\hat{\lambda}_{BC}\sigma ^{C}=0.  \label{qq26}
\end{equation}%
Relations (\ref{qq26}) ensure that equation (\ref{qq18}) is verified. The
second equation from (\ref{eqs4}) displays a solution of the form
\begin{equation}
\bar{\lambda}^{AB}=\sigma ^{\lbrack A}\hat{\lambda}^{B]C}\beta _{C},
\label{qq27}
\end{equation}%
with $\beta _{C}$ some constants. The remaining equations entering (\ref%
{eqs4}) are now identically verified. Putting together the results obtained
until now, it follows that the type II solutions to equations (\ref{eqs1})--(%
\ref{eqs4}) can be written as
\begin{gather}
W_{ab}=0,\quad M_{ab}^{c}=C_{\;\;ab}^{c}\hat{M},\quad M_{abcd}=0,\quad
M^{ab}=\mu ^{ab},  \label{qq28} \\
f_{aB}^{A}=0,\quad f_{a}^{A}=\sigma ^{A}\bar{f}_{a},\quad
f_{\;\;b}^{Aa}=\sigma ^{A}C_{\;\;bc}^{a}\frac{\partial P}{\partial \varphi
_{c}},  \label{qq29} \\
f_{abc}^{A}=-\tfrac{1}{4\cdot 4!}\sigma ^{A}C_{abc}\bar{f}_{d}\frac{\partial
\hat{M}}{\partial \varphi _{d}},\quad g_{ab}^{AB}=\tfrac{1}{2\cdot 4!}%
C_{abc}\sigma ^{\lbrack A}\hat{\lambda}^{B]C}\beta _{C}\hat{M}\frac{\partial
g}{\partial \varphi _{c}},  \label{qq30} \\
g_{AB}^{a}=\sigma _{\lbrack A}\hat{\lambda}_{B]C}\beta ^{C}\frac{\partial g}{%
\partial \varphi _{a}},\quad g_{ABC}=\sigma _{\lbrack A}\hat{\lambda}_{B]C}%
\hat{\Phi}.  \label{qq31}
\end{gather}%
We recall that $\hat{M}$, $\bar{f}_{a}$, $P$, $g$, and $\hat{\Phi}$ are
arbitrary functions of the undifferentiated scalar fields and $\beta _{C}$, $%
\hat{\lambda}_{BC}$, and $\sigma ^{C}$ are some constants. In addition, the
last two sets of constants are imposed to fulfill equation (\ref{qq26}). The
quantities $\mu ^{ab}$ are the elements of the inverse of the Killing metric
of a semi-simple Lie algebra with the structure constants $C_{\;\;ab}^{c}$,
where $C_{abc}$ must be completely antisymmetric.

\item \textbf{Type III solutions}

The third type of solutions to (\ref{eqs1}) is given by
\begin{equation}
W_{ab}=0,\quad M_{ab}^{c}=\bar{C}_{\;\;ab}^{c}w,\quad M_{abcd}=\hat{f}_{e[ab}%
\bar{C}_{\;\;cd]}^{e}q,\quad M^{ab}=0,  \label{qq32}
\end{equation}%
with $w$\ and $q$\ arbitrary functions of the undifferentiated scalar
fields, $\hat{f}_{eab}$ some arbitrary, antisymmetric constants, and $\bar{C}%
_{\;\;ab}^{c}$\ the structure constants of a Lie algebra. Let us
particularize the last solutions to the case where
\begin{equation}
\bar{C}_{\;\;ab}^{c}=\hat{k}^{c}\bar{W}_{ab},\quad w\left( \varphi \right)
=q\left( \varphi \right) =\frac{d\hat{w}\left( \hat{k}^{m}\varphi
_{m}\right) }{d\left( \hat{k}^{n}\varphi _{n}\right) },  \label{z11}
\end{equation}%
with $\hat{k}^{c}$\ some arbitrary constants, $\hat{w}$\ an arbitrary,
smooth function depending on $\hat{k}^{m}\varphi _{m}$, and $\bar{W}_{ab}$\
some antisymmetric constants satisfying the relations
\begin{equation}
\bar{W}_{a[b}\bar{W}_{cd]}=0.  \label{z12}
\end{equation}%
Obviously, equations (\ref{z12}) ensure the Jacobi identity for the
structure constants $\bar{C}_{\;\;ab}^{c}$. Replacing (\ref{z11}) back in (%
\ref{qq32}), we find
\begin{equation}
W_{ab}=0,\quad M_{ab}^{c}=\frac{\partial \hat{W}_{ab}}{\partial \varphi _{c}}%
,\quad M_{abcd}=\hat{f}_{e[ab}\frac{\partial \hat{W}_{cd]}}{\partial \varphi
_{e}},\quad M^{ab}=0,  \label{z13}
\end{equation}%
where
\begin{equation}
\hat{W}_{ab}=\bar{W}_{ab}\frac{d\hat{w}\left( \hat{k}^{m}\varphi _{m}\right)
}{d\left( \hat{k}^{n}\varphi _{n}\right) }.  \label{z16}
\end{equation}%
Due to (\ref{z12}), it is easy to see that $\hat{W}_{ab}$\ satisfy
the Jacobi identity for a Poisson manifold
\begin{equation}
\hat{W}_{e[a}\frac{\partial \hat{W}_{bc]}}{\partial \varphi _{e}}=0.
\label{z17}
\end{equation}%
Relations (\ref{z13}) and (\ref{z17}) emphasize that we
can generate solutions correlated with a Poisson manifold even if $W_{ab}=0$%
. In this situation the Poisson two-tensor results from a Lie algebra (see
the first formula in (\ref{z11}) and (\ref{z16})). It is interesting to
remark that the same equations, namely (\ref{z12}), ensure the Jacobi
identities for both the Lie algebra and the corresponding Poisson manifold.
These equations possess at least two types of solutions, namely
\begin{equation}
\bar{W}_{ab}=\varepsilon _{ijk}e_{a}^{i}e_{b}^{j}e_{c}^{k}\rho ^{c},\quad
i,j,k=1,2,3  \label{qq33}
\end{equation}%
and
\begin{equation}
\bar{W}_{ab}=\varepsilon _{\bar{a}\bar{b}\bar{c}}l_{a}^{\bar{a}}l_{b}^{\bar{b%
}}l_{c}^{\bar{c}}\bar{\rho}^{c},\quad \bar{a},\bar{b},\bar{c}=1,2,3,4,
\label{qq34}
\end{equation}%
where $e_{a}^{i}$, $\rho ^{c}$, $l_{a}^{\bar{a}}$, and $\bar{\rho}^{c}$\ are
all constants and $\varepsilon _{ijk}$\ together with $\varepsilon _{\bar{a}%
\bar{b}\bar{c}}$\ are completely antisymmetric symbols, defined via the
conventions $\varepsilon _{123}=+1$ and $\varepsilon _{124}=\varepsilon
_{134}=\varepsilon _{234}=+1$, respectively. If we tackle the remaining
equations in a manner similar to that employed at the previous cases, we
infer that the third type of solutions to (\ref{eqs1})--(\ref{eqs4}) is
expressed by (\ref{z13}) and
\begin{gather}
f_{aB}^{A}=m_{\;\;B}^{A}\hat{k}^{b}\bar{W}_{ab}\Omega ,\quad
f_{a}^{A}=0,\quad f_{\;\;b}^{Aa}=-\bar{\lambda}^{A}\tilde{k}^{c}\frac{%
\partial \hat{W}_{bc}}{\partial \varphi _{a}},  \label{qq35} \\
f_{abc}^{A}=\bar{\lambda}^{A}\left( \hat{u}_{[a}\hat{W}_{bc]}+\tfrac{1}{%
2\cdot 4!}\tilde{k}^{d}\hat{f}_{e[ab}\frac{\partial \hat{W}_{cd]}}{\partial
\varphi _{e}}\right) ,  \label{qq36} \\
g_{ab}^{AB}=\bar{\lambda}^{[A}m^{B]C}\bar{\beta}_{C}\bar{W}_{ab}\hat{Q}%
,\quad g_{AB}^{a}=0,\quad g_{ABC}=\bar{\lambda}_{[A}m_{B]C}\hat{P}.
\label{qq37}
\end{gather}%
In the above $\hat{k}^{b}$, $\tilde{k}^{a}$, $\bar{\beta}_{C}$, $\hat{f}%
_{eab}$, $\bar{\lambda}^{A}$, $\bar{W}_{ab}$
($\bar{W}_{ab}=-\bar{W}_{ba}$), and $m^{AB}$ ($m^{AB}=-m^{BA}$) are
some constants, the first four sets being arbitrary (up to the point
that $\hat{f}_{eab}$ should be completely antisymmetric) and the
last
three sets being subject to the relations (\ref{z12}) and%
\begin{equation}
m^{AB}\bar{\lambda}_{B}=0.  \label{qq38}
\end{equation}%
The quantities denoted by $\Omega $, $\hat{u}_{a}$, $\hat{Q}$, and $\hat{P}$
are arbitrary functions of the undifferentiated scalar fields. The functions
$\hat{W}_{ab}$ read as in (\ref{z16}), with $\hat{w}$\ an arbitrary, smooth
function depending on $\hat{k}^{m}\varphi _{m}$. If in particular we take $%
\Omega $ and $\hat{Q}$ to be respectively of the form of $w$ and $q$ from (%
\ref{z11}), then we obtain that the functions $f_{aB}^{A}$ and $g_{ab}^{AB}$
will be parameterized by $\hat{W}_{ab}$.
\end{itemize}

\section{Conclusion\label{concl}}

To conclude with, in this paper we have investigated the consistent
interactions that can be introduced between a finite collection of BF
theories and a finite set of two-form gauge fields (described by a sum of
Abelian Freedman-Townsend actions). Starting with the BRST differential for
the free theory, we compute the consistent first-order deformation of the
solution to the master equation with the help of standard cohomological
techniques, and obtain that it is parameterized by $11$ kinds of functions
depending on the undifferentiated scalar fields. Next, we investigate the
second-order deformation, whose existence imposes certain restrictions with
respect to these functions. Based on these restrictions, we show that we can
take all the remaining higher-order deformations to vanish. As a consequence
of our procedure, we are led to an interacting gauge theory with deformed
gauge transformations, a non-Abelian gauge algebra that only closes
on-shell, and on-shell accompanying reducibility relations. The deformed
action contains, among others, the generalized version of non-Abelian
Freedman-Townsend action. It is interesting to mention that by contrast to
the standard non-Abelian Freedman-Townsend model, where the auxiliary vector
fields are gauge-invariant, here these fields gain nonvanishing gauge
transformations, proportional with some BF gauge parameters. Finally, we
investigate the equations that restrict the functions parameterizing the
deformed solution to the master equation and give some particular classes of
solutions, which can be suggestively interpreted in terms of Poisson
manifolds and/or Lie algebras.

\section*{Acknowledgment}

This work has been supported in part by grant CEX-05-D11-49/07.10.2005 with
the Romanian Ministry of Education and Research (M.Ed.C.) and by EU contract
MRTN-CT-2004-005104.

\appendix

\section{Various notations used in subsection \protect\ref{highord} \label%
{appendixA}}

\setcounter{equation}{0} \renewcommand{\theequation}{A.\arabic{equation}}
The various notations used within formula (\ref{so2}) are listed below. The
objects denoted by $\left( K_{,m_{1}\ldots m_{p}}^{abc}\right) _{p=\overline{%
0,4}}$ are expressed by
\begin{eqnarray}
K^{abc} &=&\eta ^{a}\eta ^{b}\varphi ^{\ast c}+2\eta ^{a}A^{b\mu }H_{\mu
}^{c}+2\left( A^{a\mu }A^{b\nu }-2B^{\ast a\mu \nu }\eta ^{b}\right) C_{\mu
\nu }^{c}  \notag \\
&&+4\left( \eta ^{a}\eta ^{\ast b\mu \nu \rho }+3B^{\ast a\mu \nu }A^{b\rho
}\right) C_{\mu \nu \rho }^{c}  \notag \\
&&-4\left( \eta ^{a}\eta ^{\ast b\mu \nu \rho \lambda }+6B^{\ast a\mu \nu
}B^{\ast b\rho \lambda }-4\eta ^{\ast a\mu \nu \rho }A^{b\lambda }\right)
C_{\mu \nu \rho \lambda }^{c},  \label{xbfn39}
\end{eqnarray}%
\begin{eqnarray}
K_{,d}^{abc} &=&\left( 4H_{d}^{\ast \nu }A^{a\mu }\eta ^{b}-C_{d}^{\ast \mu
\nu }\eta ^{a}\eta ^{b}\right) C_{\mu \nu }^{c}-H_{d}^{\ast \mu }\eta
^{a}\eta ^{b}H_{\mu }^{c}  \notag \\
&&+\left( 6H_{d}^{\ast \rho }A^{a\mu }A^{b\nu }-12H_{d}^{\ast \rho }B^{\ast
a\mu \nu }\eta ^{b}+6C_{d}^{\ast \mu \nu }\eta ^{a}A^{b\rho }\right.  \notag
\\
&&\left. -C_{d}^{\ast \mu \nu \rho }\eta ^{a}\eta ^{b}\right) C_{\mu \nu
\rho }^{c}+\left( -48H_{d}^{\ast \lambda }B^{\ast a\mu \nu }A^{b\rho }\right.
\notag \\
&&+12C_{d}^{\ast \mu \nu }A^{a\rho }A^{b\lambda }+16H_{d}^{\ast \lambda
}\eta ^{\ast a\mu \nu \rho }\eta ^{b}-24C_{d}^{\ast \mu \nu }B^{\ast a\rho
\lambda }\eta ^{b}  \notag \\
&&\left. -8C_{d}^{\ast \mu \nu \rho }A^{a\lambda }\eta ^{b}-C_{d}^{\ast \mu
\nu \rho \lambda }\eta ^{a}\eta ^{b}\right) C_{\mu \nu \rho \lambda }^{c},
\label{xbfn40}
\end{eqnarray}%
\begin{eqnarray}
K_{,de}^{abc} &=&-3\left( C_{d}^{\ast \mu \nu }H_{e}^{\ast \rho }\eta
^{a}+2H_{d}^{\ast \mu }H_{e}^{\ast \nu }A^{a\rho }\right) \eta ^{b}C_{\mu
\nu \rho }^{c}  \notag \\
&&-H_{d}^{\ast \mu }H_{e}^{\ast \nu }\eta ^{a}\eta ^{b}C_{\mu \nu
}^{c}+\left( -24H_{d}^{\ast \mu }H_{e}^{\ast \nu }B^{\ast a\rho \lambda
}\eta ^{b}\right.  \notag \\
&&+12H_{d}^{\ast \mu }H_{e}^{\ast \nu }A^{a\rho }A^{b\lambda }-24C_{d}^{\ast
\mu \nu }H_{e}^{\ast \rho }A^{a\lambda }\eta ^{b}  \notag \\
&&\left. -3C_{d}^{\ast \mu \nu }C_{e}^{\ast \rho \lambda }\eta ^{a}\eta
^{b}+4C_{d}^{\ast \mu \nu \rho }H_{e}^{\ast \lambda }\eta ^{a}\eta
^{b}\right) C_{\mu \nu \rho \lambda }^{c},  \label{xbfn41}
\end{eqnarray}%
\begin{eqnarray}
K_{,def}^{abc} &=&-2\left( 4H_{d}^{\ast \mu }H_{e}^{\ast \nu }H_{f}^{\ast
\rho }A^{a\lambda }+3C_{d}^{\ast \mu \nu }H_{e}^{\ast \rho }H_{f}^{\ast
\lambda }\eta ^{a}\right) \eta ^{b}C_{\mu \nu \rho \lambda }^{c}  \notag \\
&&-H_{d}^{\ast \mu }H_{e}^{\ast \nu }H_{f}^{\ast \rho }\eta ^{a}\eta
^{b}C_{\mu \nu \rho }^{c},  \label{xbfn42}
\end{eqnarray}%
\begin{equation}
K_{,defg}^{abc}=-H_{d}^{\ast \mu }H_{e}^{\ast \nu }H_{f}^{\ast \rho
}H_{g}^{\ast \lambda }\eta ^{a}\eta ^{b}C_{\mu \nu \rho \lambda }^{c}.
\label{xbfn43}
\end{equation}%
The elements $\left( K_{d,m_{1}\ldots m_{p}}^{abc}\right) _{p=\overline{0,4}%
} $ read as%
\begin{eqnarray}
K_{d}^{abc} &=&\left( -2\eta ^{a}A_{\mu }^{b}A_{\nu }^{c}+B_{\mu \nu }^{\ast
a}\eta ^{b}\eta ^{c}\right) B_{d}^{\mu \nu }-A_{\mu }^{a}\eta ^{b}\eta
^{c}A_{d}^{\ast \mu }  \notag \\
&&+\left( -A_{\mu }^{a}A_{\nu }^{b}A_{\rho }^{c}+6\eta ^{a}B_{\mu \nu
}^{\ast b}A_{\rho }^{c}+\eta ^{b}\eta ^{c}\eta _{\mu \nu \rho }^{\ast
a}\right) \eta _{d}^{\mu \nu \rho }  \notag \\
&&-\tfrac{1}{3}\eta ^{a}\eta ^{b}\eta ^{c}\eta _{d}^{\ast }+\left( -12A_{\mu
}^{a}A_{\nu }^{b}B_{\rho \lambda }^{\ast c}+12\eta ^{a}B_{\mu \nu }^{\ast
b}B_{\rho \lambda }^{\ast c}\right.  \notag \\
&&\left. -8\eta ^{a}\eta _{\mu \nu \rho }^{\ast b}A_{\lambda }^{c}+\eta
_{\mu \nu \rho \lambda }^{\ast c}\eta ^{a}\eta ^{b}\right) \eta _{d}^{\mu
\nu \rho \lambda },  \label{xbfn44}
\end{eqnarray}%
\begin{eqnarray}
K_{d,e}^{abc} &=&\left( H_{e}^{\ast \mu }A^{a\nu }\eta ^{b}\eta ^{c}+\tfrac{1%
}{6}C_{e}^{\ast \mu \nu }\eta ^{a}\eta ^{b}\eta ^{c}\right) B_{d\mu \nu }
\notag \\
&&-\tfrac{1}{3}H_{e}^{\ast \mu }\eta ^{a}\eta ^{b}\eta ^{c}A_{d\mu }^{\ast
}+\left( -3H_{e}^{\ast \rho }\eta ^{a}A^{b\mu }A^{c\nu }\right.  \notag \\
&&-3H_{e}^{\ast \rho }\eta ^{a}\eta ^{b}B^{\ast c\mu \nu }+\tfrac{3}{2}%
C_{e}^{\ast \mu \nu }\eta ^{a}\eta ^{b}A^{c\rho }  \notag \\
&&\left. +\tfrac{1}{6}C_{e}^{\ast \mu \nu \rho }\eta ^{a}\eta ^{b}\eta
^{c}\right) \eta _{d\mu \nu \rho }+\left( 24A^{a\mu }H_{e}^{\ast \nu }\eta
^{b}B^{\ast c\rho \lambda }\right.  \notag \\
&&+4H_{e}^{\ast \lambda }A^{a\mu }A^{b\nu }A^{c\rho }-4H_{e}^{\ast \lambda
}\eta ^{a}\eta ^{b}\eta ^{\ast c\mu \nu \rho }  \notag \\
&&+6C_{e}^{\ast \mu \nu }\eta ^{a}\eta ^{b}B^{\ast c\rho \lambda
}-6C_{e}^{\ast \mu \nu }\eta ^{a}A^{b\rho }A^{c\lambda }  \notag \\
&&\left. +8C_{e}^{\ast \mu \nu \rho }\eta ^{a}\eta ^{b}A^{c\lambda }+\tfrac{1%
}{6}C_{e}^{\ast \mu \nu \rho \lambda }\eta ^{a}\eta ^{b}\eta ^{c}\right)
\eta _{d\mu \nu \rho \lambda },  \label{xbfn45}
\end{eqnarray}%
\begin{eqnarray}
K_{d,ef}^{abc} &=&\tfrac{1}{6}H_{e}^{\ast \mu }H_{f}^{\ast \nu }\eta
^{a}\eta ^{b}\eta ^{c}B_{d\mu \nu }+\tfrac{3}{2}H_{e}^{\ast \mu }H_{f}^{\ast
\nu }\eta ^{a}\eta ^{b}A^{c\rho }\eta _{d\mu \nu \rho }  \notag \\
&&+\tfrac{1}{2}C_{e}^{\ast \mu \nu }H_{f}^{\ast \rho }\eta ^{a}\eta ^{b}\eta
^{c}\eta _{d\mu \nu \rho }+\left( 6H_{e}^{\ast \mu }H_{f}^{\ast \nu }\eta
^{a}\eta ^{b}B^{\ast c\rho \lambda }\right.  \notag \\
&&-6H_{e}^{\ast \mu }H_{f}^{\ast \nu }\eta ^{a}A^{b\rho }A^{c\lambda
}+6C_{e}^{\ast \mu \nu }H_{f}^{\ast \rho }\eta ^{a}\eta ^{b}A^{c\lambda }
\notag \\
&&\left. +\tfrac{2}{3}C_{e}^{\ast \mu \nu \rho }H_{f}^{\ast \lambda }\eta
^{a}\eta ^{b}\eta ^{c}+\tfrac{1}{2}C_{e}^{\ast \mu \nu }C_{f}^{\ast \rho
\lambda }\eta ^{a}\eta ^{b}\eta ^{c}\right) \eta _{d\mu \nu \rho \lambda },
\label{xbfn46}
\end{eqnarray}%
\begin{eqnarray}
K_{d,efg}^{abc} &=&\left( 2H_{e}^{\ast \mu }H_{f}^{\ast \nu }H_{g}^{\ast
\rho }\eta ^{a}\eta ^{b}A^{c\lambda }+C_{e}^{\ast \mu \nu }H_{f}^{\ast \rho
}H_{g}^{\ast \lambda }\eta ^{a}\eta ^{b}\eta ^{c}\right) \eta _{d\mu \nu
\rho \lambda }  \notag \\
&&+\tfrac{1}{6}H_{e}^{\ast \mu }H_{f}^{\ast \nu }H_{g}^{\ast \rho }\eta
^{a}\eta ^{b}\eta ^{c}\eta _{d\mu \nu \rho },  \label{xbfn47}
\end{eqnarray}%
\begin{equation}
K_{d,efgh}^{abc}=\tfrac{1}{6}H_{e}^{\ast \mu }H_{f}^{\ast \nu }H_{g}^{\ast
\rho }H_{h}^{\ast \lambda }\eta ^{a}\eta ^{b}\eta ^{c}\eta _{d\mu \nu \rho
\lambda }.  \label{xbfn48}
\end{equation}%
The quantities $\left( K_{m_{1}\ldots m_{p}}^{abcdf}\right) _{p=\overline{0,4%
}}$, $\left( K_{b,m_{1}\ldots m_{p}}^{a}\right) _{p=\overline{0,4}}$, and $%
\left( K_{ab,m_{1}\ldots m_{p}}^{c}\right) _{p=\overline{0,4}}$ are given by
\begin{eqnarray}
K^{abcdf} &=&\tfrac{1}{8}\varepsilon ^{\mu \nu \rho \lambda }\left[ \left(
\tfrac{1}{3!}A_{\mu }^{a}A_{\nu }^{b}-B_{\mu \nu }^{\ast a}\eta ^{b}\right)
A_{\rho }^{c}A_{\lambda }^{d}+\tfrac{1}{3}\left( B_{\mu \nu }^{\ast
a}B_{\rho \lambda }^{\ast b}\right. \right.  \notag \\
&&\left. \left. -\tfrac{2}{3}\eta _{\mu \nu \rho }^{\ast a}A_{\lambda }^{b}+%
\tfrac{1}{4!}\eta _{\mu \nu \rho \lambda }^{\ast a}\eta ^{b}\right) \eta
^{c}\eta ^{d}\right] \eta ^{f},  \label{bfk1}
\end{eqnarray}%
\begin{eqnarray}
K_{e}^{abcdf} &=&\tfrac{1}{4!}\varepsilon ^{\mu \nu \rho \lambda }\left[
\tfrac{1}{2}\left( \tfrac{1}{5!}C_{e\mu \nu \rho \lambda }^{\ast }\eta ^{a}+%
\tfrac{1}{3!}C_{e\mu \nu \rho }^{\ast }A_{\lambda }^{a}+\tfrac{1}{2}C_{e\mu
\nu }^{\ast }B_{\rho \lambda }^{\ast a}\right. \right.  \notag \\
&&\left. +\tfrac{1}{3}H_{e\mu }^{\ast }\eta _{\nu \rho \lambda }^{\ast
a}\right) \eta ^{b}\eta ^{c}-H_{e\mu }^{\ast }\left( A_{\nu }^{a}A_{\rho
}^{b}-2B_{\nu \rho }^{\ast a}\eta ^{b}\right) A_{\lambda }^{c}  \notag \\
&&\left. -\tfrac{1}{2}C_{e\mu \nu }^{\ast }A_{\rho }^{a}A_{\lambda }^{b}\eta
^{c}\right] \eta ^{d}\eta ^{f},  \label{bfk2}
\end{eqnarray}%
\begin{eqnarray}
K_{eg}^{abcdf} &=&\tfrac{1}{2\cdot 4!}\varepsilon ^{\mu \nu \rho \lambda }%
\left[ \tfrac{1}{2}\left( \tfrac{1}{15}H_{e\mu }^{\ast }C_{g\nu \rho \lambda
}^{\ast }\eta ^{a}+\tfrac{1}{20}C_{e\mu \nu }^{\ast }C_{g\rho \lambda
}^{\ast }\eta ^{a}+H_{e\mu }^{\ast }C_{g\nu \rho }^{\ast }A_{\lambda
}^{a}\right) \eta ^{b}\right.  \notag \\
&&\left. -H_{e\mu }^{\ast }H_{g\nu }^{\ast }\left( A_{\rho }^{a}A_{\lambda
}^{b}-2B_{\rho \lambda }^{\ast a}\eta ^{b}\right) \right] \eta ^{c}\eta
^{d}\eta ^{f},  \label{bfk3}
\end{eqnarray}%
\begin{equation}
K_{egh}^{abcdf}=\tfrac{1}{4\cdot 4!}\varepsilon ^{\mu \nu \rho \lambda
}H_{e\mu }^{\ast }H_{g\nu }^{\ast }\left( \tfrac{1}{10}C_{h\rho \lambda
}^{\ast }\eta ^{a}+\tfrac{1}{3}H_{h\rho }^{\ast }A_{\lambda }^{a}\right)
\eta ^{b}\eta ^{c}\eta ^{d}\eta ^{f},  \label{bfk4}
\end{equation}%
\begin{equation}
K_{eghl}^{abcdf}=\tfrac{1}{2\cdot 4!\cdot 5!}\varepsilon ^{\mu \nu \rho
\lambda }H_{e\mu }^{\ast }H_{g\nu }^{\ast }H_{h\rho }^{\ast }H_{l\lambda
}^{\ast }\eta ^{a}\eta ^{b}\eta ^{c}\eta ^{d}\eta ^{f},  \label{bfk5}
\end{equation}%
\begin{eqnarray}
K_{b}^{a} &=&4\varepsilon ^{\mu \nu \rho \lambda }\left[ 2\left( -C_{\mu \nu
\rho \lambda }^{a}\eta _{b}^{\ast }+C_{\mu \nu \rho }^{a}A_{b\lambda }^{\ast
}\right) +C_{\mu \nu }^{a}B_{b\rho \lambda }\right.  \notag \\
&&\left. -\left( \varphi ^{\ast a}\eta _{b\mu \nu \rho \lambda }-H_{\mu
}^{a}\eta _{b\nu \rho \lambda }\right) \right] ,  \label{bfk7}
\end{eqnarray}%
\begin{eqnarray}
K_{b,c}^{a} &=&4\varepsilon ^{\mu \nu \rho \lambda }\left[ \eta _{b\mu \nu
\rho \lambda }\left( C_{\sigma \tau \kappa \varsigma }^{a}C_{c}^{\ast \sigma
\tau \kappa \varsigma }+C_{\sigma \tau \kappa }^{a}C_{c}^{\ast \sigma \tau
\kappa }+C_{\sigma \tau }^{a}C_{c}^{\ast \sigma \tau }\right. \right.  \notag
\\
&&\left. +H_{\sigma }^{a}H_{c}^{\ast \sigma }\right) +C_{\mu \nu \rho
\lambda }^{a}\left( \eta _{b\sigma \tau \kappa }C_{c}^{\ast \sigma \tau
\kappa }+B_{b\sigma \tau }C_{c}^{\ast \sigma \tau }-2A_{b\sigma }^{\ast
}H_{c}^{\ast \sigma }\right)  \notag \\
&&\left. +\eta _{b\nu \rho \lambda }\left( 3C_{\mu \sigma \tau
}^{a}C_{c}^{\ast \sigma \tau }-2C_{\mu \sigma }^{a}H_{c}^{\ast \sigma
}\right) +3B_{b\rho \lambda }C_{\mu \nu \sigma }^{a}H_{c}^{\ast \sigma }
\right] ,  \label{bfk8}
\end{eqnarray}%
\begin{eqnarray}
K_{b,cd}^{a} &=&4\varepsilon ^{\mu \nu \rho \lambda }\left[ \eta _{b\mu \nu
\rho \lambda }\left( C_{\sigma \tau \kappa \varsigma }^{a}\left(
4H_{c}^{\ast \sigma }C_{d}^{\ast \tau \kappa \varsigma }+3C_{c}^{\ast \sigma
\tau }C_{d}^{\ast \kappa \varsigma }\right) \right. \right.  \notag \\
&&\left. +3C_{\sigma \tau \kappa }^{a}H_{c}^{\ast \sigma }C_{d}^{\ast \tau
\kappa }+C_{\sigma \tau }^{a}H_{c}^{\ast \sigma }H_{d}^{\ast \tau }\right)
\notag \\
&&\left. +C_{\mu \nu \rho \lambda }^{a}\left( 3\eta _{b\sigma \tau \kappa
}H_{c}^{\ast \sigma }C_{d}^{\ast \tau \kappa }+B_{b\sigma \tau }H_{c}^{\ast
\sigma }H_{d}^{\ast \tau }\right) \right] ,  \label{bfk9}
\end{eqnarray}%
\begin{eqnarray}
K_{b,cde}^{a} &=&4\varepsilon ^{\mu \nu \rho \lambda }\left[ \eta _{b\mu \nu
\rho \lambda }\left( 6C_{\sigma \tau \kappa \varsigma }^{a}H_{c}^{\ast
\sigma }H_{d}^{\ast \tau }C_{e}^{\ast \kappa \varsigma }+C_{\sigma \tau
\kappa }^{a}H_{c}^{\ast \sigma }H_{d}^{\ast \tau }H_{e}^{\ast \kappa
}\right) \right.  \notag \\
&&\left. +C_{\mu \nu \rho \lambda }^{a}\eta _{b\sigma \tau \kappa
}H_{c}^{\ast \sigma }H_{d}^{\ast \tau }H_{e}^{\ast \kappa }\right] ,
\label{bfk10}
\end{eqnarray}%
\begin{equation}
K_{b,cdef}^{a}=4\varepsilon _{\mu \nu \rho \lambda }\eta _{b}^{\mu \nu \rho
\lambda }C_{\sigma \tau \kappa \varsigma }^{a}H_{c}^{\ast \sigma
}H_{d}^{\ast \tau }H_{e}^{\ast \kappa }H_{f}^{\ast \varsigma },
\label{bfk11}
\end{equation}%
\begin{eqnarray}
K_{ab}^{c} &=&\varepsilon _{\mu \nu \rho \lambda }\left[ -6\left( \eta
_{a}^{\mu \nu \sigma }B_{b}^{\rho \lambda }A_{\sigma }^{c}+3\eta _{a}^{\mu
\sigma \tau }\eta _{b\sigma \tau }^{\nu }B^{\ast c\rho \lambda }\right)
\right.  \notag \\
&&-2\eta _{a}^{\mu \nu \rho \lambda }\left( \eta _{b}^{\sigma \tau \kappa
\varsigma }\eta _{\sigma \tau \kappa \varsigma }^{\ast c}+2\eta _{b}^{\sigma
\tau \kappa }\eta _{\sigma \tau \kappa }^{\ast c}+2B_{b}^{\sigma \tau
}B_{\sigma \tau }^{\ast c}\right.  \notag \\
&&\left. \left. -2A_{b}^{\ast \sigma }A_{\sigma }^{c}-2\eta _{b}^{\ast }\eta
^{c}\right) +4\eta _{a}^{\mu \nu \rho }A_{b}^{\ast \lambda }\eta
^{c}-B_{a}^{\mu \nu }B_{b}^{\rho \lambda }\eta ^{c}\right] ,  \label{bfk12}
\end{eqnarray}%
\begin{eqnarray}
K_{ab,d}^{c} &=&\varepsilon _{\mu \nu \rho \lambda }\left[ -9\eta _{a}^{\mu
\sigma \tau }\eta _{b\sigma \tau }^{\nu }\left( \eta ^{c}C_{d}^{\ast \rho
\lambda }-2A^{c\rho }H_{d}^{\ast \lambda }\right) \right.  \notag \\
&&-\eta _{a}^{\sigma \tau \kappa \varsigma }\eta _{b\sigma \tau \kappa
\varsigma }\left( \eta ^{c}C_{d}^{\ast \mu \nu \rho \lambda }+4C_{d}^{\ast
\mu \nu \rho }A^{c\lambda }\right.  \notag \\
&&\left. +12C_{d}^{\ast \mu \nu }B^{\ast c\rho \lambda }+8H_{d}^{\ast \mu
}\eta ^{\ast c\nu \rho \lambda }\right) +6\eta _{a}^{\mu \nu \sigma
}B_{b}^{\rho \lambda }\eta ^{c}H_{d\sigma }^{\ast }  \notag \\
&&-2\eta _{a}^{\mu \nu \rho \lambda }\left( \eta _{b}^{\sigma \tau \kappa
}\left( \eta ^{c}C_{d\sigma \tau \kappa }^{\ast }+3A_{\kappa }^{c}C_{d\sigma
\tau }^{\ast }-6B_{\tau \kappa }^{\ast c}H_{d\sigma }^{\ast }\right) \right.
\notag \\
&&\left. \left. +2A_{b}^{\ast \sigma }\eta ^{c}H_{d\sigma }^{\ast
}+B_{b}^{\sigma \tau }\left( \eta ^{c}C_{d\sigma \tau }^{\ast }+2A_{\tau
}^{c}H_{d\sigma }^{\ast }\right) \right) \right] ,  \label{bfk13}
\end{eqnarray}%
\begin{eqnarray}
K_{ab,de}^{c} &=&-\varepsilon _{\mu \nu \rho \lambda }\left[ 2\eta _{a}^{\mu
\nu \rho \lambda }\left( 3\eta _{b}^{\sigma \tau \kappa }\left( H_{d\sigma
}^{\ast }C_{e\tau \kappa }^{\ast }\eta ^{c}+H_{d\sigma }^{\ast }H_{e\tau
}^{\ast }A_{\kappa }^{c}\right) \right. \right.  \notag \\
&&\left. +B_{b}^{\sigma \tau }H_{d\sigma }^{\ast }H_{e\tau }^{\ast }\eta
^{c}\right) +\eta _{a}^{\sigma \tau \kappa \varsigma }\eta _{b\sigma \tau
\kappa \varsigma }\left( \left( 4H_{d}^{\ast \mu }C_{e}^{\ast \nu \rho
\lambda }\right. \right.  \notag \\
&&\left. \left. +3C_{d}^{\ast \mu \nu }C_{e}^{\ast \rho \lambda }\right)
\eta ^{c}+12H_{d}^{\ast \mu }C_{e}^{\ast \nu \rho }A^{c\lambda
}+12H_{d}^{\ast \mu }H_{e}^{\ast \nu }B^{\ast c\rho \lambda }\right)  \notag
\\
&&\left. +9\eta _{a}^{\mu \sigma \tau }\eta _{b\sigma \tau }^{\nu
}H_{d}^{\ast \rho }H_{e}^{\ast \lambda }\eta ^{c}\right] ,  \label{bfk14}
\end{eqnarray}%
\begin{eqnarray}
K_{ab,def}^{c} &=&-2\varepsilon _{\mu \nu \rho \lambda }\left[ \eta
_{a}^{\sigma \tau \kappa \varsigma }\eta _{b\sigma \tau \kappa \varsigma
}\left( 3H_{d}^{\ast \mu }H_{e}^{\ast \nu }C_{f}^{\ast \rho \lambda }\eta
^{c}+2H_{d}^{\ast \mu }H_{e}^{\ast \nu }H_{f}^{\ast \rho }A^{c\lambda
}\right) \right.  \notag \\
&&\left. +\eta _{a}^{\mu \nu \rho \lambda }\eta _{b}^{\sigma \tau \kappa
}H_{d\sigma }^{\ast }H_{e\tau }^{\ast }H_{f\kappa }^{\ast }\eta ^{c}\right] ,
\label{bfk15}
\end{eqnarray}%
\begin{equation}
K_{ab,defg}^{c}=-\varepsilon _{\mu \nu \rho \lambda }\eta _{a}^{\sigma \tau
\kappa \varsigma }\eta _{b\sigma \tau \kappa \varsigma }H_{d}^{\ast \mu
}H_{e}^{\ast \nu }H_{f}^{\ast \rho }H_{g}^{\ast \lambda }\eta ^{c}.
\label{bfk16}
\end{equation}

Next, we identify the various notations employed in formula (\ref{so3}). The
polynomials $X_{A,m_{1}\ldots m_{p}}^{abB}$, $X_{A,m_{1}\ldots m_{p}}^{abcd}$%
, $X_{A,m_{1}\ldots m_{p}}^{ab}$, $X_{Ac,m_{1}\ldots m_{p}}^{ab}$, $%
X_{Aab,m_{1}\ldots m_{p}}$, and $X_{a,m_{1}\ldots m_{p}}^{AB}$, with $p=%
\overline{0,3}$, can be written as
\begin{eqnarray}
X_{A}^{abB} &=&\left( C_{A}^{\ast }\eta ^{a}-2C_{A}^{\ast \mu }A_{\mu
}^{a}\right) \eta ^{b}C^{B}+C_{A}^{\ast \mu }\eta ^{a}\eta ^{b}C_{\mu }^{B}
\notag \\
&&+\tfrac{2}{3}\left( V_{A}^{\mu }\eta ^{\ast a\nu \rho \lambda
}-3V_{A}^{\ast \mu \nu }B^{\ast a\rho \lambda }\right) \eta
^{b}C^{B}\varepsilon _{\mu \nu \rho \lambda }  \notag \\
&&-2\left( V_{A}^{\mu }B^{\ast a\nu \rho }-V_{A}^{\ast \mu \nu }A^{a\rho
}\right) A^{b\lambda }C^{B}\varepsilon _{\mu \nu \rho \lambda }+2V_{A}^{\ast
\mu \nu }A^{a\rho }\eta ^{b}C^{B\lambda }\varepsilon _{\mu \nu \rho \lambda }
\notag \\
&&+\left( V_{A}^{\ast \mu \nu }V_{\mu \nu }^{B}+V_{A}^{\ast \mu }V_{\mu
}^{B}\right) \eta ^{a}\eta ^{b}-2V_{A}^{\mu }B^{\ast a\nu \rho }\eta
^{b}C^{B\lambda }\varepsilon _{\mu \nu \rho \lambda }  \notag \\
&&+V_{A}^{\mu }A^{a\nu }\left( A^{b\rho }C^{B\lambda }\varepsilon _{\mu \nu
\rho \lambda }-2\eta ^{b}V_{\mu \nu }^{B}\right) ,  \label{f26}
\end{eqnarray}%
\begin{eqnarray}
X_{A,m_{1}}^{abB} &=&-\tfrac{1}{2}\left( 2H_{m_{1}}^{\ast \mu }C_{A\mu
}^{\ast }+C_{m_{1}}^{\ast \mu \nu }V_{A}^{\ast \rho \lambda }\varepsilon
_{\mu \nu \rho \lambda }+\tfrac{1}{3}C_{m_{1}}^{\ast \mu \nu \rho
}V_{A}^{\lambda }\varepsilon _{\mu \nu \rho \lambda }\right) \eta ^{a}\eta
^{b}C^{B}  \notag \\
&&+\left[ \left( C_{m_{1}}^{\ast \mu \nu }V_{A}^{\rho }+H_{m_{1}}^{\ast \mu
}V_{A}^{\ast \nu \rho }\right) A^{a\lambda }-2H_{m_{1}}^{\ast \mu
}V_{A}^{\nu }B^{\ast a\rho \lambda }\right] \eta ^{b}C^{B}\varepsilon _{\mu
\nu \rho \lambda }  \notag \\
&&-\tfrac{1}{2}\left( C_{m_{1}}^{\ast \mu \nu }V_{A}^{\rho
}+2H_{m_{1}}^{\ast \mu }V_{A}^{\ast \nu \rho }\right) \eta ^{a}\eta
^{b}C^{B\lambda }\varepsilon _{\mu \nu \rho \lambda }+H_{m_{1}}^{\ast \mu
}V_{A}^{\nu }\eta ^{a}\eta ^{b}V_{\mu \nu }^{B}  \notag \\
&&+H_{m_{1}}^{\ast \mu }V_{A}^{\nu }A^{a\rho }\left( A^{b\lambda
}C^{B}+2\eta ^{b}C^{B\lambda }\right) \varepsilon _{\mu \nu \rho
\lambda } ,  \label{f27}
\end{eqnarray}%
\begin{eqnarray}
X_{A,m_{1}m_{2}}^{abB} &=&-\tfrac{1}{6}\left( 3H_{m_{1}}^{\ast \mu
}H_{m_{2}}^{\ast \nu }V_{A}^{\ast \rho \lambda }+C_{m_{1}}^{\ast \lbrack \mu
\nu }H_{m_{2}}^{\ast \rho ]}V_{A}^{\lambda }\right) \eta ^{a}\eta
^{b}C^{B}\varepsilon _{\mu \nu \rho \lambda }  \notag \\
&&+\tfrac{1}{2}H_{m_{1}}^{\ast \mu }H_{m_{2}}^{\ast \nu }V_{A}^{\rho }\left(
2A^{a\lambda }\eta ^{b}C^{B}-\eta ^{a}\eta ^{b}C^{B\lambda }\right)
\varepsilon _{\mu \nu \rho \lambda },  \label{f28}
\end{eqnarray}%
\begin{equation}
X_{A,m_{1}m_{2}m_{3}}^{abB}=-\tfrac{1}{6}H_{m_{1}}^{\ast \mu
}H_{m_{2}}^{\ast \nu }H_{m_{3}}^{\ast \rho }V_{A}^{\lambda }\eta ^{a}\eta
^{b}C^{B}\varepsilon _{\mu \nu \rho \lambda },  \label{f29}
\end{equation}%
\begin{eqnarray}
X_{A}^{abcd} &=&\tfrac{1}{12}C_{A}^{\ast }\eta ^{a}\eta ^{b}\eta ^{c}\eta
^{d}-\tfrac{1}{3}V_{A}^{\mu }A^{a\nu }A^{b\rho }A^{c\lambda }\eta
^{d}\varepsilon _{\mu \nu \rho \lambda }  \notag \\
&&-\tfrac{1}{3}\left[ C_{A}^{\ast \mu }A_{\mu }^{a}+\left( V_{A}^{\ast \mu
\nu }B^{\ast a\rho \lambda }-\tfrac{1}{3}V_{A}^{\mu }\eta ^{\ast a\nu \rho
\lambda }\right) \varepsilon _{\mu \nu \rho \lambda }\right] \eta ^{b}\eta
^{c}\eta ^{d}  \notag \\
&&+\tfrac{1}{2}\left( V_{A}^{\ast \mu \nu }A^{a\rho }-2V_{A}^{\mu }B^{\ast
a\nu \rho }\right) A^{b\lambda }\eta ^{c}\eta ^{d}\varepsilon _{\mu \nu \rho
\lambda },  \label{f19}
\end{eqnarray}%
\begin{eqnarray}
X_{A,m_{1}}^{abcd} &=&-\tfrac{1}{4!}\left[ 2H_{m_{1}}^{\ast \mu }C_{A\mu
}^{\ast }+\left( C_{m_{1}}^{\ast \mu \nu }V_{A}^{\ast \rho \lambda }+\tfrac{1%
}{3}C_{m_{1}}^{\ast \mu \nu \rho }V_{A}^{\lambda }\right) \varepsilon _{\mu
\nu \rho \lambda }\right] \eta ^{a}\eta ^{b}\eta ^{c}\eta ^{d}  \notag \\
&&+\tfrac{1}{6}\left[ \left( C_{m_{1}}^{\ast \mu \nu }V_{A}^{\rho
}+2H_{m_{1}}^{\ast \mu }V_{A}^{\ast \nu \rho }\right) A^{a\lambda
}-2H_{m_{1}}^{\ast \mu }V_{A}^{\nu }B^{\ast a\rho \lambda }\right] \eta
^{b}\eta ^{c}\eta ^{d}\varepsilon _{\mu \nu \rho \lambda }  \notag \\
&&+\tfrac{1}{2}H_{m_{1}}^{\ast \mu }V_{A}^{\nu }A^{a\rho }A^{b\lambda }\eta
^{c}\eta ^{d}\varepsilon _{\mu \nu \rho \lambda },  \label{f20}
\end{eqnarray}%
\begin{eqnarray}
X_{A,m_{1}m_{2}}^{abcd} &=&-\tfrac{1}{4!}\left( H_{m_{1}}^{\ast \mu
}H_{m_{2}}^{\ast \nu }V_{A}^{\ast \rho \lambda }+\tfrac{1}{3}C_{m_{1}}^{\ast
\lbrack \mu \nu }H_{m_{2}}^{\ast \rho ]}V_{A}^{\lambda }\right) \eta
^{a}\eta ^{b}\eta ^{c}\eta ^{d}\varepsilon _{\mu \nu \rho \lambda }  \notag
\\
&&+\tfrac{1}{6}H_{m_{1}}^{\ast \mu }H_{m_{2}}^{\ast \nu }V_{A}^{\rho
}A^{a\lambda }\eta ^{b}\eta ^{c}\eta ^{d}\varepsilon _{\mu \nu \rho \lambda
},  \label{f21}
\end{eqnarray}%
\begin{equation}
X_{A,m_{1}m_{2}m_{3}}^{abcd}=-\tfrac{1}{3\cdot 4!}H_{m_{1}}^{\ast
\mu }H_{m_{2}}^{\ast \nu }H_{m_{3}}^{\ast \rho }V_{A}^{\lambda }\eta
^{a}\eta ^{b}\eta ^{c}\eta ^{d}\varepsilon _{\mu \nu \rho \lambda },
\label{f22}
\end{equation}%
\begin{eqnarray}
X_{A}^{ab} &=&\tfrac{1}{12}\left( C_{A}^{\ast }\eta ^{a}-C_{A\mu }^{\ast
}A^{a\mu }\right) C_{\alpha \beta \gamma \delta }^{b}\varepsilon ^{\alpha
\beta \gamma \delta }-\tfrac{1}{12}C_{A\mu }^{\ast }\eta ^{a}C_{\nu \rho
\lambda }^{b}\varepsilon ^{\mu \nu \rho \lambda }  \notag \\
&&+\tfrac{1}{6}V_{A}^{\ast \mu \nu }\left( 12B^{\ast a\rho \lambda }C_{\mu
\nu \rho \lambda }^{b}+\eta ^{a}C_{\mu \nu }^{b}-3A^{a\rho }C_{\mu \nu \rho
}^{b}\right) -\tfrac{1}{12}V_{A}^{\mu }\left( 2A^{a\nu }C_{\mu \nu
}^{b}\right.  \notag \\
&&\left. +8\eta ^{\ast a\nu \rho \lambda }C_{\mu \nu \rho \lambda
}^{b}-6B^{\ast a\nu \rho }C_{\mu \nu \rho }^{b}-\eta ^{a}H_{\mu }^{b}\right)
,  \label{f1}
\end{eqnarray}%
\begin{eqnarray}
X_{A,m_{1}}^{ab} &=&\left( -\tfrac{1}{12}H_{m_{1}}^{\ast \alpha }C_{A\alpha
}^{\ast }\varepsilon ^{\mu \nu \rho \lambda }+C_{m_{1}}^{\ast \mu \nu
}V_{A}^{\ast \rho \lambda }+\tfrac{1}{3}C_{m_{1}}^{\ast \mu \nu \rho
}V_{A}^{\lambda }\right) \eta ^{a}C_{\mu \nu \rho \lambda }^{b}  \notag \\
&&+\tfrac{1}{4}\left( 2H_{m_{1}}^{\ast \mu }V_{A}^{\ast \nu \rho
}+C_{m_{1}}^{\ast \mu \nu }V_{A}^{\rho }\right) \eta ^{a}C_{\mu \nu \rho
}^{b}  \notag \\
&&-\tfrac{1}{2}\left( 2H_{m_{1}}^{\ast \mu }V_{A}^{\ast \nu \rho
}+C_{m_{1}}^{\ast \mu \nu }V_{A}^{\rho }\right) A^{a\lambda }C_{\mu \nu \rho
\lambda }^{b}  \notag \\
&&+\tfrac{1}{2}H_{m_{1}}^{\ast \mu }V_{A}^{\nu }\left( 4B^{\ast a\rho
\lambda }C_{\mu \nu \rho \lambda }^{b}+\tfrac{1}{3}\eta ^{a}C_{\mu \nu
}^{b}-A^{a\rho }C_{\mu \nu \rho }^{b}\right) ,  \label{f2}
\end{eqnarray}%
\begin{eqnarray}
X_{A,m_{1}m_{2}}^{ab} &=&\tfrac{1}{3}\left( 3H_{m_{1}}^{\ast \mu
}H_{m_{2}}^{\ast \nu }V_{A}^{\ast \rho \lambda }+C_{m_{1}}^{\ast \lbrack \mu
\nu }H_{m_{2}}^{\rho ]}V_{A}^{\lambda }\right) \eta ^{a}C_{\mu \nu \rho
\lambda }^{b}  \notag \\
&&+\tfrac{1}{4}H_{m_{1}}^{\ast \mu }H_{m_{2}}^{\ast \nu }\left( V_{A}^{\rho
}\eta ^{a}C_{\mu \nu \rho }^{b}-4V_{A}^{\rho }A^{a\lambda }C_{\mu \nu \rho
\lambda }^{b}\right) ,  \label{f3}
\end{eqnarray}%
\begin{equation}
X_{A,m_{1}m_{2}m_{3}}^{ab}=\tfrac{1}{3}H_{m_{1}}^{\ast \mu }H_{m_{2}}^{\ast
\nu }H_{m_{3}}^{\rho }V_{A}^{\lambda }\eta ^{a}C_{\mu \nu \rho \lambda }^{b},
\label{f4}
\end{equation}%
\begin{eqnarray}
X_{Ac}^{ab} &=&\tfrac{1}{4!}\left( C_{A}^{\ast }\eta ^{a}-2C_{A\alpha
}^{\ast }A^{a\alpha }\right) \eta ^{b}\eta _{c\mu \nu \rho \lambda
}\varepsilon ^{\mu \nu \rho \lambda }-\tfrac{1}{4!}C_{A\mu }^{\ast }\eta
^{a}\eta ^{b}\eta _{c\nu \rho \lambda }\varepsilon ^{\mu \nu \rho \lambda }
\notag \\
&&+V_{A}^{\ast \mu \nu }\left( 2B^{\ast a\rho \lambda }\eta ^{b}-A^{a\rho
}A^{b\lambda }\right) \eta _{c\mu \nu \rho \lambda }  \notag \\
&&+\tfrac{1}{12}\left( V_{A}^{\ast \mu \nu }\eta ^{a}-2V_{A}^{\mu }A^{a\nu
}\right) \eta ^{b}B_{c\mu \nu }  \notag \\
&&-\tfrac{1}{2}\left( V_{A}^{\ast \mu \nu }A^{a\rho }-V_{A}^{\mu }B^{\ast
a\nu \rho }\right) \eta ^{b}\eta _{c\mu \nu \rho }  \notag \\
&&-2V_{A}^{\mu }\left( \tfrac{1}{3}\eta ^{\ast a\nu \rho \lambda }\eta
^{b}-B^{\ast a\nu \rho }A^{b\lambda }\right) \eta _{c\mu \nu \rho \lambda }
\notag \\
&&-\tfrac{1}{12}V_{A}^{\mu }\eta ^{a}\eta ^{b}A_{c\mu }^{\ast }-\tfrac{1}{4}%
V_{A}^{\mu }A^{a\nu }A^{b\rho }\eta _{c\mu \nu \rho },  \label{f5}
\end{eqnarray}%
\begin{eqnarray}
X_{Ac,m_{1}}^{ab} &=&-\tfrac{1}{4!}\left( H_{m_{1}}^{\ast \alpha }C_{A\alpha
}^{\ast }\varepsilon ^{\mu \nu \rho \lambda }-12C_{m_{1}}^{\ast \mu \nu
}V_{A}^{\ast \rho \lambda }-4C_{m_{1}}^{\ast \mu \nu \rho }V_{A}^{\lambda
}\right) \eta ^{a}\eta ^{b}\eta _{c\mu \nu \rho \lambda }  \notag \\
&&+\tfrac{1}{4}\left( 2H_{m_{1}}^{\ast \mu }V_{A}^{\ast \nu \rho
}+C_{m_{1}}^{\ast \mu \nu }V_{A}^{\rho }\right) \eta ^{a}\eta ^{b}\eta
_{c\mu \nu \rho }+\tfrac{1}{12}H_{m_{1}}^{\ast \mu }V_{A}^{\nu }\eta
^{a}\eta ^{b}B_{c\mu \nu }  \notag \\
&&-\left[ \left( 2H_{m_{1}}^{\ast \mu }V_{A}^{\ast \nu \rho
}+C_{m_{1}}^{\ast \mu \nu }V_{A}^{\rho }\right) A^{a\lambda
}-2H_{m_{1}}^{\ast \mu }V_{A}^{\nu }B^{\ast a\rho \lambda }\right] \eta
^{b}\eta _{c\mu \nu \rho \lambda }  \notag \\
&&-\tfrac{1}{2}H_{m_{1}}^{\ast \mu }V_{A}^{\nu }A^{a\rho }\left( \eta
^{b}\eta _{c\mu \nu \rho }+2A^{b\lambda }\eta _{c\mu \nu \rho \lambda
}\right) ,  \label{f6}
\end{eqnarray}%
\begin{eqnarray}
X_{Ac,m_{1}m_{2}}^{ab} &=&\tfrac{1}{6}\left( 3H_{m_{1}}^{\ast \mu
}H_{m_{2}}^{\ast \nu }V_{A}^{\ast \rho \lambda }+C_{m_{1}}^{\ast \lbrack \mu
\nu }H_{m_{2}}^{\ast \rho ]}V_{A}^{\lambda }\right) \eta ^{a}\eta ^{b}\eta
_{c\mu \nu \rho \lambda }  \notag \\
&&+\tfrac{1}{8}H_{m_{1}}^{\ast \mu }H_{m_{2}}^{\ast \nu }V_{A}^{\rho }\left(
\eta ^{a}\eta ^{b}\eta _{c\mu \nu \rho }-8A^{a\lambda }\eta ^{b}\eta _{c\mu
\nu \rho \lambda }\right) ,  \label{f7}
\end{eqnarray}%
\begin{equation}
X_{Ac,m_{1}m_{2}m_{3}}^{ab}=\tfrac{1}{6}H_{m_{1}}^{\ast \mu }H_{m_{2}}^{\ast
\nu }H_{m_{3}}^{\ast \rho }V_{A}^{\lambda }\eta ^{a}\eta ^{b}\eta _{c\mu \nu
\rho \lambda },  \label{f8}
\end{equation}%
\begin{eqnarray}
X_{Aab} &=&-\left( C_{A}^{\ast }\eta _{a\mu \nu \rho \lambda }+2C_{A\mu
}^{\ast }\eta _{a\nu \rho \lambda }\right) \eta _{b}^{\mu \nu \rho \lambda }+%
\tfrac{3}{4}V_{A}^{\ast \mu \nu }\eta _{a\alpha \beta }^{\;\rho
}\eta
_{b}^{\lambda \alpha \beta }\varepsilon _{\mu \nu \rho \lambda }  \notag \\
&&+2\left( V_{A}^{\ast \alpha \beta }B_{a\alpha \beta }-\tfrac{1}{12}%
V_{A}^{\alpha }A_{a\alpha }^{\ast }\right) \eta _{b\mu \nu \rho \lambda
}\varepsilon ^{\mu \nu \rho \lambda }  \notag \\
&&-\tfrac{1}{12}V_{A}^{\alpha }B_{a\alpha \mu }\eta _{b\nu \rho \lambda
}\varepsilon ^{\mu \nu \rho \lambda },  \label{f9}
\end{eqnarray}%
\begin{eqnarray}
X_{Aab,m_{1}} &=&\tfrac{1}{6}\left( 6H_{m_{1}}^{\ast \mu }C_{A\mu }^{\ast
}+3C_{m_{1}}^{\ast \mu \nu }V_{A}^{\ast \rho \lambda }\varepsilon _{\mu \nu
\rho \lambda }+C_{m_{1}}^{\ast \mu \nu \rho }V_{A}^{\lambda }\varepsilon
_{\mu \nu \rho \lambda }\right) \eta _{a\alpha \beta \gamma \delta }\eta
_{b}^{\alpha \beta \gamma \delta }  \notag \\
&&+\tfrac{1}{4}\left( 2H_{m_{1}}^{\ast \mu }V_{A}^{\ast \nu \rho
}+C_{m_{1}}^{\ast \mu \nu }V_{A}^{\rho }\right) \eta _{a\mu \nu \rho }\eta
_{b\alpha \beta \gamma \delta }\varepsilon ^{\alpha \beta \gamma \delta }
\notag \\
&&+\tfrac{3}{4}H_{m_{1}}^{\ast \mu }V_{A}^{\nu }\eta _{a\alpha \beta
}^{\;\rho }\eta _{b}^{\lambda \alpha \beta }\varepsilon _{\mu \nu
\rho \lambda }+\tfrac{1}{6}H_{m_{1}}^{\ast \mu }V_{A}^{\nu }B_{a\mu
\nu }\eta _{b\alpha \beta \gamma \delta }\varepsilon ^{\alpha \beta
\gamma \delta }, \label{f10}
\end{eqnarray}%
\begin{eqnarray}
X_{Aab,m_{1}m_{2}} &=&\tfrac{1}{6}\left( 3H_{m_{1}}^{\ast \mu
}H_{m_{2}}^{\ast \nu }V_{A}^{\ast \rho \lambda }+C_{m_{1}}^{\ast \lbrack \mu
\nu }H_{m_{2}}^{\ast \rho ]}V_{A}^{\lambda }\right) \eta _{a\alpha \beta
\gamma \delta }\eta _{b}^{\alpha \beta \gamma \delta }\varepsilon _{\mu \nu
\rho \lambda }  \notag \\
&&+\tfrac{1}{4}H_{m_{1}}^{\ast \mu }H_{m_{2}}^{\ast \nu }V_{A}^{\rho }\eta
_{a\mu \nu \rho }\eta _{b\alpha \beta \gamma \delta }\varepsilon ^{\alpha
\beta \gamma \delta },  \label{f11}
\end{eqnarray}%
\begin{equation}
X_{Aab,m_{1}m_{2}m_{3}}=\tfrac{1}{6}H_{m_{1}}^{\ast \mu
}H_{m_{2}}^{\ast \nu }H_{m_{3}}^{\ast \rho }V_{A}^{\lambda }\eta
_{a\alpha \beta \gamma \delta }\eta _{b}^{\alpha \beta \gamma \delta
}\varepsilon _{\mu \nu \rho \lambda }, \label{f12}
\end{equation}%
\begin{eqnarray}
X_{a}^{AB} &=&-4\left( C^{\ast A}C^{B}+C_{\alpha }^{\ast A}C^{B\alpha
}\right) \eta _{a\mu \nu \rho \lambda }\varepsilon ^{\mu \nu \rho \lambda }
\notag \\
&&+4\left( C_{\mu }^{\ast A}C^{B}\varepsilon ^{\mu \nu \rho \lambda
}-6V^{\ast A\nu \rho }C^{B\lambda }\right) \eta _{a\nu \rho \lambda
}-8V^{\ast A\mu \nu }C^{B}B_{a\mu \nu }  \notag \\
&&+8V_{\mu }^{A}C^{B}A_{a}^{\ast \mu }-\left( V_{\alpha }^{\ast A}V^{B\alpha
}+4V^{\ast A\alpha \beta }V_{\alpha \beta }^{B}\right) \eta _{a\mu \nu \rho
\lambda }\varepsilon ^{\mu \nu \rho \lambda }  \notag \\
&&+4V^{A\alpha }V_{\alpha \mu }^{B}\varepsilon ^{\mu \nu \rho \lambda }\eta
_{a\nu \rho \lambda }-8V^{A\mu }C^{B\nu }B_{a\mu \nu },  \label{f30}
\end{eqnarray}%
\begin{eqnarray}
X_{a,m_{1}}^{AB} &=&4\left( H_{m_{1}}^{\ast \alpha }C_{\alpha }^{\ast
A}\varepsilon ^{\mu \nu \rho \lambda }-12C_{m_{1}}^{\ast \mu \nu }V^{\ast
A\rho \lambda }-4C_{m_{1}}^{\ast \mu \nu \rho }V^{A\lambda }\right)
C^{B}\eta _{a\mu \nu \rho \lambda }  \notag \\
&&-12\left( 2H_{m_{1}}^{\ast \mu }V^{\ast A\nu \rho
}+C_{m_{1}}^{\ast \mu \nu }V^{A\rho }\right) \left( C^{B}\eta _{a\mu
\nu \rho }+4C^{B\lambda }\eta
_{a\mu \nu \rho \lambda }\right)  \notag \\
&&-4H_{m_{1}}^{\ast \mu }V^{A\nu }\left( 6C^{B\rho }\eta _{a\mu \nu \rho
}+V_{\mu \nu }^{B}\eta _{a\alpha \beta \gamma \delta }\varepsilon ^{\alpha
\beta \gamma \delta }+2C^{B}B_{a\mu \nu }\right) ,  \label{f31}
\end{eqnarray}%
\begin{eqnarray}
X_{a,m_{1}m_{2}}^{AB} &=&-16\left( 3H_{m_{1}}^{\ast \mu }H_{m_{2}}^{\ast \nu
}V^{\ast A\rho \lambda }+C_{m_{1}}^{\ast \lbrack \mu \nu }H_{m_{2}}^{\ast
\rho ]}V^{A\lambda }\right) C^{B}\eta _{a\mu \nu \rho \lambda }  \notag \\
&&-12H_{m_{1}}^{\ast \mu }H_{m_{2}}^{\ast \nu }V^{A\rho }\left( C^{B}\eta
_{a\mu \nu \rho }+4C^{B\lambda }\eta _{a\mu \nu \rho \lambda }\right) ,
\label{f32}
\end{eqnarray}%
\begin{equation}
X_{a,m_{1}m_{2}m_{3}}^{AB}=-16H_{m_{1}}^{\ast \mu }H_{m_{2}}^{\ast \nu
}H_{m_{3}}^{\ast \rho }V_{A}^{\lambda }C^{B}\eta _{a\mu \nu \rho \lambda }.
\label{f33}
\end{equation}%
The objects denoted by $X_{m_{1}\ldots m_{p}}^{aABC}$, $X_{AB,m_{1}\ldots
m_{p}}^{abc}$, $X_{AB,m_{1}\ldots m_{p}}^{a}$, and $X_{ABa,m_{1}\ldots
m_{p}}^{b}$, with $p=\overline{0,2}$, read as
\begin{eqnarray}
X^{aABC} &=&-\tfrac{1}{4!}\left( 2C^{\ast A\mu }V_{\mu }^{B}+V^{\ast A\mu
\nu }V^{\ast B\rho \lambda }\varepsilon _{\mu \nu \rho \lambda }\right)
C^{C}\eta ^{a}  \notag \\
&&+\tfrac{1}{12}V^{\ast A\mu \nu }V^{B\rho }\left( C^{C}A^{a\lambda
}-C^{C\lambda }\eta ^{a}\right) \varepsilon _{\mu \nu \rho \lambda }  \notag
\\
&&+\tfrac{1}{4!}V^{A\mu }V^{B\nu }C^{C}B^{\ast a\rho \lambda }\varepsilon
_{\mu \nu \rho \lambda }  \notag \\
&&-\tfrac{1}{4!}V^{A\mu }V^{B\nu }\left( C^{C\rho }A^{a\lambda }\varepsilon
_{\mu \nu \rho \lambda }+V_{\mu \nu }^{C}\eta ^{a}\right) ,  \label{f23}
\end{eqnarray}%
\begin{eqnarray}
X_{m_{1}}^{aABC} &=&\tfrac{1}{2\cdot 4!}\left( C_{m_{1}}^{\ast \mu \nu
}V^{A\rho }+4H_{m_{1}}^{\ast \mu }V^{\ast A\nu \rho }\right) V^{B\lambda
}C^{C}\eta ^{a}\varepsilon _{\mu \nu \rho \lambda }  \notag \\
&&+\tfrac{1}{4!}H_{m_{1}}^{\ast \mu }V^{A\nu }V^{B\rho }\left(
C^{C}A^{a\lambda }-C^{C\lambda }\eta ^{a}\right) \varepsilon _{\mu \nu \rho
\lambda },  \label{f24}
\end{eqnarray}%
\begin{equation}
X_{m_{1}m_{2}}^{aABC}=\tfrac{1}{2\cdot 4!}H_{m_{1}}^{\ast \mu
}H_{m_{2}}^{\ast \nu }V^{A\rho }V^{B\lambda }C^{C}\eta ^{a}\varepsilon _{\mu
\nu \rho \lambda },  \label{f25}
\end{equation}%
\begin{eqnarray}
X_{AB}^{abc} &=&\tfrac{1}{2\cdot 4!}\left[ \left( V_{A}^{\ast \mu
\nu }V_{B}^{\rho }A^{a\lambda }+V_{A}^{\mu }V_{B}^{\nu }B^{\ast
a\rho \lambda }\right) \eta ^{b}-V_{A}^{\mu }V_{B}^{\nu }A^{a\rho
}A^{b\lambda }\right]
\eta ^{c}\varepsilon _{\mu \nu \rho \lambda }  \notag \\
&&-\tfrac{1}{6\cdot 4!}\left( 2C_{A\mu }^{\ast }V_{B}^{\mu }+V_{A}^{\ast \mu
\nu }V_{B}^{\ast \rho \lambda }\varepsilon _{\mu \nu \rho \lambda }\right)
\eta ^{a}\eta ^{b}\eta ^{c},  \label{f34}
\end{eqnarray}%
\begin{eqnarray}
X_{AB,m_{1}}^{abc} &=&\tfrac{1}{12\cdot 4!}\left[ \left( 4H_{m_{1}}^{\ast
\mu }V_{A}^{\ast \nu \rho }V_{B}^{\lambda }+C_{m_{1}}^{\ast \mu \nu
}V_{A}^{\rho }V_{B}^{\lambda }\right) \eta ^{a}\right.  \notag \\
&&\left. +6H_{m_{1}}^{\ast \mu }V_{A}^{\nu }V_{B}^{\rho }A^{a\lambda }\right]
\eta ^{b}\eta ^{c}\varepsilon _{\mu \nu \rho \lambda },  \label{f35}
\end{eqnarray}%
\begin{equation}
X_{AB,m_{1}m_{2}}^{abc}=\tfrac{1}{12\cdot 4!}H_{m_{1}}^{\ast \mu
}H_{m_{2}}^{\ast \nu }V_{A}^{\rho }V_{B}^{\lambda }\eta ^{a}\eta
^{b}\eta ^{c}\varepsilon _{\mu \nu \rho \lambda },  \label{f36}
\end{equation}%
\begin{eqnarray}
X_{AB}^{a} &=&-\tfrac{1}{12\cdot 4!}\left( C_{A\alpha }^{\ast }V_{B}^{\alpha
}\varepsilon ^{\mu \nu \rho \lambda }-12V_{A}^{\ast \mu \nu }V_{B}^{\ast
\rho \lambda }\right) C_{\mu \nu \rho \lambda }^{a}  \notag \\
&&-\tfrac{1}{2\cdot 4!}V_{A}^{\ast \mu \nu }V_{B}^{\rho }C_{\mu \nu \rho
}^{a}-\tfrac{1}{2\cdot 4!}V_{A}^{\mu }V_{B}^{\nu }C_{\mu \nu }^{a},
\label{f13}
\end{eqnarray}%
\begin{eqnarray}
X_{AB,m_{1}}^{a} &=&-\tfrac{1}{12}\left( H_{m_{1}}^{\ast \mu }V_{A}^{\ast
\nu \rho }V_{B}^{\lambda }+\tfrac{1}{4}C_{m_{1}}^{\ast \mu \nu }V_{A}^{\rho
}V_{B}^{\lambda }\right) C_{\mu \nu \rho \lambda }^{a}  \notag \\
&&-\tfrac{1}{4\cdot 4!}H_{m_{1}}^{\ast \mu }V_{A}^{\nu }V_{B}^{\rho }C_{\mu
\nu \rho }^{a},  \label{f14}
\end{eqnarray}%
\begin{equation}
X_{AB,m_{1}m_{2}}^{a}=-\tfrac{1}{2\cdot 4!}H_{m_{1}}^{\ast \mu
}H_{m_{2}}^{\ast \nu }V_{A}^{\rho }V_{B}^{\lambda }C_{\mu \nu \rho \lambda
}^{a},  \label{f15}
\end{equation}%
\begin{eqnarray}
X_{ABa}^{b} &=&-\tfrac{1}{12\cdot 4!}\left( C_{A\alpha }^{\ast
}V_{B}^{\alpha }\varepsilon ^{\mu \nu \rho \lambda }-12V_{A}^{\ast \mu \nu
}V_{B}^{\ast \rho \lambda }\right) \eta _{a\mu \nu \rho \lambda }\eta
^{b} \notag \\
&&+\tfrac{1}{2\cdot 4!}V_{A}^{\ast \mu \nu }V_{B}^{\rho }\left( \eta _{a\mu
\nu \rho }\eta ^{b}-4\eta _{a\mu \nu \rho \lambda }A^{b\lambda }\right) -%
\tfrac{1}{4!}V_{A}^{\mu }V_{B}^{\nu }\eta _{a\mu \nu \rho \lambda }B^{\ast
b\rho \lambda }  \notag \\
&&-\tfrac{1}{12\cdot 4!}V_{A}^{\mu }V_{B}^{\nu }\left( B_{a\mu \nu }\eta
^{b}+3\eta _{a\mu \nu \rho }A^{b\rho }\right) ,  \label{f16}
\end{eqnarray}%
\begin{eqnarray}
X_{ABa,m_{1}}^{b} &=&-\tfrac{1}{12}\left( H_{m_{1}}^{\ast \mu }V_{A}^{\ast
\nu \rho }+\tfrac{1}{4}C_{m_{1}}^{\ast \mu \nu }V_{A}^{\rho }\right)
V_{B}^{\lambda }\eta _{a\mu \nu \rho \lambda }\eta ^{b}  \notag \\
&&+\tfrac{1}{4\cdot 4!}H_{m_{1}}^{\ast \mu }V_{A}^{\nu }V_{B}^{\rho }\left(
\eta _{a\mu \nu \rho }\eta ^{b}-4\eta _{a\mu \nu \rho \lambda }A^{b\lambda
}\right) ,  \label{f17}
\end{eqnarray}%
\begin{equation}
X_{ABa,m_{1}m_{2}}^{b}=-\tfrac{1}{2\cdot 4!}H_{m_{1}}^{\ast \mu
}H_{m_{2}}^{\ast \nu }V_{A}^{\rho }V_{B}^{\lambda }\eta _{a\mu \nu \rho
\lambda }\eta ^{b}.  \label{f18}
\end{equation}%
In the end of this section we list the remaining type-$X$ objects from (\ref%
{so3}), namely $X_{ABCD,m_{1}\ldots m_{p}}$, $X_{ABC,m_{1}\ldots m_{p}}^{ab}$%
, and $X_{a,m_{1}\ldots m_{p}}^{ABC}$, with $p=\overline{0,1}$, as well as $%
X_{ABCD}^{a}$:
\begin{equation}
X_{ABCD}=\tfrac{1}{12}\left( V_{A}^{\ast \mu \nu }V_{B}^{\rho
}V_{C}^{\lambda }C_{D}+\tfrac{1}{3}V_{A}^{\mu }V_{B}^{\nu
}V_{C}^{\rho }C_{D}^{\lambda }\right) \varepsilon _{\mu \nu \rho
\lambda }, \label{f37}
\end{equation}%
\begin{equation}
X_{ABCD,m_{1}}=\tfrac{2}{3\cdot 4!}H_{m_{1}}^{\ast \mu }V_{A}^{\nu
}V_{B}^{\rho }V_{C}^{\lambda }C_{D}\varepsilon _{\mu \nu \rho \lambda },
\label{f38}
\end{equation}%
\begin{equation}
X_{ABC}^{ab}=\tfrac{1}{4!}\left( V_{A}^{\ast \mu \nu }V_{B}^{\rho
}V_{C}^{\lambda }\eta ^{a}\eta ^{b}-\tfrac{2}{3}V_{A}^{\mu
}V_{B}^{\nu }V_{C}^{\rho }A^{a\lambda }\eta ^{b}\right) \varepsilon
_{\mu \nu \rho \lambda },  \label{f42}
\end{equation}%
\begin{equation}
X_{ABC,m_{1}}^{ab}=\tfrac{1}{3\cdot 4!}H_{m_{1}}^{\ast \mu }V_{A}^{\nu
}V_{B}^{\rho }V_{C}^{\lambda }\eta ^{a}\eta ^{b}\varepsilon _{\mu \nu \rho
\lambda },  \label{f43}
\end{equation}%
\begin{equation}
X_{a}^{ABC}=-\tfrac{1}{12}V^{\ast A\mu \nu }V^{B\rho }V^{C\lambda }\eta
_{a\mu \nu \rho \lambda }-\tfrac{1}{6\cdot 4!}V^{A\mu }V^{B\nu }V^{C\rho
}\eta _{a\mu \nu \rho },  \label{f40}
\end{equation}%
\begin{equation}
X_{a,m_{1}}^{ABC}=-\tfrac{2}{3\cdot 4!}H_{m_{1}}^{\ast \mu }V^{A\nu
}V^{B\rho }V^{C\lambda }\eta _{a\mu \nu \rho \lambda },  \label{f41}
\end{equation}%
\begin{equation}
X_{ABCD}^{a}=-\tfrac{1}{3\cdot 4!}V_{A}^{\mu }V_{B}^{\nu
}V_{C}^{\rho }V_{D}^{\lambda }\eta ^{a}\varepsilon _{\mu \nu \rho
\lambda}. \label{f44}
\end{equation}

\section{Gauge generators of the deformed model \label{appendixB}}

\setcounter{equation}{0} \renewcommand{\theequation}{B.\arabic{equation}}

From the terms of antighost number $1$ present in (\ref{defsolmast}) we
determine the deformed gauge generators that produce the deformed gauge
transformations (\ref{gaugeA})--(\ref{gaugeV1}). We added a supplementary
index between parentheses to the gauge generators such as to distinguish
among the fields to which the gauge generators are associated with. We list
below only the nonvanishing generators of the various fields, which read as:
\begin{equation}
(\bar{Z}_{a(\varphi )})_{b}=-\lambda W_{ab},  \label{g1}
\end{equation}%
\begin{eqnarray}
(\bar{Z}_{\mu (H)}^{a})_{b} &=&\tfrac{\lambda }{2}\varepsilon _{\mu \nu \rho
\lambda }\left[ \left( -\tfrac{1}{12}\frac{\partial M_{bcde}}{\partial
\varphi _{a}}A^{c\nu }+\frac{\partial f_{bde}^{A}}{\partial \varphi _{a}}%
V_{A}^{\nu }\right) A^{d\rho }\right.  \notag \\
&&\left. +\frac{\partial g_{be}^{AB}}{\partial \varphi _{a}}V_{A}^{\nu
}V_{B}^{\rho }\right] A^{e\lambda }+\lambda \left[ -\frac{\partial W_{bc}}{%
\partial \varphi _{a}}H_{\mu }^{c}+\frac{\partial f_{bB}^{A}}{\partial
\varphi _{a}}V_{A}^{\nu }V_{\mu \nu }^{B}\right.  \notag \\
&&\left. +\left( \frac{\partial M_{bc}^{d}}{\partial \varphi _{a}}A^{c\nu }+%
\tfrac{1}{12}\frac{\partial f_{\;\;b}^{Ad}}{\partial \varphi _{a}}V_{A}^{\nu
}\right) B_{d\mu \nu }\right] ,  \label{g2a}
\end{eqnarray}%
\begin{equation}
(\bar{Z}_{\mu (H)}^{a})_{b}^{\alpha \beta }=-\delta _{b}^{a}\partial
_{\left. {}\right. }^{\left[ \alpha \right. }\delta _{\mu }^{\left. \beta %
\right] }+\lambda \left( \frac{\partial W_{bc}}{\partial \varphi _{a}}%
A_{\left. {}\right. }^{c[\alpha }\delta _{\mu }^{\beta ]}-\tfrac{1}{12}\frac{%
\partial f_{Ab}}{\partial \varphi _{a}}V_{\left. {}\right. }^{A[\alpha
}\delta _{\mu }^{\beta ]}\right) ,  \label{g2b}
\end{equation}%
\begin{eqnarray}
(\bar{Z}_{\mu (H)}^{a})_{\alpha \beta \gamma }^{b} &=&-\tfrac{\lambda }{2}%
\frac{\partial M_{cd}^{b}}{\partial \varphi _{a}}\sigma _{\mu
\lbrack \alpha }^{\left. {}\right. }A_{\beta }^{c}A_{\gamma
]}^{d}+2\lambda \frac{\partial M^{bc}}{\partial \varphi _{a}}\sigma
_{\mu \rho }B_{c}^{\rho \lambda
}\varepsilon _{\lambda \alpha \beta \gamma }  \notag \\
&&+\tfrac{\lambda }{4!}\left( \frac{\partial f_{Ac}^{b}}{\partial \varphi
_{a}}\sigma _{\mu \lbrack \alpha }V_{\beta }^{A}A_{\gamma ]}^{c}-\frac{%
\partial g_{AB}^{b}}{\partial \varphi _{a}}\sigma _{\mu \lbrack \alpha
}V_{\beta }^{A}V_{\gamma ]}^{B}\right) ,  \label{g2c}
\end{eqnarray}%
\begin{equation}
(\bar{Z}_{\mu (H)}^{a})_{A}^{\sigma }=\lambda \varepsilon _{\mu \nu \rho
\lambda }\sigma ^{\lambda \sigma }\left( \frac{\partial f_{bAB}}{\partial
\varphi _{a}}V^{B\nu }A^{b\rho }+\tfrac{1}{2}\frac{\partial g_{\quad A}^{BC}%
}{\partial \varphi _{a}}V_{B}^{\nu }V_{C}^{\rho }\right) ,  \label{g2d}
\end{equation}%
\begin{equation}
(\bar{Z}_{\mu (A)}^{a})_{b}=\delta _{b}^{a}\partial _{\mu }-\lambda
M_{bc}^{a}A_{\mu }^{c}-\tfrac{\lambda }{12}f_{Ab}^{a}V_{\mu }^{A},
\label{g3a}
\end{equation}%
\begin{equation}
(\bar{Z}_{\mu (A)}^{a})_{\alpha \beta \gamma }^{b}=-2\lambda
M^{ab}\varepsilon _{\mu \alpha \beta \gamma },  \label{g3b}
\end{equation}%
\begin{eqnarray}
(\bar{Z}_{a(B)}^{\mu \nu })_{b} &=&\lambda \varepsilon ^{\mu \nu \rho
\lambda }\left( \tfrac{1}{8}M_{abcd}A_{\rho }^{c}A_{\lambda
}^{d}+f_{Aabc}V_{\rho }^{A}A_{\lambda }^{c}-\tfrac{1}{2}g_{ABab}V_{\rho
}^{A}V_{\lambda }^{B}\right)  \notag \\
&&-\lambda M_{ab}^{c}B_{c}^{\mu \nu },  \label{g4a}
\end{eqnarray}%
\begin{equation}
(\bar{Z}_{a(B)}^{\mu \nu })_{b}^{\alpha \beta }=\lambda W_{ab}\sigma ^{\mu
\lbrack \alpha }\sigma ^{\beta ]\nu },  \label{g4b}
\end{equation}%
\begin{equation}
(\bar{Z}_{a(B)}^{\mu \nu })_{\alpha \beta \gamma }^{b}=-\tfrac{1}{2}\delta
_{a}^{b}\partial _{\left[ \alpha \right. }^{\left. {}\right. }\delta _{\beta
}^{\mu }\delta _{\left. \gamma \right] }^{\nu }-\tfrac{\lambda }{2}\left(
M_{ac}^{b}\delta _{\lbrack \alpha }^{\mu }\delta _{\beta }^{\nu }A_{\gamma
]}^{c}+\tfrac{1}{12}f_{Aa}^{b}\delta _{\lbrack \alpha }^{\mu }\delta _{\beta
}^{\nu }V_{\gamma ]}^{A}\right) ,  \label{g4c}
\end{equation}%
\begin{equation}
(\bar{Z}_{a(B)}^{\mu \nu })_{A}^{\lambda }=-\lambda \varepsilon ^{\mu \nu
\rho \lambda }f_{aAB}V_{\rho }^{B},  \label{g4d}
\end{equation}%
\begin{equation}
(\bar{Z}_{\mu (V)}^{A})_{a}=\lambda f_{aB}^{A}V_{\mu }^{B},  \label{g5}
\end{equation}%
\begin{equation}
(\bar{Z}_{\mu \nu (V)}^{A})_{a}=\lambda f_{aB}^{A}V_{\mu \nu }^{B}+\tfrac{%
\lambda }{12}f_{\;\;a}^{Ab}B_{b\mu \nu }+\lambda \varepsilon _{\mu \nu \rho
\lambda }\left( \tfrac{1}{2}f_{abc}^{A}A^{b\rho }+g_{ac}^{AB}V_{B}^{\rho
}\right) A^{c\lambda },  \label{g6a}
\end{equation}%
\begin{equation}
(\bar{Z}_{\mu \nu (V)}^{A})_{a}^{\alpha \beta }=\tfrac{\lambda }{4!}%
f_{a}^{A}\delta _{\mu }^{[\alpha }\delta _{\nu }^{\beta ]},  \label{g6b}
\end{equation}%
\begin{equation}
(\bar{Z}_{\mu \nu (V)}^{A})_{\alpha \beta \gamma }^{a}=\tfrac{\lambda }{4!}%
\left( f_{\;\;b}^{Aa}A_{\sigma }^{b}-g^{aAB}V_{B\sigma }\right) \sigma _{\mu
\rho }\sigma _{\nu \lambda }\delta _{\lbrack \alpha }^{\rho }\delta _{\beta
}^{\lambda }\delta _{\gamma ]}^{\sigma },  \label{g6c}
\end{equation}%
\begin{equation}
(\bar{Z}_{\mu \nu (V)}^{A})_{B\lambda }=\varepsilon _{\mu \nu \rho \lambda
}\left( \delta _{B}^{A}\partial ^{\rho }-\lambda f_{aB}^{A}A^{a\rho
}+\lambda g_{\quad B}^{AC}V_{C}^{\rho }\right) .  \label{g6d}
\end{equation}

\section{Reducibility of the deformed gauge transformations \label{appendixC}%
}

\setcounter{equation}{0} \renewcommand{\theequation}{C.\arabic{equation}}

From the terms of antighost number $2$ in (\ref{defsolmast}) that are
simultaneously linear in the ghosts for ghosts and in the antifields of the
ghosts we identify the first-order reducibility functions for the coupled
model as%
\begin{equation}
(\bar{Z}_{\alpha \beta }^{(1)a})_{b}^{\mu \nu \rho }=-\tfrac{1}{2}\left(
\delta _{b}^{a}\partial _{\left. {}\right. }^{[\mu }\delta _{\alpha }^{\nu
}\delta _{\beta }^{\rho ]}-\lambda \frac{\partial W_{bc}}{\partial \varphi
_{a}}A_{\left. {}\right. }^{c[\mu }\delta _{\alpha }^{\nu }\delta _{\beta
}^{\rho ]}\right) -\tfrac{\lambda }{2\cdot 4!}\frac{\partial f_{b}^{A}}{%
\partial \varphi _{a}}\delta _{\alpha }^{[\mu }\delta _{\beta }^{\nu }\delta
_{\gamma }^{\rho ]}V_{A}^{\gamma },  \label{r1}
\end{equation}%
\begin{eqnarray}
(\bar{Z}_{\alpha \beta }^{(1)a})_{\mu \nu \rho \lambda }^{b} &=&\tfrac{%
\lambda }{8}\frac{\partial M_{cd}^{b}}{\partial \varphi _{a}}\sigma _{\alpha
^{\prime }[\alpha }\sigma _{\beta ]\beta ^{\prime }}\delta _{\lbrack \mu
}^{\alpha ^{\prime }}\delta _{\nu }^{\beta ^{\prime }}A_{\rho
}^{c}A_{\lambda ]}^{d}+\lambda \varepsilon _{\mu \nu \rho \lambda }\frac{%
\partial M^{bc}}{\partial \varphi _{a}}B_{c\alpha \beta }  \notag \\
&&-\tfrac{\lambda }{4\cdot 4!}\varepsilon _{\mu \nu \rho \lambda
}\varepsilon _{\alpha \beta \gamma \delta }\left( \frac{\partial g^{bAB}}{%
\partial \varphi _{a}}V_{A}^{\gamma }V_{B}^{\delta }-2\frac{\partial
f_{\;\;c}^{Ab}}{\partial \varphi _{a}}V_{A}^{\gamma }A^{c\delta }\right) ,
\label{r2}
\end{eqnarray}%
\begin{equation}
(\bar{Z}_{\alpha \beta }^{(1)a})_{A}=\tfrac{\lambda }{2}
\varepsilon _{\alpha \beta \rho \lambda }
\left( \frac{\partial f_{bA}^{B}}{\partial \varphi _{a}}%
V_{B}^{\rho }A^{b\lambda }-\tfrac{1}{2}\frac{\partial g_{\quad A}^{BC}}{%
\partial \varphi _{a}}V_{B}^{\rho }V_{C}^{\lambda }\right) ,  \label{r2a}
\end{equation}%
\begin{eqnarray}
(\bar{Z}_{a}^{(1)\alpha \beta \gamma })_{\mu \nu \rho \lambda }^{b} &=&-%
\tfrac{1}{6}\left( \delta _{a}^{b}\partial _{\lbrack \mu }^{\left. {}\right.
}\delta _{\nu }^{\alpha }\delta _{\rho }^{\beta }\delta _{\lambda ]}^{\gamma
}+\lambda M_{ac}^{b}A_{[\mu }^{c}\delta _{\nu }^{\alpha }\delta _{\rho
}^{\beta }\delta _{\lambda ]}^{\gamma }\right)  \notag \\
&&+\tfrac{\lambda }{3\cdot 4!}f_{\;\;a}^{Ab}\delta _{\lbrack \mu }^{\alpha
}\delta _{\nu }^{\beta }\delta _{\rho }^{\gamma }\delta _{\lambda ]}^{\delta
}V_{A\delta },  \label{r3}
\end{eqnarray}%
\begin{equation}
(\bar{Z}_{a}^{(1)\alpha \beta \gamma })_{b}^{\mu \nu \rho }=-\tfrac{\lambda
}{3}W_{ab}\left( \sigma ^{\alpha \lbrack \mu }\sigma ^{\nu ]\beta }\sigma
^{\rho \gamma }+\sigma ^{\alpha \lbrack \nu }\sigma ^{\rho ]\beta }\sigma
^{\mu \gamma }+\sigma ^{\alpha \lbrack \rho }\sigma ^{\mu ]\beta }\sigma
^{\nu \gamma }\right) ,  \label{r4}
\end{equation}%
\begin{equation}
(\bar{Z}_{a}^{(1)\alpha \beta \gamma })_{A}=-\tfrac{\lambda }{3}\varepsilon
^{\alpha \beta \gamma \delta }f_{aA}^{B}V_{B\delta },  \label{r4a}
\end{equation}%
\begin{equation}
(\bar{Z}_{\mu }^{(1)A})_{B}=\delta _{B}^{A}\partial _{\mu }-\lambda
f_{aB}^{A}A_{\mu }^{a}+\lambda g_{\quad B}^{AC}V_{C\mu },  \label{r5}
\end{equation}%
\begin{equation}
(\bar{Z}_{\mu }^{(1)A})_{a}^{\alpha \beta \gamma }=\tfrac{\lambda }{4!}%
f_{a}^{A}\sigma _{\mu \nu }\varepsilon ^{\nu \alpha \beta \gamma },
\label{r6}
\end{equation}%
\begin{equation}
(\bar{Z}_{\mu }^{(1)A})_{\alpha \beta \gamma \delta }^{a}=-\tfrac{\lambda }{%
4!}\varepsilon _{\alpha \beta \gamma \delta }\left( f_{\;\;b}^{Aa}A_{\mu
}^{b}-g^{aAB}V_{B\mu }\right) ,  \label{r7}
\end{equation}%
\begin{equation}
(\bar{Z}^{(1)a})_{\alpha \beta \gamma \delta }^{b}=-2\lambda \varepsilon
_{\alpha \beta \gamma \delta }M^{ab}.  \label{r8}
\end{equation}%
The first-order reducibility relations of the coupled theory result from the
components of (\ref{defsolmast}) with the antighost number equal to $2$ that
are simultaneously linear in the ghosts for ghosts and quadratic in the
antifields of the original fields, being expressed in De Witt condensed form
as
\begin{equation}
(\bar{Z}_{\mu (A)}^{a})_{e}(\bar{Z}^{(1)e})_{\alpha \beta \gamma \delta
}^{b}+(\bar{Z}_{\mu (A)}^{a})_{\nu \rho \lambda }^{e}(\bar{Z}_{e}^{(1)\nu
\rho \lambda })_{\alpha \beta \gamma \delta }^{b}=-2\lambda \varepsilon
_{\alpha \beta \gamma \delta }\frac{\partial M^{ab}}{\partial \varphi _{c}}%
\frac{\delta S^{\mathrm{L}}}{\delta H^{c\mu }},  \label{r10}
\end{equation}%
\begin{eqnarray}
&&(\bar{Z}_{a(B)}^{\mu \nu })_{\rho \lambda \sigma }^{e}(\bar{Z}%
_{e}^{(1)\rho \lambda \sigma })_{b}^{\alpha \beta \gamma }+(\bar{Z}%
_{a(B)}^{\mu \nu })_{e}^{\rho \lambda }(\bar{Z}_{\rho \lambda
}^{(1)e})_{b}^{\alpha \beta \gamma }+(\bar{Z}_{a(B)}^{\mu \nu })_{A}^{\sigma
}(\bar{Z}_{\sigma }^{(1)A})_{b}^{\alpha \beta \gamma }  \notag \\
&=&\lambda \frac{\partial W_{ab}}{\partial \varphi _{c}}\frac{\delta S^{%
\mathrm{L}}}{\delta H_{\rho }^{c}}\sigma ^{\mu \mu ^{\prime }}\sigma ^{\nu
\nu ^{\prime }}\delta _{\mu ^{\prime }}^{[\alpha }\delta _{\nu ^{\prime
}}^{\beta }\delta _{\rho }^{\gamma ]},  \label{r12}
\end{eqnarray}%
\begin{eqnarray}
&&(\bar{Z}_{a(B)}^{\mu \nu })_{e}(\bar{Z}^{(1)e})_{\alpha \beta \gamma
\delta }^{b}+(\bar{Z}_{a(B)}^{\mu \nu })_{\rho \lambda \sigma }^{e}(\bar{Z}%
_{e}^{(1)\rho \lambda \sigma })_{\alpha \beta \gamma \delta }^{b}  \notag \\
&&+(\bar{Z}_{a(B)}^{\mu \nu })_{e}^{\rho \lambda }(\bar{Z}_{\rho \lambda
}^{(1)e})_{\alpha \beta \gamma \delta }^{b}+(\bar{Z}_{a(B)}^{\mu \nu
})_{A}^{\sigma }(\bar{Z}_{\sigma }^{(1)A})_{\alpha \beta \gamma \delta }^{b}
\notag \\
&=&-\tfrac{\lambda }{2}\delta _{\lbrack \alpha }^{\mu }\delta _{\beta }^{\nu
}\delta _{\gamma }^{\rho }\delta _{\delta ]}^{\lambda }\left( \frac{\partial
M_{ac}^{b}}{\partial \varphi _{d}}\frac{\delta S^{\mathrm{L}}}{\delta
H^{d\rho }}A_{\lambda }^{c}+M_{ac}^{b}\frac{\delta S^{\mathrm{L}}}{\delta
B_{c}^{\rho \lambda }}\right)  \notag \\
&&-\tfrac{\lambda }{4!}\delta _{\lbrack \alpha }^{\mu }\delta _{\beta }^{\nu
}\delta _{\gamma }^{\rho }\delta _{\delta ]}^{\lambda }\left( f_{\;\;a}^{Ab}%
\frac{\delta S^{\mathrm{L}}}{\delta V^{A\rho \lambda }}+\frac{\partial
f_{\;\;a}^{Ab}}{\partial \varphi _{c}}\frac{\delta S^{\mathrm{L}}}{\delta
H^{c\rho }}V_{A\lambda }\right) ,  \label{r13}
\end{eqnarray}%
\begin{eqnarray}
&&(\bar{Z}_{a(B)}^{\mu \nu })_{\rho \lambda \sigma }^{e}(\bar{Z}%
_{e}^{(1)\rho \lambda \sigma })_{A}+(\bar{Z}_{a(B)}^{\mu \nu })_{e}^{\rho
\lambda }(\bar{Z}_{\rho \lambda }^{(1)e})_{A}+(\bar{Z}_{a(B)}^{\mu \nu
})_{B}^{\sigma }(\bar{Z}_{\sigma }^{(1)B})_{A}  \notag \\
&=&\lambda \varepsilon ^{\mu \nu \rho \lambda }\left( f_{aA}^{B}\frac{\delta
S^{\mathrm{L}}}{\delta V^{B\rho \lambda }}+\frac{\partial f_{aA}^{B}}{%
\partial \varphi _{c}}\frac{\delta S^{\mathrm{L}}}{\delta H^{c\rho }}%
V_{B\lambda }\right) ,  \label{r13a}
\end{eqnarray}%
\begin{eqnarray}
&&(\bar{Z}_{\mu \nu (V)}^{A})_{C}^{\sigma }(\bar{Z}_{\sigma }^{(1)C})_{B}+(%
\bar{Z}_{\mu \nu (V)}^{A})_{e}^{\rho \lambda }(\bar{Z}_{\rho \lambda
}^{(1)e})_{B}+(\bar{Z}_{\mu \nu (V)}^{A})_{\rho \lambda \sigma }^{e}(\bar{Z}%
_{e}^{(1)\rho \lambda \sigma })_{B}  \notag \\
&=&-\lambda \varepsilon _{\mu \nu \rho \lambda }\left( f_{aB}^{A}\frac{%
\delta S^{\mathrm{L}}}{\delta B_{a\rho \lambda }}+\frac{\partial f_{aB}^{A}}{%
\partial \varphi _{c}}\frac{\delta S^{\mathrm{L}}}{\delta H_{\rho }^{c}}%
A^{a\lambda }\right)  \notag \\
&&+\lambda \varepsilon _{\mu \nu \rho \lambda }\left( g_{\quad B}^{AC}\frac{%
\delta S^{\mathrm{L}}}{\delta V_{\rho \lambda }^{C}}+\frac{\partial g_{\quad
B}^{AC}}{\partial \varphi _{c}}\frac{\delta S^{\mathrm{L}}}{\delta H_{\rho
}^{c}}V_{C}^{\lambda }\right) ,  \label{r14}
\end{eqnarray}%
\begin{eqnarray}
&&(\bar{Z}_{\mu \nu (V)}^{A})_{B}^{\sigma }(\bar{Z}_{\sigma
}^{(1)B})_{a}^{\alpha \beta \gamma }+(\bar{Z}_{\mu \nu (V)}^{A})_{e}^{\rho
\lambda }(\bar{Z}_{\rho \lambda }^{(1)e})_{a}^{\alpha \beta \gamma }+(\bar{Z}%
_{\mu \nu (V)}^{A})_{\rho \lambda \sigma }^{e}(\bar{Z}_{e}^{(1)\rho \lambda
\sigma })_{a}^{\alpha \beta \gamma }  \notag \\
&=&\tfrac{\lambda }{4!}\delta _{\mu }^{[\alpha }\delta _{\nu }^{\beta
}\delta _{\rho }^{\gamma ]}\frac{\partial f_{a}^{A}}{\partial \varphi _{b}}%
\frac{\delta S^{\mathrm{L}}}{\delta H_{\rho }^{b}},  \label{r15}
\end{eqnarray}%
\begin{eqnarray}
&&(\bar{Z}_{\mu \nu (V)}^{A})_{B}^{\sigma }(\bar{Z}_{\sigma
}^{(1)B})_{\alpha \beta \gamma \delta }^{a}+(\bar{Z}_{\mu \nu
(V)}^{A})_{e}^{\rho \lambda }(\bar{Z}_{\rho \lambda }^{(1)e})_{\alpha \beta
\gamma \delta }^{a}  \notag \\
&&+(\bar{Z}_{\mu \nu (V)}^{A})_{e}(\bar{Z}^{(1)e})_{\alpha \beta \gamma
\delta }^{a}+(\bar{Z}_{\mu \nu (V)}^{A})_{\rho \lambda \sigma }^{e}(\bar{Z}%
_{e}^{(1)\rho \lambda \sigma })_{\alpha \beta \gamma \delta }^{a}  \notag \\
&=&\tfrac{\lambda }{4!}\sigma _{\mu \mu ^{\prime }}\sigma _{\nu \nu ^{\prime
}}\delta _{\lbrack \alpha }^{\mu ^{\prime }}\delta _{\beta }^{\nu ^{\prime
}}\delta _{\gamma }^{\rho }\delta _{\delta ]}^{\lambda }\left( f_{\;\;b}^{Aa}%
\frac{\delta S^{\mathrm{L}}}{\delta B_{b}^{\rho \lambda }}+\frac{\partial
f_{\;\;b}^{Aa}}{\partial \varphi _{c}}\frac{\delta S^{\mathrm{L}}}{\delta
H^{c\rho }}A_{\lambda }^{b}\right)  \notag \\
&&-\tfrac{\lambda }{4!}\sigma _{\mu \mu ^{\prime }}\sigma _{\nu \nu ^{\prime
}}\delta _{\lbrack \alpha }^{\mu ^{\prime }}\delta _{\beta }^{\nu ^{\prime
}}\delta _{\gamma }^{\rho }\delta _{\delta ]}^{\lambda }\left( g^{aAB}\frac{%
\delta S^{\mathrm{L}}}{\delta V^{B\rho \lambda }}+\frac{\partial g^{aAB}}{%
\partial \varphi _{b}}\frac{\delta S^{\mathrm{L}}}{\delta H^{b\rho }}%
V_{B\lambda }\right) ,  \label{r15a}
\end{eqnarray}%
\begin{eqnarray}
&&(\bar{Z}_{\mu (H)}^{a})_{\rho \lambda \sigma }^{e}(\bar{Z}_{e}^{(1)\rho
\lambda \sigma })_{b}^{\alpha \beta \gamma }+(\bar{Z}_{\mu
(H)}^{a})_{e}^{\rho \lambda }(\bar{Z}_{\rho \lambda }^{(1)e})_{b}^{\alpha
\beta \gamma }+(\bar{Z}_{\mu (H)}^{a})_{B}^{\sigma }(\bar{Z}_{\sigma
}^{(1)B})_{b}^{\alpha \beta \gamma }  \notag \\
&=&\lambda \delta _{\mu }^{[\alpha }\delta _{\nu }^{\beta }\delta
_{\rho }^{\gamma ]}\left( \frac{\partial W_{bc}}{\partial \varphi
_{a}}\frac{\delta S^{\mathrm{L}}}{\delta B_{c\nu \rho
}}+\frac{\partial ^{2}W_{bc}}{\partial \varphi _{a}\partial \varphi
_{e}}\frac{\delta S^{\mathrm{L}}}{\delta H_{\nu
}^{e}}A^{c\rho }\right)  \notag \\
&&-\tfrac{\lambda }{4!}\delta _{\mu }^{[\alpha }\delta _{\nu }^{\beta
}\delta _{\rho }^{\gamma ]}\left( \frac{\partial f_{b}^{A}}{\partial \varphi
_{a}}\frac{\delta S^{\mathrm{L}}}{\delta V_{\nu \rho }^{A}}+\frac{\partial
^{2}f_{b}^{A}}{\partial \varphi _{a}\partial \varphi _{c}}\frac{\delta S^{%
\mathrm{L}}}{\delta H_{\nu }^{c}}V_{A}^{\rho }\right) ,  \label{r17}
\end{eqnarray}%
\begin{eqnarray}
&&(\bar{Z}_{\mu (H)}^{a})_{e}(\bar{Z}^{(1)e})_{\alpha \beta \gamma \delta
}^{b}+(\bar{Z}_{\mu (H)}^{a})_{\rho \lambda \sigma }^{e}(\bar{Z}%
_{e}^{(1)\rho \lambda \sigma })_{\alpha \beta \gamma \delta }^{b}  \notag \\
&&+(\bar{Z}_{\mu (H)}^{a})_{e}^{\rho \lambda }(\bar{Z}_{\rho \lambda
}^{(1)e})_{\alpha \beta \gamma \delta }^{b}+(\bar{Z}_{\mu
(H)}^{a})_{B}^{\sigma }(\bar{Z}_{\sigma }^{(1)B})_{\alpha \beta \gamma
\delta }^{b}  \notag \\
&=&\tfrac{\lambda }{2}\sigma _{\mu \lbrack \alpha }^{\left.
{}\right. }\delta _{\beta }^{\nu }\delta _{\gamma }^{\rho }\delta
_{\delta ]}^{\lambda
}\left( \frac{\partial M_{cd}^{b}}{\partial \varphi _{a}}\frac{\delta S^{%
\mathrm{L}}}{\delta B_{c}^{\nu \rho }}A_{\lambda
}^{d}+\tfrac{1}{2}\frac{\partial
M_{cd}^{b}}{\partial \varphi _{a}\partial \varphi _{e}}\frac{\delta S^{%
\mathrm{L}}}{\delta H^{e\nu }}A_{\rho }^{c}A_{\lambda }^{d}\right)  \notag \\
&&+2\lambda \varepsilon _{\alpha \beta \gamma \delta }\left( \frac{\partial
M^{bc}}{\partial \varphi _{a}}\frac{\delta S^{\mathrm{L}}}{\delta A^{c\mu }}+%
\frac{\partial ^{2}M^{bc}}{\partial \varphi _{a}\partial \varphi _{d}}\frac{%
\delta S^{\mathrm{L}}}{\delta H_{\nu }^{d}}B_{c\mu \nu }\right)  \notag \\
&&-\tfrac{\lambda }{4!}\sigma _{\mu \lbrack \alpha }^{\left. {}\right.
}\delta _{\beta }^{\nu }\delta _{\gamma }^{\rho }\delta _{\delta ]}^{\lambda
}\left[ \frac{\partial ^{2}f_{\;\;c}^{Ab}}{\partial \varphi _{a}\partial
\varphi _{d}}\frac{\delta S^{\mathrm{L}}}{\delta H^{d\nu }}V_{A\rho
}A_{\lambda }^{c}+\frac{\partial f_{\;\;c}^{Ab}}{\partial \varphi _{a}}%
\left( \frac{\delta S^{\mathrm{L}}}{\delta V^{A\nu \rho }}A_{\lambda
}^{c}\right. \right.  \notag \\
&&\left. \left. -\frac{\delta S^{\mathrm{L}}}{\delta B_{c}^{\nu \rho }}%
V_{A\lambda }\right) -\left( \frac{\partial ^{2}g^{bAB}}{\partial \varphi
_{a}\partial \varphi _{c}}\frac{\delta S^{\mathrm{L}}}{\delta H^{c\nu }}%
V_{A\rho }+\frac{\partial g^{bAB}}{\partial \varphi _{a}}\frac{\delta S^{%
\mathrm{L}}}{\delta V^{A\nu \rho }}\right) V_{B\lambda }\right] ,
\label{r18}
\end{eqnarray}%
\begin{eqnarray}
&&(\bar{Z}_{\mu (H)}^{a})_{C}^{\sigma }(\bar{Z}_{\sigma }^{(1)C})_{A}+(\bar{Z%
}_{\mu (H)}^{a})_{e}^{\rho \lambda }(\bar{Z}_{\rho \lambda }^{(1)e})_{A}+(%
\bar{Z}_{\mu (H)}^{a})_{\rho \lambda \sigma }^{e}(\bar{Z}_{e}^{(1)\rho
\lambda \sigma })_{A}  \notag \\
&=&\lambda \varepsilon _{\mu \nu \rho \lambda }\left[ \frac{\delta S^{%
\mathrm{L}}}{\delta V_{\nu \rho }^{B}}\left( \frac{\partial f_{bA}^{B}}{%
\partial \varphi _{a}}A^{b\lambda }-\frac{\partial g_{\quad A}^{BC}}{%
\partial \varphi _{a}}V_{C}^{\lambda }\right) -\frac{\partial f_{bA}^{B}}{%
\partial \varphi _{a}}\frac{\delta S^{\mathrm{L}}}{\delta B_{b\nu \rho }}%
V_{B}^{\lambda }\right.  \notag \\
&&\left. +\frac{\delta S^{\mathrm{L}}}{\delta H_{\nu }^{c}}\left( \frac{%
\partial ^{2}f_{bA}^{B}}{\partial \varphi _{a}\partial \varphi _{c}}%
V_{B}^{\rho }A^{b\lambda }-\tfrac{1}{2}\frac{\partial ^{2}g_{\quad A}^{BC}}{%
\partial \varphi _{a}\partial \varphi _{c}}V_{B}^{\rho }V_{C}^{\lambda
}\right) \right] .  \label{r18a}
\end{eqnarray}%
The deformed gauge generators are given in (\ref{g1})--(\ref{g6d}) and $S^{%
\mathrm{L}}$ represents the deformed Lagrangian action (\ref{ldef}).

The pieces of antighost number $3$ from (\ref{defsolmast}) that are
simultaneously linear in the ghosts for ghosts for ghosts and in the
antifields of the ghosts for ghosts offer us the second-order reducibility
functions for the interacting model of the form
\begin{equation}
(\bar{Z}^{(2)A})_{a}^{\mu \nu \rho \lambda }=\tfrac{\lambda }{4!}%
f_{a}^{A}\varepsilon ^{\mu \nu \rho \lambda },  \label{r19}
\end{equation}%
\begin{eqnarray}
(\bar{Z}_{\alpha \beta \gamma }^{(2)a})_{b}^{\mu \nu \rho \lambda } &=&-%
\tfrac{1}{6}\left( \delta _{b}^{a}\partial _{\left. {}\right. }^{[\mu
}\delta _{\alpha }^{\nu }\delta _{\beta }^{\rho }\delta _{\gamma }^{\lambda
]}+\lambda \frac{\partial W_{cb}}{\partial \varphi _{a}}A_{\left. {}\right.
}^{c[\mu }\delta _{\alpha }^{\nu }\delta _{\beta }^{\rho }\delta _{\gamma
}^{\lambda ]}\right)  \notag \\
&&+\tfrac{\lambda }{3!\cdot 4!}\delta _{\alpha }^{[\mu }\delta _{\beta
}^{\nu }\delta _{\gamma }^{\rho }\delta _{\delta }^{\lambda ]}\frac{\partial
f_{b}^{A}}{\partial \varphi _{a}}V_{A}^{\delta },  \label{r20}
\end{eqnarray}%
\begin{equation}
(\bar{Z}_{a}^{(2)\mu _{1}\mu _{2}\mu _{3}\mu _{4}})_{b}^{\mu \nu \rho
\lambda }=\tfrac{\lambda }{12}W_{ab}\sum\limits_{\pi \in S_{4}}\left(
-\right) ^{\pi }\sigma ^{\mu _{\pi (1)}\mu }\sigma ^{\mu _{\pi (2)}\nu
}\sigma ^{\mu _{\pi (3)}\rho }\sigma ^{\mu _{\pi (4)}\lambda },  \label{r21}
\end{equation}%
where $S_{4}$ denotes the set of permutations of $\left\{ 1,2,3,4\right\} $
and $\left( -\right) ^{\pi }$ is the signature of a given permutation $\pi $%
. By means of the terms with the antighost number equal to $3$ present in (%
\ref{defsolmast}) that are linear in the ghosts for ghosts for ghosts and
also quadratic in antifields we infer the second-order reducibility
relations for the interacting model in condensed De Witt form, which read as
\begin{eqnarray}
&&(\bar{Z}_{\mu }^{(1)A})_{B}(\bar{Z}^{(2)B})_{a}^{\alpha \beta \gamma
\delta }+(\bar{Z}_{\mu }^{(1)A})_{b}^{\nu \rho \lambda }(\bar{Z}_{\nu \rho
\lambda }^{(2)b})_{a}^{\alpha \beta \gamma \delta }  \notag \\
&&+(\bar{Z}_{\mu }^{(1)A})_{\nu \rho \lambda \sigma }^{b}(\bar{Z}%
_{b}^{(2)\nu \rho \lambda \sigma })_{a}^{\alpha \beta \gamma \delta }  \notag
\\
&=&\tfrac{\lambda }{4!}\varepsilon ^{\alpha \beta \gamma \delta }\frac{%
\partial f_{a}^{A}}{\partial \varphi _{b}}\frac{\delta S^{\mathrm{L}}}{%
\delta H^{b\mu }},  \label{r22}
\end{eqnarray}%
\begin{eqnarray}
&&(\bar{Z}_{a}^{(1)\alpha \beta \gamma })_{A}(\bar{Z}^{(2)A})_{b}^{\mu \nu
\rho \lambda }+(\bar{Z}_{a}^{(1)\alpha \beta \gamma })_{e}^{\delta \sigma
\varepsilon }(\bar{Z}_{\delta \sigma \varepsilon }^{(2)e})_{b}^{\mu \nu \rho
\lambda }  \notag \\
&&+(\bar{Z}_{a}^{(1)\alpha \beta \gamma })_{\delta \sigma \varepsilon \eta
}^{e}(\bar{Z}_{e}^{(2)\delta \sigma \varepsilon \eta })_{b}^{\mu \nu \rho
\lambda }  \notag \\
&=&\tfrac{\lambda }{3}\delta _{\alpha ^{\prime }}^{[\mu }\delta _{\beta
^{\prime }}^{\nu }\delta _{\gamma ^{\prime }}^{\rho }\delta _{\delta
^{\prime }}^{\lambda ]}\sigma ^{\alpha \alpha ^{\prime }}\sigma ^{\beta
\beta ^{\prime }}\sigma ^{\gamma \gamma ^{\prime }}\frac{\partial W_{ab}}{%
\partial \varphi _{c}}\frac{\delta S^{\mathrm{L}}}{\delta H_{\delta ^{\prime
}}^{c}},  \label{r23}
\end{eqnarray}%
\begin{eqnarray}
&&(\bar{Z}_{\mu \nu }^{(1)a})_{A}(\bar{Z}^{(2)A})_{b}^{\alpha \beta \gamma
\delta }+(\bar{Z}_{\mu \nu }^{(1)a})_{e}^{\delta \sigma \varepsilon }(\bar{Z}%
_{\delta \sigma \varepsilon }^{(2)e})_{b}^{\alpha \beta \gamma \delta }
\notag \\
&&+(\bar{Z}_{\mu \nu }^{(1)a})_{\delta \sigma \varepsilon \eta }^{e}(\bar{Z}%
_{e}^{(2)\delta \sigma \varepsilon \eta })_{b}^{\alpha \beta \gamma \delta }
\notag \\
&=&\tfrac{\lambda }{2}\delta _{\mu }^{[\alpha }\delta _{\nu }^{\beta
}\delta _{\rho }^{\gamma }\delta _{\lambda }^{\delta ]}\left[
\frac{\delta
S^{\mathrm{L}}}{\delta H_{\rho }^{d}}\left( \frac{\partial ^{2}W_{bc}}{%
\partial \varphi _{a}\partial \varphi _{d}}A^{c\lambda }-\tfrac{1}{4!}\frac{%
\partial ^{2}f_{b}^{A}}{\partial \varphi _{a}\partial \varphi _{d}}%
V_{A}^{\lambda }\right) \right.  \notag \\
&&\left. +\frac{\partial W_{bc}}{\partial \varphi _{a}}\frac{\delta S^{%
\mathrm{L}}}{\delta B_{c\rho \lambda }}-\tfrac{1}{4!}\frac{\partial f_{b}^{A}%
}{\partial \varphi _{a}}\frac{\delta S^{\mathrm{L}}}{\delta V_{\rho \lambda
}^{A}}\right] .  \label{r25}
\end{eqnarray}

\section{Gauge algebra of the deformed model\label{appendixD}}

\setcounter{equation}{0} \renewcommand{\theequation}{D.\arabic{equation}}

The nonvanishing commutators among the deformed gauge transformations (\ref%
{gaugeA})--(\ref{gaugeV1}) result from the terms quadratic in the ghosts
with pure ghost number $1$ present in (\ref{defsolmast}). By analyzing these
terms and taking into account the expressions (\ref{g1})--(\ref{g6d}), we
deduce the following nonvanishing relations:
\begin{equation}
(\bar{Z}_{e(\varphi )})_{b}\frac{\delta (\bar{Z}_{a(\varphi )})_{c}}{\delta
\varphi _{e}}-(\bar{Z}_{e(\varphi )})_{c}\frac{\delta (\bar{Z}_{a(\varphi
)})_{b}}{\delta \varphi _{e}}=\lambda M_{bc}^{e}(\bar{Z}_{a(\varphi )})_{e},
\label{co1}
\end{equation}%
\begin{eqnarray}
&&(\bar{Z}_{e(\varphi )})_{b}\frac{\delta (\bar{Z}_{\mu (A)}^{a})_{c}}{%
\delta \varphi _{e}}+(\bar{Z}_{\sigma (A)}^{m})_{b}\frac{\delta (\bar{Z}%
_{\mu (A)}^{a})_{c}}{\delta A_{\sigma }^{m}}+(\bar{Z}_{\sigma (V)}^{A})_{b}%
\frac{\delta (\bar{Z}_{\mu (A)}^{a})_{c}}{\delta V_{\sigma }^{A}}  \notag \\
&&-(\bar{Z}_{e(\varphi )})_{c}\frac{\delta (\bar{Z}_{\mu (A)}^{a})_{b}}{%
\delta \varphi _{e}}-(\bar{Z}_{\sigma (A)}^{m})_{c}\frac{\delta (\bar{Z}%
_{\mu (A)}^{a})_{b}}{\delta A_{\sigma }^{m}}-(\bar{Z}_{\sigma (V)}^{A})_{c}%
\frac{\delta (\bar{Z}_{\mu (A)}^{a})_{b}}{\delta V_{\sigma }^{A}}  \notag \\
&=&\lambda \left[ M_{bc}^{d}(\bar{Z}_{\mu (A)}^{a})_{d}+\tfrac{1}{12}%
M_{dbce}\varepsilon ^{\alpha \beta \gamma \delta }A_{\delta }^{e}(\bar{Z}%
_{\mu (A)}^{a})_{\alpha \beta \gamma }^{d}\right.  \notag \\
&&\left. -\tfrac{1}{3}f_{Abcd}\varepsilon ^{\alpha \beta \gamma \delta
}V_{\delta }^{A}(\bar{Z}_{\mu (A)}^{a})_{\alpha \beta \gamma }^{d}-\frac{%
\delta S^{\mathrm{L}}}{\delta H^{d\mu }}\frac{\partial M_{bc}^{a}}{\partial
\varphi _{d}}\right] ,  \label{co2}
\end{eqnarray}%
\begin{equation}
(\bar{Z}_{e(\varphi )})_{b}\frac{\delta (\bar{Z}_{\mu (A)}^{a})_{\alpha
\beta \gamma }^{c}}{\delta \varphi _{e}}-(\bar{Z}_{\sigma (A)}^{m})_{\alpha
\beta \gamma }^{c}\frac{\delta (\bar{Z}_{\mu (A)}^{a})_{b}}{\delta A_{\sigma
}^{m}}=-\lambda M_{bd}^{c}(\bar{Z}_{\mu (A)}^{a})_{\alpha \beta \gamma }^{d},
\label{co3}
\end{equation}%
\begin{eqnarray}
&&(\bar{Z}_{e(\varphi )})_{b}\frac{\delta (\bar{Z}_{a(B)}^{\mu \nu })_{c}}{%
\delta \varphi _{e}}+(\bar{Z}_{\sigma (A)}^{m})_{b}\frac{\delta (\bar{Z}%
_{a(B)}^{\mu \nu })_{c}}{\delta A_{\sigma }^{m}}+(\bar{Z}_{m(B)}^{\sigma
\varepsilon })_{b}\frac{\delta (\bar{Z}_{a(B)}^{\mu \nu })_{c}}{\delta
B_{m}^{\sigma \varepsilon }}  \notag \\
&&+(\bar{Z}_{\sigma (V)}^{A})_{b}\frac{\delta (\bar{Z}_{a(B)}^{\mu \nu })_{c}%
}{\delta V_{\sigma }^{A}}-(\bar{Z}_{e(\varphi )})_{c}\frac{\delta (\bar{Z}%
_{a(B)}^{\mu \nu })_{b}}{\delta \varphi _{e}}-(\bar{Z}_{\sigma (A)}^{m})_{c}%
\frac{\delta (\bar{Z}_{a(B)}^{\mu \nu })_{b}}{\delta A_{\sigma }^{m}}  \notag
\\
&&-(\bar{Z}_{m(B)}^{\sigma \varepsilon })_{c}\frac{\delta (\bar{Z}%
_{a(B)}^{\mu \nu })_{b}}{\delta B_{m}^{\sigma \varepsilon }}-(\bar{Z}%
_{\sigma (V)}^{A})_{c}\frac{\delta (\bar{Z}_{a(B)}^{\mu \nu })_{b}}{\delta
V_{\sigma }^{A}}  \notag \\
&=&\lambda \left\{ M_{bc}^{d}(\bar{Z}_{a(B)}^{\mu \nu })_{d}-\tfrac{1}{3}%
f_{Abcd}\varepsilon ^{\alpha \beta \gamma \delta }V_{\delta }^{A}(\bar{Z}%
_{a(B)}^{\mu \nu })_{\alpha \beta \gamma }^{d}\right.  \notag \\
&&+\tfrac{1}{12}M_{dbce}\varepsilon ^{\alpha \beta \gamma \delta }A_{\delta
}^{e}(\bar{Z}_{a(B)}^{\mu \nu })_{\alpha \beta \gamma }^{d}  \notag \\
&&-\tfrac{1}{2}\left[ \frac{\partial M_{bc}^{d}}{\partial \varphi _{e}}%
B_{d\alpha \beta }-\varepsilon _{\alpha \beta \gamma \delta }\left( \tfrac{1%
}{8}\frac{\partial M_{bcdf}}{\partial \varphi _{e}}A^{d\gamma }+\frac{%
\partial f_{bcf}^{A}}{\partial \varphi _{e}}V_{A}^{\gamma }\right)
A^{f\delta }\right.  \notag \\
&&\left. +\tfrac{1}{2}\varepsilon _{\alpha \beta \gamma \delta }\frac{%
\partial g_{bc}^{AB}}{\partial \varphi _{e}}V_{A}^{\gamma }V_{B}^{\delta }%
\right] (\bar{Z}_{a(B)}^{\mu \nu })_{e}^{\alpha \beta }+\left(
g_{bc}^{AB}V_{B\lambda }-f_{bcd}^{A}A_{\lambda }^{d}\right) (\bar{Z}%
_{a(B)}^{\mu \nu })_{A}^{\lambda }  \notag \\
&&-\lambda \varepsilon ^{\mu \nu \rho \lambda }\left[ \frac{\delta S^{%
\mathrm{L}}}{\delta H^{m\rho }}\left( \frac{\partial f_{abc}^{A}}{\partial
\varphi _{m}}V_{A\lambda }-\tfrac{1}{4}\frac{\partial M_{abcd}}{\partial
\varphi _{m}}A_{\lambda }^{d}\right) +f_{abc}^{A}\frac{\delta S^{\mathrm{L}}%
}{\delta V^{A\rho \lambda }}\right.  \notag \\
&&\left. \left. -\tfrac{1}{4}M_{abcd}\frac{\delta S^{\mathrm{L}}}{\delta
B_{d}^{\rho \lambda }}\right] \right\} ,  \label{co5}
\end{eqnarray}%
\begin{equation}
(\bar{Z}_{e(\varphi )})_{b}\frac{\delta (\bar{Z}_{a(B)}^{\mu \nu
})_{c}^{\alpha \beta }}{\delta \varphi _{e}}-(\bar{Z}_{m(B)}^{\sigma
\varepsilon })_{c}^{\alpha \beta }\frac{\delta (\bar{Z}_{a(B)}^{\mu \nu
})_{b}}{\delta B_{m}^{\sigma \varepsilon }}=\lambda \frac{\partial W_{bc}}{%
\partial \varphi _{d}}(\bar{Z}_{a(B)}^{\mu \nu })_{d}^{\alpha \beta },
\label{co6}
\end{equation}%
\begin{eqnarray}
&&(\bar{Z}_{e(\varphi )})_{b}\frac{\delta (\bar{Z}_{a(B)}^{\mu \nu
})_{\alpha \beta \gamma }^{c}}{\delta \varphi _{e}}+(\bar{Z}_{\sigma
(A)}^{m})_{b}\frac{\delta (\bar{Z}_{a(B)}^{\mu \nu })_{\alpha \beta \gamma
}^{c}}{\delta A_{\sigma }^{m}}+(\bar{Z}_{\sigma (V)}^{A})_{b}\frac{\delta (%
\bar{Z}_{a(B)}^{\mu \nu })_{\alpha \beta \gamma }^{c}}{\delta V_{\sigma }^{A}%
}  \notag \\
&&-(\bar{Z}_{\sigma (A)}^{m})_{\alpha \beta \gamma }^{c}\frac{\delta (\bar{Z}%
_{a(B)}^{\mu \nu })_{b}}{\delta A_{\sigma }^{m}}-(\bar{Z}_{m(B)}^{\sigma
\varepsilon })_{\alpha \beta \gamma }^{c}\frac{\delta (\bar{Z}_{a(B)}^{\mu
\nu })_{b}}{\delta B_{m}^{\sigma \varepsilon }}  \notag \\
&=&\lambda \left[ -\tfrac{1}{4}\left( \frac{\partial M_{bd}^{c}}{\partial
\varphi _{e}}A_{[\alpha }^{d}\delta _{\beta }^{\rho }\delta _{\gamma
]}^{\lambda }+\tfrac{1}{12}\frac{\partial f_{Ab}^{c}}{\partial \varphi _{e}}%
V_{[\alpha }^{A}\delta _{\beta }^{\rho }\delta _{\gamma ]}^{\lambda }\right)
(\bar{Z}_{a(B)}^{\mu \nu })_{e\rho \lambda }\right.  \notag \\
&&+M_{eb}^{c}(\bar{Z}_{a(B)}^{\mu \nu })_{\alpha \beta \gamma }^{e}+\tfrac{1%
}{2}\frac{\partial M_{ab}^{c}}{\partial \varphi _{m}}\frac{\delta S^{\mathrm{%
L}}}{\delta H^{m\rho }}\delta _{\lbrack \alpha }^{\mu }\delta _{\beta }^{\nu
}\delta _{\gamma ]}^{\rho }  \notag \\
&&\left. +\tfrac{1}{4!}\varepsilon _{\lambda \alpha \beta \gamma
}f_{\;\;b}^{Mc}(\bar{Z}_{a(B)}^{\mu \nu })_{M}^{\lambda }\right] ,
\label{c7}
\end{eqnarray}%
\begin{eqnarray}
&&(\bar{Z}_{\sigma (A)}^{m})_{\alpha \beta \gamma }^{b}\frac{\delta (\bar{Z}%
_{a(B)}^{\mu \nu })_{\alpha ^{\prime }\beta ^{\prime }\gamma ^{\prime }}^{c}%
}{\delta A_{\sigma }^{m}}-(\bar{Z}_{\sigma (A)}^{m})_{\alpha ^{\prime }\beta
^{\prime }\gamma ^{\prime }}^{c}\frac{\delta (\bar{Z}_{a(B)}^{\mu \nu
})_{\alpha \beta \gamma }^{b}}{\delta A_{\sigma }^{m}}  \notag \\
&=&-\tfrac{\lambda }{2}\frac{\partial M^{bc}}{\partial \varphi _{e}}%
\varepsilon ^{\rho \lambda \delta \varepsilon }\varepsilon _{\delta \alpha
\beta \gamma }\varepsilon _{\varepsilon \alpha ^{\prime }\beta ^{\prime
}\gamma ^{\prime }}(\bar{Z}_{a(B)}^{\mu \nu })_{e\rho \lambda },  \label{co8}
\end{eqnarray}%
\begin{eqnarray}
&&(\bar{Z}_{e(\varphi )})_{b}\frac{\delta (\bar{Z}_{a(B)}^{\mu \nu
})_{A}^{\lambda }}{\delta \varphi _{e}}+(\bar{Z}_{\sigma (V)}^{B})_{b}%
\frac{\delta (\bar{Z}_{a(B)}^{\mu \nu })_{A}^{\lambda }}{\delta V_{\sigma
}^{B}}-(\bar{Z}_{m(B)}^{\sigma \varepsilon })_{A}^{\lambda }\frac{\delta (%
\bar{Z}_{a(B)}^{\mu \nu })_{b}}{\delta B_{m}^{\sigma \varepsilon }}  \notag
\\
&=&-\lambda f_{bA}^{B}(\bar{Z}_{a(B)}^{\mu \nu })_{B}^{\lambda }+\tfrac{%
\lambda }{2}\varepsilon ^{\alpha \beta \rho \lambda }\frac{\partial f_{bMA}}{%
\partial \varphi _{e}}V_{\rho }^{M}(\bar{Z}_{a(B)}^{\mu \nu })_{e\alpha
\beta },  \label{co9}
\end{eqnarray}%
\begin{eqnarray}
&&(\bar{Z}_{e(\varphi )})_{b}\frac{\delta (\bar{Z}_{\mu (V)}^{A})_{c}}{%
\delta \varphi _{e}}+(\bar{Z}_{\sigma (V)}^{B})_{b}\frac{\delta (\bar{Z}%
_{\mu (V)}^{A})_{c}}{\delta V_{\sigma }^{B}}-(\bar{Z}_{e(\varphi )})_{c}%
\frac{\delta (\bar{Z}_{\mu (V)}^{A})_{b}}{\delta \varphi _{e}}  \notag \\
&&-(\bar{Z}_{\sigma (V)}^{B})_{c}\frac{\delta (\bar{Z}_{\mu (V)}^{A})_{b}}{%
\delta V_{\sigma }^{B}}=\lambda M_{bc}^{d}(\bar{Z}_{\mu (V)}^{A})_{d},
\label{co10}
\end{eqnarray}%
\begin{eqnarray}
&&(\bar{Z}_{e(\varphi )})_{b}\frac{\delta (\bar{Z}_{\mu \nu (V)}^{A})_{c}}{%
\delta \varphi _{e}}+(\bar{Z}_{\sigma (A)}^{m})_{b}\frac{\delta (\bar{Z}%
_{\mu \nu (V)}^{A})_{c}}{\delta A_{\sigma }^{m}}+(\bar{Z}_{m(B)}^{\sigma
\varepsilon })_{b}\frac{\delta (\bar{Z}_{\mu \nu (V)}^{A})_{c}}{\delta
B_{m}^{\sigma \varepsilon }}  \notag \\
&&+(\bar{Z}_{\sigma (V)}^{B})_{b}\frac{\delta (\bar{Z}_{\mu \nu (V)}^{A})_{c}%
}{\delta V_{\sigma }^{B}}+(\bar{Z}_{\sigma \varepsilon (V)}^{B})_{b}\frac{%
\delta (\bar{Z}_{\mu \nu (V)}^{A})_{c}}{\delta V_{\sigma \varepsilon }^{B}}-(%
\bar{Z}_{e(\varphi )})_{c}\frac{\delta (\bar{Z}_{\mu \nu (V)}^{A})_{b}}{%
\delta \varphi _{e}}  \notag \\
&&-(\bar{Z}_{\sigma (A)}^{m})_{c}\frac{\delta (\bar{Z}_{\mu \nu (V)}^{A})_{b}%
}{\delta A_{\sigma }^{m}}-(\bar{Z}_{m(B)}^{\sigma \varepsilon })_{c}\frac{%
\delta (\bar{Z}_{\mu \nu (V)}^{A})_{b}}{\delta B_{m}^{\sigma \varepsilon }}-(%
\bar{Z}_{\sigma (V)}^{B})_{c}\frac{\delta (\bar{Z}_{\mu \nu (V)}^{A})_{b}}{%
\delta V_{\sigma }^{B}}  \notag \\
&&-(\bar{Z}_{\sigma \varepsilon (V)}^{B})_{c}\frac{\delta (\bar{Z}_{\mu \nu
(V)}^{A})_{b}}{\delta V_{\sigma \varepsilon }^{B}}  \notag \\
&=&\lambda \left\{ M_{bc}^{d}(\bar{Z}_{\mu \nu (V)}^{A})_{d}-\tfrac{1}{3}%
\varepsilon ^{\alpha \beta \gamma \delta }\left[ f_{Mbcd}V_{\delta }^{M}-%
\tfrac{1}{4}M_{dbce}A_{\delta }^{e}\right] (\bar{Z}_{\mu \nu
(V)}^{A})_{\alpha \beta \gamma }^{d}\right.  \notag \\
&&-\tfrac{1}{2}\left[ \frac{\partial M_{bc}^{d}}{\partial \varphi _{e}}%
B_{d\alpha \beta }-\varepsilon _{\alpha \beta \gamma \delta }\left( \tfrac{1%
}{8}\frac{\partial M_{bcdf}}{\partial \varphi _{e}}A^{d\gamma }+\frac{%
\partial f_{bcf}^{M}}{\partial \varphi _{e}}V_{M}^{\gamma }\right)
A^{f\delta }\right.  \notag \\
&&\left. +\tfrac{1}{2}\varepsilon _{\alpha \beta \gamma \delta }\frac{%
\partial g_{bc}^{BC}}{\partial \varphi _{e}}V_{B}^{\gamma }V_{C}^{\delta }%
\right] (\bar{Z}_{\mu \nu (V)}^{A})_{e}^{\alpha \beta }+\left(
g_{bc}^{MB}V_{B\lambda }-f_{bcd}^{M}A_{\lambda }^{d}\right) (\bar{Z}_{\mu
\nu (V)}^{A})_{M}^{\lambda }  \notag \\
&&+\varepsilon _{\mu \nu \rho \lambda }\left[ \frac{\delta S^{\mathrm{L}}}{%
\delta H_{\rho }^{m}}\left( \frac{\partial f_{bcd}^{A}}{\partial \varphi _{m}%
}A^{d\lambda }-\frac{\partial g_{bc}^{AB}}{\partial \varphi _{m}}%
V_{B}^{\lambda }\right) +f_{bcd}^{A}\frac{\delta S^{\mathrm{L}}}{\delta
B_{d\rho \lambda }}\right.  \notag \\
&&\left. \left. -g_{bc}^{AB}\frac{\delta S^{\mathrm{L}}}{\delta V_{\rho
\lambda }^{B}}\right] \right\} ,  \label{c11}
\end{eqnarray}%
\begin{eqnarray}
&&(\bar{Z}_{e(\varphi )})_{b}\frac{\delta (\bar{Z}_{\mu \nu
(V)}^{A})_{\alpha \beta \gamma }^{c}}{\delta \varphi _{e}}+(\bar{Z}_{\sigma
(A)}^{m})_{b}\frac{\delta (\bar{Z}_{\mu \nu (V)}^{A})_{\alpha \beta \gamma
}^{c}}{\delta A_{\sigma }^{m}}+(\bar{Z}_{\sigma (V)}^{B})_{b}\frac{\delta (%
\bar{Z}_{\mu \nu (V)}^{A})_{\alpha \beta \gamma }^{c}}{\delta V_{\sigma }^{B}%
}  \notag \\
&&-(\bar{Z}_{\sigma (A)}^{m})_{\alpha \beta \gamma }^{c}\frac{\delta (\bar{Z}%
_{\mu \nu (V)}^{A})_{b}}{\delta A_{\sigma }^{m}}-(\bar{Z}_{m(B)}^{\sigma
\varepsilon })_{\alpha \beta \gamma }^{c}\frac{\delta (\bar{Z}_{\mu \nu
(V)}^{A})_{b}}{\delta B_{m}^{\sigma \varepsilon }}-(\bar{Z}_{\sigma
\varepsilon (V)}^{B})_{\alpha \beta \gamma }^{c}\frac{\delta (\bar{Z}_{\mu
\nu (V)}^{A})_{b}}{\delta V_{\sigma \varepsilon }^{B}}  \notag \\
&=&\lambda \left[ -\tfrac{1}{4}\left( \frac{\partial M_{bd}^{c}}{\partial
\varphi _{e}}A_{[\alpha }^{d}\delta _{\beta }^{\rho }\delta _{\gamma
]}^{\lambda }+\tfrac{1}{12}\frac{\partial f_{Mb}^{c}}{\partial \varphi _{e}}%
V_{[\alpha }^{M}\delta _{\beta }^{\rho }\delta _{\gamma ]}^{\lambda
}\right)
(\bar{Z}_{\mu \nu (V)}^{A})_{e\rho \lambda }\right.  \notag \\
&&+M_{eb}^{c}(\bar{Z}_{\mu \nu (V)}^{A})_{\alpha \beta \gamma }^{e}+\tfrac{1%
}{4!}\varepsilon _{\lambda \alpha \beta \gamma }f_{\;\;b}^{Mc}(\bar{Z}_{\mu
\nu (V)}^{A})_{M}^{\lambda }  \notag \\
&&\left. -\tfrac{1}{4!}\varepsilon _{\mu \nu \rho \lambda }\varepsilon
_{\sigma \alpha \beta \gamma }\sigma ^{\lambda \sigma }\frac{\partial
f_{\;\;b}^{Ac}}{\partial \varphi _{m}}\frac{\delta S^{\mathrm{L}}}{\delta
H_{\rho }^{m}}\right] ,  \label{co11}
\end{eqnarray}%
\begin{eqnarray}
&&(\bar{Z}_{e(\varphi )})_{b}\frac{\delta (\bar{Z}_{\mu \nu
(V)}^{A})_{c}^{\alpha \beta }}{\delta \varphi _{e}}-(\bar{Z}_{\sigma
\varepsilon (V)}^{B})_{c}^{\alpha \beta }\frac{\delta (\bar{Z}_{\mu \nu
(V)}^{A})_{b}}{\delta V_{\sigma \varepsilon }^{B}}-(\bar{Z}_{m(B)}^{\sigma
\varepsilon })_{c}^{\alpha \beta }\frac{\delta (\bar{Z}_{\mu \nu (V)})_{b}}{%
\delta B_{m}^{\sigma \varepsilon }}  \notag \\
&=&\lambda \frac{\partial W_{bc}}{\partial \varphi _{d}}(\bar{Z}_{\mu \nu
(V)}^{A})_{d}^{\alpha \beta },  \label{c12}
\end{eqnarray}%
\begin{eqnarray}
&&(\bar{Z}_{\sigma (A)}^{m})_{\alpha \beta \gamma }^{b}\frac{\delta (\bar{Z}%
_{\mu \nu (V)}^{A})_{\alpha ^{\prime }\beta ^{\prime }\gamma ^{\prime }}^{c}%
}{\delta A_{\sigma }^{m}}-(\bar{Z}_{\sigma (A)}^{m})_{\alpha ^{\prime }\beta
^{\prime }\gamma ^{\prime }}^{c}\frac{\delta (\bar{Z}_{\mu \nu
(V)}^{A})_{\alpha \beta \gamma }^{b}}{\delta A_{\sigma }^{m}}  \notag \\
&=&-\tfrac{\lambda }{2}\frac{\partial M^{bc}}{\partial \varphi _{e}}%
\varepsilon ^{\rho \lambda \delta \varepsilon }\varepsilon _{\delta \alpha
\beta \gamma }\varepsilon _{\varepsilon \alpha ^{\prime }\beta ^{\prime
}\gamma ^{\prime }}(\bar{Z}_{\mu \nu (V)}^{A})_{e\rho \lambda },
\label{co13}
\end{eqnarray}%
\begin{eqnarray}
&&(\bar{Z}_{e(\varphi )})_{b}\frac{\delta (\bar{Z}_{\mu \nu
(V)}^{A})_{B}^{\lambda }}{\delta \varphi _{e}}+(\bar{Z}_{\sigma (A)}^{m})_{b}%
\frac{\delta (\bar{Z}_{\mu \nu (V)}^{A})_{B}^{\lambda }}{\delta A_{\sigma
}^{m}}+(\bar{Z}_{\sigma (V)}^{C})_{b}\frac{\delta (\bar{Z}_{\mu \nu
(V)}^{A})_{B}^{\lambda }}{\delta V_{\sigma }^{C}}  \notag \\
&&-(\bar{Z}_{m(B)}^{\sigma \varepsilon })_{B}^{\lambda }\frac{\delta (\bar{Z}%
_{\mu \nu (V)}^{A})_{b}}{\delta B_{m}^{\sigma \varepsilon }}-(\bar{Z}%
_{\sigma \varepsilon (V)}^{C})_{B}^{\lambda }\frac{\delta (\bar{Z}_{\mu \nu
(V)}^{A})_{b}}{\delta V_{\sigma \varepsilon }^{C}}  \notag \\
&=&-\lambda f_{bB}^{M}(\bar{Z}_{\mu \nu (V)}^{A})_{M}^{\lambda }+\tfrac{%
\lambda }{2}\varepsilon ^{\alpha \beta \rho \lambda }\frac{\partial f_{bMB}}{%
\partial \varphi _{e}}V_{\rho }^{M}(\bar{Z}_{\mu \nu (V)}^{A})_{e\alpha
\beta }  \notag \\
&&+\lambda \sigma ^{\lambda \sigma }\varepsilon _{\mu \nu \rho \sigma }\frac{%
\partial f_{bB}^{A}}{\partial \varphi _{m}}\frac{\delta S^{\mathrm{L}}}{%
\delta H_{\rho }^{m}},  \label{c14a}
\end{eqnarray}%
\begin{eqnarray}
&&(\bar{Z}_{e(\varphi )})_{b}\frac{\delta (\bar{Z}_{\mu (H)}^{a})_{c}}{%
\delta \varphi _{e}}+(\bar{Z}_{\sigma (A)}^{m})_{b}\frac{\delta (\bar{Z}%
_{\mu (H)}^{a})_{c}}{\delta A_{\sigma }^{m}}+(\bar{Z}_{\sigma (H)}^{m})_{b}%
\frac{\delta (\bar{Z}_{\mu (H)}^{a})_{c}}{\delta H_{\sigma }^{m}}  \notag \\
&&+(\bar{Z}_{m(B)}^{\sigma \varepsilon })_{b}\frac{\delta (\bar{Z}_{\mu
(H)}^{a})_{c}}{\delta B_{m}^{\sigma \varepsilon }}+(\bar{Z}_{\sigma
(V)}^{A})_{b}\frac{\delta (\bar{Z}_{\mu (H)}^{a})_{c}}{\delta V_{\sigma }^{A}%
}+(\bar{Z}_{\sigma \varepsilon (V)}^{A})_{b}\frac{\delta (\bar{Z}_{\mu
(H)}^{a})_{c}}{\delta V_{\sigma \varepsilon }^{A}}  \notag \\
&&-(\bar{Z}_{e(\varphi )})_{c}\frac{\delta (\bar{Z}_{\mu (H)}^{a})_{b}}{%
\delta \varphi _{e}}-(\bar{Z}_{\sigma (A)}^{m})_{c}\frac{\delta (\bar{Z}%
_{\mu (H)}^{a})_{b}}{\delta A_{\sigma }^{m}}-(\bar{Z}_{\sigma (H)}^{m})_{c}%
\frac{\delta (\bar{Z}_{\mu (H)}^{a})_{b}}{\delta H_{\sigma }^{m}}  \notag \\
&&-(\bar{Z}_{m(B)}^{\sigma \varepsilon })_{c}\frac{\delta (\bar{Z}_{\mu
(H)}^{a})_{b}}{\delta B_{m}^{\sigma \varepsilon }}-(\bar{Z}_{\sigma
(V)}^{A})_{c}\frac{\delta (\bar{Z}_{\mu (H)}^{a})_{b}}{\delta V_{\sigma }^{A}%
}-(\bar{Z}_{\sigma \varepsilon (V)}^{A})_{c}\frac{\delta (\bar{Z}_{\mu
(H)}^{a})_{b}}{\delta V_{\sigma \varepsilon }^{A}}  \notag \\
&=&\lambda \left\{ M_{bc}^{d}(\bar{Z}_{\mu (H)}^{a})_{d}-\tfrac{1}{3}%
\varepsilon ^{\alpha \beta \gamma \delta }\left[ f_{Mbcd}V_{\delta }^{M}-%
\tfrac{1}{4}M_{dbce}A_{\delta }^{e}\right] (\bar{Z}_{\mu (H)}^{a})_{\alpha
\beta \gamma }^{d}\right.  \notag \\
&&-\tfrac{1}{2}\left[ \frac{\partial M_{bc}^{d}}{\partial \varphi _{e}}%
B_{d\alpha \beta }-\varepsilon _{\alpha \beta \gamma \delta }\left( \tfrac{1%
}{8}\frac{\partial M_{bcdf}}{\partial \varphi _{e}}A^{d\gamma }+\frac{%
\partial f_{bcf}^{M}}{\partial \varphi _{e}}V_{M}^{\gamma }\right)
A^{f\delta }\right.  \notag \\
&&\left. +\tfrac{1}{2}\varepsilon _{\alpha \beta \gamma \delta }\frac{%
\partial g_{bc}^{BC}}{\partial \varphi _{e}}V_{B}^{\gamma }V_{C}^{\delta }%
\right] (\bar{Z}_{\mu (H)}^{a})_{e}^{\alpha \beta }+\left(
g_{bc}^{MB}V_{B\lambda }-f_{bcd}^{M}A_{\lambda }^{d}\right) (\bar{Z}_{\mu
(H)}^{a})_{M}^{\lambda }  \notag \\
&&+\frac{\delta S^{\mathrm{L}}}{\delta H_{\nu }^{m}}\left[ \frac{\partial
^{2}M_{bc}^{d}}{\partial \varphi _{m}\partial \varphi _{a}}B_{d\mu \nu }+%
\tfrac{1}{2}\varepsilon _{\mu \nu \rho \lambda }\left( \frac{\partial
^{2}g_{bc}^{AB}}{\partial \varphi _{m}\partial \varphi _{a}}V_{A}^{\rho
}V_{B}^{\lambda }\right. \right.  \notag \\
&&\left. \left. -2\frac{\partial ^{2}f_{bcd}^{A}}{\partial \varphi
_{m}\partial \varphi _{a}}V_{A}^{\rho }A^{d\lambda }-\tfrac{1}{4}\frac{%
\partial ^{2}M_{bcde}}{\partial \varphi _{m}\partial \varphi _{a}}A^{d\rho
}A^{e\lambda }\right) \right]  \notag \\
&&+\varepsilon _{\mu \nu \rho \lambda }\frac{\delta S^{\mathrm{L}}}{\delta
B_{d\rho \lambda }}\left( \frac{\partial f_{bcd}^{A}}{\partial \varphi _{a}}%
V_{A}^{\nu }-\tfrac{1}{8}\frac{\partial M_{bcde}}{\partial \varphi _{a}}%
A^{e\nu }\right)  \notag \\
&&\left. +\varepsilon _{\mu \nu \rho \lambda }\frac{\delta S^{\mathrm{L}}}{%
\delta V_{\rho \lambda }^{A}}\left( \frac{\partial
g_{bc}^{AB}}{\partial \varphi _{a}}V_{B}^{\nu }-\frac{\partial
f_{bcd}^{A}}{\partial \varphi
_{a}}A^{d\nu }\right) +\frac{\partial M_{bc}^{d}}{\partial \varphi _{a}}%
\frac{\delta S^{\mathrm{L}}}{\delta A^{d\mu }}\right\} ,  \label{c14}
\end{eqnarray}%
\begin{eqnarray}
&&(\bar{Z}_{e(\varphi )})_{b}\frac{\delta (\bar{Z}_{\mu (H)}^{a})_{\alpha
\beta \gamma }^{c}}{\delta \varphi _{e}}+(\bar{Z}_{\sigma (A)}^{m})_{b}\frac{%
\delta (\bar{Z}_{\mu (H)}^{a})_{\alpha \beta \gamma }^{c}}{\delta A_{\sigma
}^{m}}+(\bar{Z}_{m(B)}^{\sigma \varepsilon })_{b}\frac{\delta (\bar{Z}_{\mu
(H)}^{a})_{\alpha \beta \gamma }^{c}}{\delta B_{m}^{\sigma \varepsilon }}
\notag \\
&&+(\bar{Z}_{\varepsilon (V)}^{A})_{b}\frac{\delta (\bar{Z}_{\mu
(H)}^{a})_{\alpha \beta \gamma }^{c}}{\delta V_{\varepsilon }^{A}}-(\bar{Z}%
_{\sigma (A)}^{m})_{\alpha \beta \gamma }^{c}\frac{\delta (\bar{Z}_{\mu
(H)}^{a})_{b}}{\delta A_{\sigma }^{m}}  \notag \\
&&-(\bar{Z}_{\sigma (H)}^{m})_{\alpha \beta \gamma }^{c}\frac{\delta (\bar{Z}%
_{\mu (H)}^{a})_{b}}{\delta H_{\sigma }^{m}}-(\bar{Z}_{m(B)}^{\sigma
\varepsilon })_{\alpha \beta \gamma }^{c}\frac{\delta (\bar{Z}_{\mu
(H)}^{a})_{b}}{\delta B_{m}^{\sigma \varepsilon }}  \notag \\
&=&\lambda \left\{ -\tfrac{1}{4}\left( \frac{\partial M_{bd}^{c}}{\partial
\varphi _{e}}A_{[\alpha }^{d}\delta _{\beta }^{\rho }\delta _{\gamma
]}^{\lambda }+\tfrac{1}{12}\frac{\partial f_{Ab}^{c}}{\partial \varphi _{e}}%
V_{[\alpha }^{A}\delta _{\beta }^{\rho }\delta _{\gamma ]}^{\lambda }\right)
(\bar{Z}_{\mu (H)}^{a})_{e\rho \lambda }\right.  \notag \\
&&+M_{eb}^{c}(\bar{Z}_{\mu (H)}^{a})_{\alpha \beta \gamma }^{e}+\tfrac{1}{4!}%
\varepsilon _{\lambda \alpha \beta \gamma }f_{\;\;b}^{Mc}(\bar{Z}_{\mu
(H)}^{a})_{M}^{\lambda }  \notag \\
&&+\tfrac{1}{2}\sigma _{\mu \lbrack \alpha }^{\left. {}\right. }\delta
_{\beta }^{\nu }\delta _{\gamma ]}^{\rho }\left[ \frac{\delta S^{\mathrm{L}}%
}{\delta H^{m\nu }}\left( \frac{\partial ^{2}M_{bd}^{c}}{\partial \varphi
_{a}\partial \varphi _{m}}A_{\rho }^{d}+\tfrac{1}{12}\frac{\partial
^{2}f_{Ab}^{c}}{\partial \varphi _{a}\partial \varphi _{m}}V_{\rho
}^{A}\right) \right.  \notag \\
&&\left. \left. +\frac{\partial M_{bd}^{c}}{\partial \varphi _{a}}\frac{%
\delta S^{\mathrm{L}}}{\delta B_{d}^{\nu \rho }}+\tfrac{1}{12}\frac{\partial
f_{Ab}^{c}}{\partial \varphi _{a}}\frac{\delta S^{\mathrm{L}}}{\delta
V_{A}^{\nu \rho }}\right] \right\} ,  \label{co15}
\end{eqnarray}%
\begin{eqnarray}
&&(\bar{Z}_{e(\varphi )})_{b}\frac{\delta (\bar{Z}_{\mu
(H)}^{a})_{c}^{\alpha \beta }}{\delta \varphi _{e}}+(\bar{Z}_{\sigma
(A)}^{m})_{b}\frac{\delta (\bar{Z}_{\mu (H)}^{a})_{c}^{\alpha \beta }}{%
\delta A_{\sigma }^{m}}+(\bar{Z}_{\sigma (V)}^{A})_{b}\frac{\delta (\bar{Z}%
_{\mu (H)}^{a})_{c}^{\alpha \beta }}{\delta V_{\sigma }^{A}}  \notag \\
&&-(\bar{Z}_{\sigma (H)}^{m})_{c}^{\alpha \beta }\frac{\delta (\bar{Z}_{\mu
(H)}^{a})_{b}}{\delta H_{\sigma }^{m}}-(\bar{Z}_{m(B)}^{\sigma \varepsilon
})_{c}^{\alpha \beta }\frac{\delta (\bar{Z}_{\mu (H)}^{a})_{b}}{\delta
B_{m}^{\sigma \varepsilon }}-(\bar{Z}_{\sigma \varepsilon
(V)}^{A})_{c}^{\alpha \beta }\frac{\delta (\bar{Z}_{\mu (H)}^{a})_{b}}{%
\delta V_{\sigma \varepsilon }^{A}}  \notag \\
&=&\lambda \left[ \frac{\partial W_{bc}}{\partial \varphi _{d}}(\bar{Z}_{\mu
(H)}^{a})_{d}^{\alpha \beta }-\frac{\partial ^{2}W_{bc}}{\partial \varphi
_{a}\partial \varphi _{d}}\frac{\delta S^{\mathrm{L}}}{\delta H_{\nu }^{d}}%
\delta _{\mu }^{[\alpha }\delta _{\nu }^{\beta ]}\right] ,  \label{co16}
\end{eqnarray}%
\begin{eqnarray}
&&(\bar{Z}_{\sigma (A)}^{m})_{\alpha \beta \gamma }^{b}\frac{\delta (\bar{Z}%
_{\mu (H)}^{a})_{\alpha ^{\prime }\beta ^{\prime }\gamma ^{\prime }}^{c}}{%
\delta A_{\sigma }^{m}}+(\bar{Z}_{m(B)}^{\sigma \varepsilon })_{\alpha \beta
\gamma }^{b}\frac{\delta (\bar{Z}_{\mu (H)}^{a})_{\alpha ^{\prime }\beta
^{\prime }\gamma ^{\prime }}^{c}}{\delta B_{m}^{\sigma \varepsilon }}  \notag
\\
&&-(\bar{Z}_{\sigma (A)}^{m})_{\alpha ^{\prime }\beta ^{\prime }\gamma
^{\prime }}^{c}\frac{\delta (\bar{Z}_{\mu (H)}^{a})_{\alpha \beta \gamma
}^{b}}{\delta A_{\sigma }^{m}}-(\bar{Z}_{m(B)}^{\sigma \varepsilon
})_{\alpha ^{\prime }\beta ^{\prime }\gamma ^{\prime }}^{c}\frac{\delta (%
\bar{Z}_{\mu (H)}^{a})_{\alpha \beta \gamma }^{b}}{\delta B_{m}^{\sigma
\varepsilon }}  \notag \\
&=&-\tfrac{\lambda }{2}\frac{\partial M^{bc}}{\partial \varphi _{e}}%
\varepsilon ^{\rho \lambda \delta \varepsilon }\varepsilon _{\delta \alpha
\beta \gamma }\varepsilon _{\varepsilon \alpha ^{\prime }\beta ^{\prime
}\gamma ^{\prime }}(\bar{Z}_{\mu (H)}^{a})_{e\rho \lambda }  \notag \\
&&+\lambda \varepsilon _{\mu \nu \rho \lambda }\varepsilon _{\delta \alpha
\beta \gamma }\varepsilon _{\varepsilon \alpha ^{\prime }\beta ^{\prime
}\gamma ^{\prime }}\sigma ^{\rho \delta }\sigma ^{\varepsilon \lambda }\frac{%
\partial ^{2}M^{bc}}{\partial \varphi _{a}\partial \varphi _{d}}\frac{\delta
S^{\mathrm{L}}}{\delta H_{\nu }^{d}},  \label{co17}
\end{eqnarray}%
\begin{eqnarray}
&&(\bar{Z}_{e(\varphi )})_{b}\frac{\delta (\bar{Z}_{\mu
(H)}^{a})_{B}^{\lambda }}{\delta \varphi _{e}}+(\bar{Z}_{\sigma (A)}^{m})_{b}%
\frac{\delta (\bar{Z}_{\mu (H)}^{a})_{B}^{\lambda }}{\delta A_{\sigma }^{m}}%
+(\bar{Z}_{\sigma (V)}^{C})_{b}\frac{\delta (\bar{Z}_{\mu
(H)}^{a})_{B}^{\lambda }}{\delta V_{\sigma }^{C}}  \notag \\
&&-(\bar{Z}_{\sigma \varepsilon (V)}^{C})_{B}^{\lambda }\frac{\delta (\bar{Z}%
_{\mu (H)}^{a})_{b}}{\delta V_{\sigma \varepsilon }^{C}}-(\bar{Z}_{\sigma
(H)}^{m})_{B}^{\lambda }\frac{\delta (\bar{Z}_{\mu (H)}^{a})_{b}}{\delta
H_{\sigma }^{m}}-(\bar{Z}_{m(B)}^{\sigma \varepsilon })_{B}^{\lambda }\frac{%
\delta (\bar{Z}_{\mu (H)}^{a})_{b}}{\delta B_{m}^{\sigma \varepsilon }}
\notag \\
&=&-\lambda f_{bB}^{M}(\bar{Z}_{\mu (H)}^{a})_{M}^{\lambda }+\tfrac{\lambda
}{2}\varepsilon ^{\alpha \beta \rho \lambda }\frac{\partial f_{bMB}}{%
\partial \varphi _{e}}V_{\rho }^{M}(\bar{Z}_{\mu (H)}^{a})_{e\alpha \beta }
\notag \\
&&-\lambda \varepsilon _{\mu \nu \rho \sigma }\sigma ^{\sigma \lambda
}\left( \frac{\partial ^{2}f_{bMB}}{\partial \varphi _{a}\partial \varphi
_{e}}V^{M\rho }\frac{\delta S^{\mathrm{L}}}{\delta H_{\nu }^{e}}+\frac{%
\partial f_{bMB}}{\partial \varphi _{a}}\frac{\delta S^{\mathrm{L}}}{\delta
V_{M\nu \rho }}\right) .  \label{co19}
\end{eqnarray}


\begin{thebibliography}{99}
\bibitem{4a} B. Voronov and I. V. Tyutin, \textit{Formulation of gauge
theories of general form. I}, Theor. Math. Phys. \textbf{50} (1982) 218.

\bibitem{4b} B. Voronov and I. V. Tyutin, \textit{Formulation of gauge
theories of general form. II. Gauge invariant renormalizability and
renormalization structure}, Theor. Math. Phys. \textbf{52} (1982) 628.

\bibitem{4c} J. Gomis and S. Weinberg, \textit{Are nonrenormalizable gauge
theories renormalizable?}, Nucl. Phys. \textbf{B469} (1996) 473
[hep-th/9510087].

\bibitem{4d} S. Weinberg, \textit{The Quantum Theory of Fields}, Cambridge
University Press, Cambridge (1996).

\bibitem{5} O. Piguet and S. P. Sorella, \textit{Algebraic Renormalization:
Perturbative Renormalization, Symmetries and Anomalies}, Lecture Notes in
Physics, Springer Verlag, Berlin, Vol. \textbf{28} (1995).

\bibitem{6a} P. S. Howe, V. Lindstr\"{o}m and P. White, \textit{Anomalies
and renormalization in the BRST-BV framework}, Phys. Lett. \textbf{B246}
(1990) 430.

\bibitem{6b} W. Troost, P. van Nieuwenhuizen and A. van Proeyen, \textit{%
Anomalies and the Batalin-Vilkovisky lagrangian formalism}, Nucl.
Phys. \textbf{B333} (1990) 727.

\bibitem{6c} G. Barnich and M. Henneaux, \textit{Renormalization of gauge
invariant operators and anomalies in Yang-Mills theory}, Phys. Rev. Lett.
\textbf{72} (1994) 1588 [hep-th/9312206].

\bibitem{6d} G. Barnich, \textit{Perturbative gauge anomalies in the
Hamiltonian formalism: a cohomological analysis}, Mod. Phys. Lett. \textbf{A9%
} (1994) 665 [hep-th/9310167].

\bibitem{6e} G. Barnich, \textit{Higher order cohomological restrictions on
anomalies and counterterms}, Phys. Lett. \textbf{B419} (1998) 211
[hep-th/9710162].

\bibitem{7} F. Brandt, M. Henneaux and A. Wilch, \textit{Global symmetries
in the antifield formalism}, Phys. Lett. \textbf{B387} (1996) 320
[hep-th/9606172].

\bibitem{7a1} R. Arnowitt and S. Deser, \textit{Interaction between gauge
vector fields}, Nucl. Phys. \textbf{49} (1963) 133.

\bibitem{7a2} J. Fang and C. Fronsdal, \textit{Deformation of gauge groups.
Gravitation}, J. Math. Phys. \textbf{20} (1979) 2264.

\bibitem{7a3} F. A. Berends, G. H. Burgers and H. Van Dam, \textit{On spin
three selfinteractions}, Z. Phys. \textbf{C24} (1984) 247.

\bibitem{7a4} F. A. Berends, G. H. Burgers and H. Van Dam, \textit{On the
theoretical problems in constructing interactions involving higher spin
massless particles}, Nucl. Phys. \textbf{B260} (1985) 295.

\bibitem{7a5} A. K. H. Bengtsson, \textit{On gauge invariance for spin-3
fields}, Phys. Rev. \textbf{D32} (1985) 2031.

\bibitem{8b1} J. D. Stasheff, \textit{Deformation theory and the
Batalin-Vilkovisky master equation}, in Deformation theory and symplectic
geometry, Proceedings of Ascona meeting, June 1996, Eds. D. Sternheimer, J.
Rawnsley, and S. Gutt, Math. Physics Studies \textbf{20}, 271-284, Kluwer
Acad. Publ., Dordrecht (1997) [q-alg/9702012]

\bibitem{8b2} J. D. Stasheff, \textit{The (secret?) homological algebra of
the Batalin-Vilkovisky approach}, in Secondary Calculus and Cohomological
Physics, Proceedings of Moscow meeting, August 1997, Eds. M. Henneaux, J.
Krasil'shchik, A. Vinogradov, Contemporary Mathematics, vol. \textbf{219},
American Mathematical Society (1998) [hep-th/9712157].

\bibitem{8b3} J. A. Garcia and B. Knaepen, \textit{Couplings between
generalized gauge fields}, Phys. Lett. \textbf{B441} (1998) 198
[hep-th/9807016].

\bibitem{def} G. Barnich and M. Henneaux, \textit{Consistent couplings
between fields with a gauge freedom and deformations of the master equation}%
, Phys. Lett. \textbf{B311} (1993) 123 [hep-th/9304057].

\bibitem{contempmath} M. Henneaux, \textit{Consistent interactions between
gauge fields: the cohomological approach}, Contemp. Math. \textbf{219}
(1998) 93 [hep-th/9712226].

\bibitem{birmingham91} D. Birmingham, M. Blau, M. Rakowski and G. Thompson,
\textit{Topological field theory}, Phys. Rept. \textbf{209} (1991) 129.

\bibitem{stroblspec} P. Schaller and T. Strobl, \textit{Poisson structure
induced (topological) field theories}, Mod. Phys. Lett. \textbf{A9} (1994)
3129 [hep-th/9405110].

\bibitem{psmikeda94} N. Ikeda, \textit{Two-dimensional gravity and nonlinear
gauge theory}, Annals Phys. \textbf{235} (1994) 435 [hep-th/9312059].

\bibitem{psmstrobl95} A. Yu. Alekseev, P. Schaller and T. Strobl, \textit{%
The topological G/G WZW model in the generalized momentum representation},
Phys. Rev. \textbf{D52} (1995) 7146 [hep-th/9505012].

\bibitem{psmstroblCQG961} T. Kl\"{o}sch and T. Strobl, \textit{Classical and
quantum gravity in 1+1 dimensions: I. A unifying approach}, Class.
Quantum Grav. \textbf{13} (1996) 965 [gr-qc/9508020]; Erratum-ibid.
\textbf{14} (1997) 825.

\bibitem{psmstroblCQG962} T. Kl\"{o}sch and T. Strobl, \textit{Classical and
quantum gravity in 1+1 dimensions: II. The universal coverings}, Class.
Quantum Grav. \textbf{13} (1996) 2395 [gr-qc/9511081].

\bibitem{psmstrobl97} T. Kl\"{o}sch and T. Strobl, \textit{Classical and
quantum gravity in 1+1 dimensions: III. Solutions of arbitrary topology},
Class. Quantum Grav. \textbf{14} (1997) 1689 [hep-th/9607226].

\bibitem{psmcattaneo2000} A. S. Cattaneo and G. Felder, \textit{A path
integral approach to the Kontsevich quantization formula}, Commun. Math.
Phys. \textbf{212} (2000) 591 [math.QA/9902090].

\bibitem{psmcattaneo2001} A. S. Cattaneo and G. Felder, \textit{Poisson
Sigma models and deformation quantization}, Mod. Phys. Lett. \textbf{A16}
(2001) 179 [hep-th/0102208].

\bibitem{grav2teit83} C. Teitelboim, \textit{Gravitation and hamiltonian
structure in two spacetime dimensions}, Phys. Lett. \textbf{B126}
(1983) 41.

\bibitem{grav2jackiw85} R. Jackiw, \textit{Lower dimensional gravity}, Nucl.
Phys. \textbf{B252} (1985) 343.

\bibitem{grav2katanaev86} M. O. Katanayev and I. V. Volovich, \textit{String
model with dynamical geometry and torsion}, Phys. Lett. \textbf{B175} (1986)
413.

\bibitem{grav2brown88} J. Brown, \textit{Lower Dimensional Gravity}, World
Scientific, Singapore (1988).

\bibitem{grav2katanaev90} M. O. Katanaev and I. V. Volovich, \textit{%
Two-dimensional gravity with dynamical torsion and strings}, Annals Phys.
\textbf{197} (1990) 1.

\bibitem{grav2schmidt} H.-J. Schmidt, \textit{Scale-invariant gravity in two
dimensions}, J. Math. Phys. \textbf{32} (1991) 1562.

\bibitem{grav2solod} S. N. Solodukhin, \textit{Topological 2D
Riemann-Cartan-Weyl gravity}, Class. Quantum Grav. \textbf{10} (1993) 1011.

\bibitem{grav2ikedaizawa90} N. Ikeda and K. I. Izawa, \textit{General form
of dilaton gravity and nonlinear gauge theory}, Prog. Theor. Phys. \textbf{90%
} (1993) 237 [hep-th/9304012].

\bibitem{grav2strobl94} T. Strobl, \textit{Dirac quantization of
gravity-Yang-Mills systems in 1+1 dimensions}, Phys. Rev. \textbf{D50}
(1994) 7346 [hep-th/9403121].

\bibitem{grav2grumvassil02} D. Grumiller, W. Kummer and D. V. Vassilevich,
\textit{Dilaton gravity in two dimensions}, Phys. Rept. \textbf{369} (2002)
327 [hep-th/0204253].

\bibitem{grav2strobl00} T. Strobl, \textit{Gravity in two spacetime
dimensions}, Habilitation thesis RWTH Aachen, May 1999 [hep-th/0011240].

\bibitem{ezawa} K. Ezawa, \textit{Ashtekar's formulation for $N=1,2$
supergravities as ``constrained" BF theories}, Prog. Theor. Phys. \textbf{95}
(1996) 863 [hep-th/9511047].

\bibitem{freidel} L. Freidel, K. Krasnov and R. Puzio, \textit{BF
description of higher-dimensional gravity theories}, Adv. Theor. Math. Phys.
\textbf{3} (1999) 1289 [hep-th/9901069].

\bibitem{smolin} L. Smolin, \textit{A holographic formulation of quantum
general relativity}, Phys. Rev. \textbf{D61} (2000) 084007 [hep-th/9808191].

\bibitem{ling} Y. Ling and L. Smolin, \textit{Holographic formulation of
quantum supergravity}, Phys. Rev. \textbf{D63} (2001) 064010
[hep-th/0009018].

\bibitem{defBFizawa2000} K.-I. Izawa, \textit{On nonlinear gauge theory from
a deformation theory perspective}, Prog. Theor. Phys. \textbf{103} (2000)
225 [hep-th/9910133].

\bibitem{defBFmpla} C. Bizdadea, \textit{Note on two-dimensional nonlinear
gauge theories}, Mod. Phys. Lett. \textbf{A15} (2000) 2047
[hep-th/0201059].

\bibitem{defBFikeda00} N. Ikeda, \textit{A deformation of three dimensional
BF theory}, J. High Energy Phys. \textbf{11} (2000) 009
[hep-th/0010096].

\bibitem{defBFikeda01} N. Ikeda, \textit{Deformation of BF theories,
topological open membrane and a generalization of the star
deformation}, J. High Energy Phys. \textbf{07} (2001) 037
[hep-th/0105286].

\bibitem{defBFijmpa} C. Bizdadea, E. M. Cioroianu and S. O. Saliu, \textit{%
Hamiltonian cohomological derivation of four-dimensional nonlinear gauge
theories}, Int. J. Mod. Phys. \textbf{A17} (2002) 2191 [hep-th/0206186].

\bibitem{defBFjhep} C. Bizdadea, C. C. Ciob\^{\i}rc\u{a}, E. M. Cioroianu,
S. O. Saliu and S. C. S\u{a}raru, \textit{Hamiltonian BRST deformation of a
class of n-dimensional BF-type theories}, J. High Energy Phys. \textbf{01%
} (2003) 049 [hep-th/0302037].

\bibitem{defBFijmpajuvi06} E. M. Cioroianu and S. C. S\u{a}raru, \textit{%
PT-symmetry breaking hamiltonian interactions in BF models}, Int. J.
Mod. Phys. \textbf{A21} (2006) 2573 [hep-th/0606164].

\bibitem{defBFepjc} C. Bizdadea, E. M. Cioroianu, S. O. Saliu and
S. C. S\u{a}raru, \textit{Couplings of a collection of BF models to matter theories}%
, Eur. Phys. J. \textbf{C41} (2005) 401 [hep-th/0508037].

\bibitem{defBFikeda03} N. Ikeda, \textit{Chern-Simons gauge theory coupled
with BF theory}, Int. J. Mod. Phys. \textbf{A18} (2003) 2689
[hep-th/0203043].

\bibitem{defBFijmpajuvi04} E. M. Cioroianu and S. C. S\u{a}raru, \textit{%
Two-dimensional interactions between a BF-type theory and a collection of
vector fields}, Int. J. Mod. Phys. \textbf{A19} (2004) 4101 [hep-th/0501056].

\bibitem{otherBFikeda02} N. Ikeda, \textit{Topological field theories and
geometry of Batalin-Vilkovisky algebras}, J. High Energy Phys. \textbf{10%
} (2002) 076 [hep-th/0209042].

\bibitem{otherBFikedaizawa04} N. Ikeda and K.-I. Izawa, \textit{Dimensional
reduction of nonlinear gauge theories}, J. High Energy Phys.
\textbf{09} (2004) 030 [hep-th/0407243].

\bibitem{otherBFikeda06} N. Ikeda, \textit{Deformation of Batalin-Vilkovisky
structures} [math.SG/0604157].

\bibitem{gen2} G. Barnich, F. Brandt and M. Henneaux, \textit{Local BRST
cohomology in the antifield formalism: II. Application to Yang-Mills theory}%
, Commun. Math. Phys. \textbf{174} (1995) 93 [hep-th/9405194].

\bibitem{gen1a} G. Barnich, F. Brandt and M. Henneaux, \textit{Local BRST
cohomology in the antifield formalism: I. General theorems}, Commun. Math.
Phys. \textbf{174} (1995) 57 [hep-th/9405109].

\bibitem{gen1b} G. Barnich, F. Brandt and M. Henneaux, \textit{Local BRST
cohomology in gauge theories}, Phys. Rept. \textbf{338} (2000) 439
[hep-th/0002245].

\bibitem{henfreedman} M. Henneaux, \textit{Uniqueness of the
Freedman-Townsend interaction vertex for two-form gauge fields},
Phys. Lett. \textbf{B368} (1996) 83 [hep-th/9511145].
\end{thebibliography}
\end{document}